\documentclass[11pt,a4paper]{article}

\usepackage{jheppub}
\usepackage[utf8]{inputenc}
\usepackage{amsmath}
\usepackage{amsthm}
\usepackage{amssymb}
\usepackage{enumitem}   
\usepackage{url}
\usepackage{mathtools}
\usepackage{slashed}
\usepackage{multirow}
\usepackage{xcolor}
\usepackage{slashed}
\usepackage{multirow}
\usepackage[toc,page]{appendix}
\usepackage{hyperref}

\usepackage{amsmath, amsfonts, amssymb}
\usepackage{graphicx}
\usepackage{color}
\usepackage{enumerate}
\usepackage{hyperref}
\usepackage{latexsym}
\usepackage{enumerate}
\usepackage{soul}
\usepackage[normalem]{ulem}
\usepackage{wasysym}
\usepackage{makecell}
\usepackage{slashed} 
\usepackage{comment,verbatim}
\usepackage{amsmath}
\usepackage{subfig}

\newcommand{\be}{\begin{equation}}
\newcommand{\ee}{\end{equation}}

\begin{document}



\title{
Friction on ALP domain walls and gravitational waves}

\author[a]{Simone Blasi,}
\author[a]{Alberto Mariotti,}
\author[a]{A\"aron Rase,}
\author[b,c]{Alexander Sevrin,}
\author[a,c]{Kevin Turbang}

\affiliation[a]{Theoretische Natuurkunde and IIHE/ELEM, Vrije Universiteit Brussel, \& The  International Solvay Institutes, Pleinlaan 2, B-1050 Brussels, Belgium}
\affiliation[b]{Theoretische Natuurkunde, Vrije Universiteit Brussel, \& The  International Solvay Institutes, Pleinlaan 2, B-1050 Brussels, Belgium}
\affiliation[c]{Universiteit Antwerpen, Prinsstraat 13, 2000 Antwerpen, Belgium}

\emailAdd{simone.blasi@vub.be}
\emailAdd{alberto.mariotti@vub.be}
\emailAdd{aaron.rase@vub.be}
\emailAdd{alexandre.sevrin@vub.be}
\emailAdd{kevin.turbang@vub.be}


\abstract{
We study the early Universe evolution of axion--like particle (ALP) domain walls taking into account the effect of friction from particles in the surrounding plasma, 
including 
the case of 
particles 
in thermal equilibrium and frozen out species.
We characterize the friction force from 
interactions within the ALP effective theory, providing new results for the fermion contribution as well as identifying simple conditions for friction to be relevant during the domain wall life time. When friction dominates, the domain wall network departs from the standard scaling regime and the corresponding gravitational wave emission is affected. As a relevant example, we show how this can be the case for ALP domain walls emitting at the typical frequencies of Pulsar Timing Array experiments, when the ALP couples to the SM leptons. We then move to a general exploration of the gravitational wave prospects in the ALP parameter space.
We finally illustrate how the gravitational wave signal from ALP domain walls is correlated with the quality of the underlying $U(1)$ symmetry.
}

\maketitle

\section{Introduction}
\label{sec:introduction}

The discovery of gravitational waves by the LIGO-Virgo collaboration\,\cite{LIGOScientific:2016aoc} has opened a new window of exploration of our Universe.
In this context, the detection of a stochastic gravitational wave background (SGWB) of cosmological origin would represent a milestone in understanding the early stages of the Universe (see e.g. \cite{Christensen:2018iqi} for a review).
It is therefore essential to study the possible sources of a SGWB of cosmological origin and how they can impart information about fundamental physics.

In particular, strong first--order phase transitions in the early Universe can generate a SGWB during the dynamics associated to bubble nucleation.
Furthermore, the symmetry--breaking pattern involved in the phase transition (independently of the strength) can lead to the formation of defects through the Kibble mechanism \cite{Kibble:1976sj} according to the topology of the vacuum manifold.
While the SM predicts no stable defects, they can be formed in beyond the SM (BSM) 
scenarios due to the breaking of new continuous or discrete symmetries. 
Topological defects are known to be a strong source of SGWB, the most studied example being cosmic strings (see e.g. \cite{Vilenkin:2000jqa})
arising from a non--trivial first homotopy group.

In this paper we will instead focus on domain walls (DWs), two--dimensional defects originating in scenarios with a spontaneously broken discrete symmetry leading to a disconnected vacuum manifold\,\cite{Zeldovich:1974uw,Kibble:1976sj,Vilenkin:2000jqa}.
DWs can be powerful sources of a SGWB, as confirmed by numerical simulations \cite{Hiramatsu:2012sc,Hiramatsu:2013qaa,Saikawa:2017hiv}, 
and their signatures in connection with BSM models have been subject of several recent investigations
\cite{Gelmini:2021yzu,Gelmini:2020bqg,Gelmini:2022nim,Craig:2020bnv,ZambujalFerreira:2021cte,
Babichev:2021uvl,Barman:2022yos,Wu:2022tpe,Borah:2022wdy,Wu:2022stu,Ferreira:2022zzo,Fornal:2022qim}.

DWs emerge naturally in models addressing the strong CP problem by the Peccei--Quinn (PQ) mechanism \cite{Peccei:1977hh,Peccei:1977ur,Weinberg:1977ma,Wilczek:1977pj}. The reason is that, taking into account the anomaly of the global $U(1)_\text{PQ}$ under the color group, the axion potential exhibits a discrete symmetry spontaneously broken in the vacuum. This implies the formation of DWs attached to the strings forming at the PQ--breaking scale, which can be topologically stable depending on the anomaly coefficient, or the so--called domain wall number.
More generally, one can 
consider axion--like particles (ALPs) whose mass is not tied to the QCD confinement scale.
These particles can arise in String Theory \cite{Svrcek:2006yi,Arvanitaki:2009fg}, or as heavy QCD axions (see e.g. \cite{Holdom:1982ex,Holdom:1985vx,Flynn:1987rs,Rubakov:1997vp,Choi:1998ep,Berezhiani:2000gh,Hook:2014cda,Fukuda:2015ana,Dimopoulos:2016lvn,Agrawal:2017ksf,Gaillard:2018xgk}), and they can be dark matter candidate \cite{Preskill:1982cy,Abbott:1982af,Dine:1981rt,Dine:1982ah,Arias:2012az} 
and have interesting phenomenology (see e.g. \cite{Ringwald:2012hr,Marsh:2015xka,Bauer:2017ris,Brivio:2017ije,CidVidal:2018blh,Bauer:2020jbp,Goh:2008xz,Bellazzini:2017neg,Belyaev:2015hgo}
and also \cite{DiLuzio:2020wdo,Choi:2020rgn} for recent reviews).
In ALP models one can similarly expect the formation of a string--wall network.

If absolutely stable, DWs can come to dominate the energy density of the Universe \cite{Sikivie:1982qv}, potentially constituting a cosmological problem.
However, this is not a real issue here as global symmetries are not expected to be exact
and can be explicitly broken for instance by higher-dimensional operators\,\cite{Barr:1992qq,Kamionkowski:1992ax,Kamionkowski:1992mf,Holman:1992us,Berezhiani:1992pq,Ghigna:1992iv,Senjanovic:1993uz,Dobrescu:1996jp,Banks:2010zn}. 
This naturally induces a bias for the axion potential, leading to the decay of the DW network\footnote{Notice that in the case of the QCD axion,
such operators can spoil the solution to the strong CP problem leading to a tension with experimental constraints.}.

In this paper we will consider ALP DWs as our physics case,
and investigate the impact of particle friction on the DW evolution. Friction will generically slow down the average wall velocity in the network, potentially leading to significant departure from the standard scaling regime, with several phenomenological implications for GWs,
see e.g.\,\cite{NAKAYAMA2017500}, and particle production.

Unlike the case of bubble walls during first order phase transitions, particles in the plasma will have approximately the same mass on the two sides of the DW. This is because of the defining DW symmetry relating the disconnected vacua, which is only broken by a small bias term. In addition, as the DW motion is not accelerated by vacuum pressure (up to small bias corrections) but rather by its tension force, DWs are only mildly relativistic with $\gamma \sim 1$. Therefore, the leading--order contribution to friction arises as a consequence of plasma particles reflecting on the wall surface, see e.g.\,\cite{Arnold:1993wc}. 

For ALP models, the coupling structure between the background DW and the particles in the plasma is fixed by symmetry arguments within the ALP effective field theory (EFT). 
Furthermore, the DW formation in ALP models typically occurs at temperatures much larger than the ALP mass (which sets the DW width). This means that particles scattering off the wall can have a very large momentum compared to the DW width, opening a new kinematic regime for friction.
These qualitative differences compared to generic scalar theories will lead to new features in the friction force with respect to previous studies\,\cite{Abel:1995wk,Vilenkin:2000jqa}, such as a different scaling with the temperature. Exploration of friction in QCD axion models was initiated in \cite{Huang:1985tt}. Here we present a new, detailed calculation for the friction force in the context of ALPs by considering reflection from fermions (possibly being dark matter), and comment on ALP self--reflection.

We will then determine the parameter space where friction can be relevant depending
on the size of the effective couplings, finding that friction can affect both the early and the late evolution of the DW network when the GW emission is maximal. As a relevant example of the latter, we show how friction from SM fermions can affect the DW interpretation of the signal observed at Pulsar Timing Array (PTA) experiments\,\cite{NANOGrav:2020bcs,Chen:2021rqp,Goncharov:2021oub}.

As a byproduct of our study, we will also point out that a certain quality is required for the global ALP $U(1)$ in order to generate a large SGWB signal. 
For instance, a naive dimension five Planck suppressed operator makes the ALP DWs so short-lived that the resulting GWs are undetectable, even at future experiments.
On the other hand, observable GWs are compatible with dimension six operators or larger.

The paper is organized as follows: in Section \ref{sec:rev} we review basic aspects about domain walls (their dynamics, the bias, the gravitational wave spectrum), specializing to the DWs arising in ALP models.
In Section \ref{sec:3} we explore the importance of friction for ALP domain walls. We review the formalism to compute the pressure from particle reflection off the wall, and then we 
investigate in detail the case of friction from a fermion coupled to the ALP, and from the ALP itself. We conclude this section by showing that friction from SM particles (specifically from the leptons)
can play a significant role in ALP domain walls whose gravitational wave signal peaks at the frequency relevant for PTA experiments such as NANOGrav.
In Section \ref{sec: VOS} we employ velocity-dependent one-scale (VOS) equations to estimate the gravitational wave signal during friction and we explore in detail the parameter space for ALP models, showing which portion 
can be probed by current and future gravitational wave experiments. During this analysis, we will highlight the fact that a certain quality is required for the symmetry underlying the ALP model for the gravitational wave signal to be detectable.

\section{Domain walls from ALPs}
\label{sec:rev}

Domain walls are topological defects that arise in models where a 
discrete symmetry is spontaneously broken (see\,\cite{Vilenkin:2000jqa,Vilenkin:1984ib} for comprehensive reviews).
More precisely, they appear when the vacuum manifold 
$\mathcal{M}$ has a non-trivial homotopy group $\pi_0(\mathcal{M})$.
\footnote{In fact, depending on the non-trivial homotopy group one can create domain walls ($\pi_0(\mathcal{M})$), 
strings ($\pi_1(\mathcal{M}$)), monopoles ($\pi_2(\mathcal{M}$)).}

Assuming that the discrete symmetry is restored at high temperatures,
domain walls get formed at the discrete symmetry breaking through 
the Kibble mechanism\,\cite{Kibble:1976sj}.
When the Universe cools down below the critical temperature, uncorrelated patches
in space will randomly choose one of the degenerate vacua. Once thermal
fluctuations become sufficiently suppressed, this choice cannot be undone
and domains can be considered formed. At the boundary between different domains,
the field will be trapped at the maximum of the potential 
leading to a large energy density localized in a two--dimensional surface, 
the domain wall.

A very appealing way in which
discrete symmetries can emerge is the case of anomalous global symmetries
as for DWs arising in axion models\,\cite{Sikivie:1982qv,Vilenkin:1982ks}.
In general, we can define an axion--like particle (ALP) as the 
pseudo Nambu--Goldstone boson arising from the spontaneous breaking 
of an anomalous $U(1)$ symmetry, the best motivated
example being the Peccei--Quinn axion.
Such mechanism can be understood by considering the following Lagrangian:
\be
\label{generalL}
\mathcal{L} =  \partial_{\mu} \Phi^{\dagger} 
\partial^{\mu} \Phi - \lambda \left(\Phi^{\dagger} \Phi  - \frac{v_a^2}{2} \right)^2 - V(a)\,,
\ee
where $\Phi = \rho\,\text{exp}(i a/v_a)/\sqrt{2}$ and $a$ is the axion.
The first potential term implies that the $U(1)$ symmetry is spontaneously broken
with $\langle \Phi \rangle = v_a/\sqrt{2}$
and the axion domain is $[0, 2 \pi v_a)$.
The last potential term in \eqref{generalL} is induced by the anomaly of the $U(1)$ group 
under a strongly coupled gauge theory whose dynamical scale is $\Lambda$,
and explicitly breaks the $U(1)$ symmetry to a $\mathbb{Z}_{2N}$ discrete symmetry, 
where $N$ is the anomaly coefficient. 
The typical form of such explicit breaking at zero temperature is
\be
\label{eq:pote}
V(a)=  \Lambda^4 \left[1-\cos \left(\frac{a N_\text{DW}}{v_a} \right)\right]
\ee
where $N_\text{DW} \equiv 2N$ is the domain wall number,
and the axion decay constant is $f_a \equiv v_a/N_\text{DW}$.
At temperatures $T$ around the confinement scale $\Lambda$ and above, thermal corrections to the ALP potential become important, see e.g.\,\cite{Marsh:2015xka, Villadoro} and references therein for further details. In particular, for $T \gg \Lambda$ the
overall magnitude of the potential $V(a)$ is suppressed by a factor $\propto (\Lambda/T)^n$,
where $n>0$ is a $\mathcal{O}(\text{few})$ number depending on the matter
content of the theory. 

The ALP potential possesses $N_\text{DW}$ discrete vacua
connected by a shift symmetry,
\be
\mathbb{Z}_{N_\text{DW}}: 
\frac{a}{v_a} \longmapsto \frac{a}{v_a} + \frac{2 \pi k}{N_\text{DW}}
\ee
with $k = 0,1\dots,N_\text{DW}-1$. Correspondingly,
there are $N_\text{DW}$ minima at $a = 2 \pi k v_a/N_\text{DW}$.
The DW solutions for this potential are known,
see e.g.\,\cite{Vilenkin:2000jqa}, and
for the interpolation between neighbouring $k$ and $k+1$ vacua they
read
\be
\label{DWcosine}
\frac{a(z)}{v_a} = \frac{2 \pi k}{N_\text{DW}} +
\frac{4}{N_\text{DW}} \tan^{-1} e^{m_a z},
\ee
where the $z$ coordinate is transverse to the DW 
and $m_a$ is the mass of the ALP, 
$m_a = N_\text{DW} \Lambda^2/v_a$.
The DW width is $\delta \sim 1/m_a$, and the DW tension $\sigma$ is given by
\be
\label{eq:DWtension}
\sigma =\int T^0_0 dz = \frac{8 m_a v_a^2}{N_\text{DW}^2},
\ee
where $T_0^0$ is the 00-component of the energy-momentum tensor
of the ALP field evaluated on the DW solution.

In realistic models of QCD--like confinement, the non--perturbative
potential generated for the ALP retains the same $\mathbb{Z}_{N_\text{DW}}$ discrete
symmetry of \eqref{eq:pote},
but can differ significantly from the simple cosine shape\,\cite{Villadoro}. 
For this type of potentials the DW interpolating between
neighbouring minima will be quantitatively different from the simple form
in \eqref{DWcosine}. Similarly, the tension
receives $\mathcal{O}(1)$ corrections, see App.\,\ref{darkqcd}.
The most important implication is that 
ALPs can have a non--zero reflection probability when
self--scattering off the DW, in contrast to the simple cosine
potential which is known to be exactly reflectionless, see e.g.\,\cite{Vilenkin:2000jqa}. This can possibly
lead to a relevant source of self--friction from ALP particles, as we will discuss
in Sec.\,\ref{selfreflection}. When considering the interaction
of fermions 
with the DW, we will nonetheless refer 
to the simple
cosine form, as this introduces only $\mathcal{O}(1)$ factors which
do not severely alter our predictions.

Let us now discuss the implication of the symmetry
breaking pattern outlined above for the formation
of topological defects. Throughout our discussion
we shall always assume that the reheating
temperature of the Universe is above the scale 
$v_a$ such that the $U(1)$ symmetry is unbroken
after inflation. 
Assuming a large separation of scales between
$\Lambda$ and $v_a$, we can consider the $U(1)$ symmetry to be exact at the time
of spontaneous breaking. This leads to a ``$U(1)\rightarrow$ nothing" pattern
whose topology implies the formation of a cosmic
string network. These are one--dimensional
defects around which the axion field
winds by multiples of $2\pi v_a$. 
At the string core, the field $\Phi$ has a vanishing vacuum expectation
value and the axion field is not defined.
As the field sits at the top of the potential, cosmic strings have a large energy
density which is primarily
characterized by the string tension $\mu \sim v_a^2$.
The network of cosmic strings is known to reach a
scaling solution in which,
up to logarithmic corrections in the case of global strings, the energy density
scales with the cosmic time $t$
as $\rho_s \sim \mu/t^2$, thus remaining a constant
fraction of the total energy density of the Universe, see e.g.\,\cite{PhysRevLett.60.257,PhysRevD.40.973,PhysRevLett.64.119,Martins:2018dqg,Gorghetto:2018myk,Hindmarsh:2019csc}. 
Cosmic strings can nonetheless affect crucial observables
in cosmology, such as the relic abundance of axions\,\cite{PhysRevD.35.1138,Davis:1989nj,Yamaguchi:1998gx,Hagmann:2000ja,Gorghetto:2018myk,Gorghetto:2020qws,OHare:2021zrq},
and can source gravitational waves, see e.g.\,\cite{Vachaspati:1984gt,Blanco-Pillado:2011egf,Blanco-Pillado:2013qja,Ringeval:2005kr,Lorenz:2010sm,Gorghetto:2021fsn},
even though debate remains in the community on the actual amount
of gravitational wave emission\,\cite{Daverio:2015nva,Hindmarsh:2017qff}.

As the temperature drops, however, the effect of the explicit breaking
of the $U(1)$ symmetry starts to play a role.
In particular, the axion field values can no longer
be considered equivalent as the potential 
\label{pote} starts to attract the axion towards either of
its discrete minima.
The only force preventing the axion to locally roll to the closest
minimum is given by Hubble friction, 
which however decreases with the temperature,
contrary to the axion potential which in fact increases at lower temperatures
(and eventually becomes practically constant below the dark confinement scale $\Lambda$).
Therefore, when the temperature--dependent axion mass, $m_a(T)$, equals Hubble,
domain walls stretching between the cosmic strings start to form. 
This happens approximately at the formation temperature $T_{\text{form}}$ that can be then obtained by 
solving $H(T_\text{form}) \simeq m_a(T_{\text{form}})$. For ALP decay constants below the Planck scale,
one typically has $T_\text{form} \gtrsim \Lambda$, so that at $T=\Lambda$ we can consider
the domain wall network to be formed.

Since the axion field takes on all the values from $0$ to $2\pi v_a$ in a closed path
around a string, it encounters
exactly $N_\text{DW}$ minima, such that each string is attached
to $N_\text{DW}$ domain walls.
The case in which $N_\text{DW}=1$ shows a particular behavior because there
exists a unique vacuum for the theory with consequent trivial topology,
as the two points $a=0$ and $a=2\pi v_a$ are to be identified. This means 
that the whole string--wall
network is unstable and collapses short after the domain walls 
are formed\,\cite{Vilenkin:1982ks,Barr:1986hs,Shellard:1986in,BARR1987591,Chang:1998tb}.

For $N_\text{DW} > 1$ the non--trivial vacuum structure prevents the tangled network 
from collapsing\,\cite{Vilenkin:1982ks,Sikivie:1982qv}. 
Similarly to what happens for the strings, the DW network approaches a scaling regime
in which the number of walls per Hubble patch is constant
and the average velocity is relativistic but not ultra--relativistic\,\cite{Ryden:1989vj,Hindmarsh:1996xv,Garagounis:2002kt,Oliveira:2004he,Avelino:2005pe,Leite:2011sc}.
The energy density of the DWs then scales as $\rho_w \sim \sigma/t$, corresponding
to a characteristic curvature radius and average wall--wall distance set by the horizon size.

By comparing with the energy density of strings we have
$\rho_s/\rho_w \sim \mu/(\sigma t) \sim 1/(m_a t)$, showing that shortly
after the DW formation at $H \sim 1/t \sim m_a$, the energy 
of the hybrid network is dominated by the walls (see also\,\cite{Hiramatsu:2012sc}).
Therefore, when discussing the evolution of 
the DW network we will neglect all the possible effects arising due to the presence of
the strings
and effectively consider the case of axionic DWs in isolation. 
 \medskip

If the discrete symmetry determining the DW formation is an exact symmetry of the theory, the DW network is stable for $N_\text{DW} > 1$.
Whether in a scaling or friction dominated regime, its energy density grows with respect
to the critical density,
and at some point it will come to dominate potentially leading to cosmological inconsistencies
\cite{Zeldovich:1974uw,Sikivie:1982qv,Vilenkin:1981zs}.

Assuming scaling regime, domain walls will dominate when 
$\rho_w \sim \sigma/t$ becomes of the order of the critical density, $\rho_c = 3 H^2 M_\text{Pl}^2$.
This corresponds to the temperature
\begin{equation}
\label{Tdom}
T^2_\text{dom} = 16 G \sigma \left( \sqrt{\frac{8 \pi G}{10} g_*} \right)^{-1}
\end{equation}
where we assumed radiation domination,
$M_\text{Pl} = (8\pi G)^{-1/2}$ is the reduced Planck mass, and $g_*$ the number of relativistic degrees of freedom.

Wall domination can be avoided if the underlying discrete symmetry
is biased by a small energy difference $\Delta V$ between the vacua 
interpolated by the DW solution\,\cite{Sikivie:1982qv}. 
Such energy difference generates a pressure force per unit area, $p_V \sim \Delta V$,
which tends to shrink the higher--energy domains until the whole space is filled
by the only true vacuum, eventually collapsing
the wall network\,\cite{Coulson:1995nv,Larsson:1996sp,Avelino:2008qy,Correia:2014kqa,Correia:2018tty}. 

The moment at which the collapse starts, $t_\text{ann}$, can be estimated by balancing the vacuum pressure, $p_V$ and the wall tension force, $\sigma/R$, where $R$ is the average curvature radius. As mentioned
above, $R$ is of the order of the horizon size in the scaling regime, and therefore
$t_\text{ann} \sim \sigma/\Delta V$.

The presence of the bias does not influence the formation of the walls provided that it does not affect percolation \cite{STAUFFER19791}. 
The approximate bound is that $\Delta V /\Lambda^4 < 0.795 $ for a $\mathbb{Z}_2$ symmetric potential \cite{Gelmini:1988sf, Lalak:1992px}.
We shall assume that generalizing to $N_\text{DW}>2$ does not lead to a significantly stronger percolation bound. This condition is very mild and it is largely satisfied for the ALP parameters of interest where we consider long--lived domain walls with a small bias term.

One may however wonder how natural is the existence of a bias term which introduces a small explicit breaking of the discrete symmetry. 
In the context of the ALP model, the discrete symmetry is descending from an anomalous $U(1)$, which is not expected to be exact in the first place.
In particular, we shall consider explicit breaking from higher-dimensional operators suppressed by the Planck scale \cite{Barr:1992qq,Kamionkowski:1992mf,Holman:1992us,Berezhiani:1992pq,Ghigna:1992iv,Senjanovic:1993uz,Dobrescu:1996jp,Banks:2010zn} that can be parameterized as 
\be
\label{bias_MP}
V_{M_\text{Pl}} = 
\mathcal{C}_{n,m}
\frac{\left(\Phi^{\dagger} \Phi \right)^m \Phi^n}{M_\text{Pl}^{2m+n-4}} + \,\text{h.c.}
\ee
with $\mathcal{C}_{n,m}$ Wilson coefficients.
These operators induce a bias term in the ALP potential, automatically solving the DW problem and setting the annihilation time $t_\text{ann}$ as explained above.

\subsection{Gravitational waves from domain wall dynamics}\label{sec:Gravitational waves from domain wall dynamics}

The network of DWs continuously emits gravitational waves due to the 
oscillations of the domain walls. This emission stops when the DW network annihilates due to the bias
at an annihilation time $t_\text{ann}$ that we consider unspecified for the moment.

The GW spectrum at a cosmic time $t$ is characterized by
\begin{equation}\label{eq: GW spectrum}
\Omega_\text{gw}(t,f) = \frac{1}{\rho_c(t)}\left(\frac{d\rho_\text{gw}(t)}{d\ln f}\right)\,,
\end{equation}
where $\rho_c(t) = 3H(t)^2 M_\text{Pl}^2$ is the critical density of the Universe at the time
$t$. The form of the GW energy density per logarithmic frequency, $d\rho_\text{gw}/d\ln f$, is dictated by the signal shape observed in numerical simulations during the scaling regime. Ref. \cite{Hiramatsu:2013qaa} has shown that the spectrum as a function of frequency grows as $f^3$ and decreases as $f^{-1}$ when the peak frequency $f_\text{peak}$ is reached, given by Hubble at the time $t$,
\begin{equation}
\label{eq:Omegadf}
\Omega_\text{gw}(t,f) = \frac{1}{\rho_c(t)}\left(\frac{d\rho_\text{gw}(t)}{d\ln f}\right)_\text{peak}\times\begin{cases}
      \left(\frac{f}{f_\text{peak}}\right)^{3} & \text{if $f\leq f_\text{peak}$}\\
      \left(\frac{f}{f_\text{peak}}\right)^{-1} & \text{if $f > f_\text{peak}$}
    \end{cases}\,, 
\end{equation}
where ``peak'' means that the quantity is evaluated at the peak frequency $f_\text{peak} = H(t)$. Furthermore, the peak amplitude is given by
\begin{equation}
\label{eq:Omegadfpeak}
\Omega_\text{gw,peak}(t) = \frac{1}{\rho_c(t)}\left(\frac{d\rho_\text{gw}}{d\ln f}\right)_\text{peak} = \frac{\tilde{\epsilon}_\text{gw}G \mathcal{A}^2\sigma^2}{\rho_c(t)}\,,
\end{equation}
where $\tilde{\epsilon}_\text{gw} = 0.7\pm 0.4$ and $\mathcal{A} = 0.8\pm 0.1$ according to the simulations \footnote{In fact, $\mathcal{A}$ is an $\mathcal{O}(1)$ number that increases with $N_\text{DW}$ according to the simulations performed in \cite{Kawasaki:2014sqa}. Nevertheless, we will keep this value fixed throughout our work.}.

 Redshifting the contribution
at the time $t$ to today $t_0$, we have
\begin{equation}\label{eq:OmegaGW}
 \left.\Omega_\text{gw}(t, f)\right|_{t_0}= \frac{\rho_c(t)}{\rho_c(t_0)} \left( \frac{a(t)}{a(t_0)}\right)^4
 \Omega_\text{gw}(t, f)
 = \beta \Omega_\text{R} \Omega_\text{gw}(t, f),
\end{equation}
where $\beta = (g_\ast(T)/g_{\ast 0}) (g_{\ast s0}/ g_{\ast s}(T))^{4/3}$
takes into account the change in the relativistic degrees of freedom with $g_{*0} = 3.36$ and $g_{*s0} = 3.91$,
and $\Omega_\text{R} \simeq 9.2 \times 10^{-5}$, so that the peak amplitude today is given by
\begin{equation}
\label{eq:pheno_omega}
\left.\Omega_{\text{gw,peak}}(T)\right|_{t_0}  = 2.34\times 10^{-6} \Tilde{\epsilon}_\text{gw}\mathcal{A}^2\left(\frac{g_*(T)}{10}\right)\left(\frac{g_{*s}(T)}{10}\right)^{-4/3}\left(\frac{T_\text{dom}}{T}\right)^4\,,
\end{equation}
whereas the redshifted peak frequency is
\begin{equation}\label{eq:pheno_freq}
\left.f_\text{peak}(T)\right|_{t_0}= \frac{a(t)}{a(t_0)}H(t) \simeq 1.15\times 10^{-9}\ \text{Hz}\left(\frac{g_*(T)}{10}\right)^{1/2}\left(\frac{g_{*s}(T)}{10}\right)^{-1/3}\left(\frac{T}{10\ \text{MeV}}\right)\,,
\end{equation}
The conversion from time to temperature $T$ has been made assuming a radiation dominated Universe, i.e. $H = 1/(2t) = \sqrt{\pi^2 g_*/90}\ (T^2/M_\text{Pl})$,
which will be the underlying assumption throughout the paper.

Note that \eqref{eq:OmegaGW} is the normalized energy density of gravitational waves emitted by the network at time $t$ (and redshifted to today).
The total signal emitted by the DW network should be obtained by integrating the GW emitted power along the whole time interval from the
DW formation to the DW network annihilation $t_\text{ann}$. 
We will approximate the total signal by redshifting the spectrum at annihilation solely, as the largest contribution to the signal happens at this time. 
Indeed we note from \eqref{eq:pheno_omega} that the later is the DW annihilation, namely the closer is the annihilation temperature to the temperature of DW domination, the larger is the signal.

\section{On the importance of friction}
\label{sec:3}
In this section we investigate the regimes under which friction is important for the evolution of the ALP DWs. 
We first define this condition in a model--independent way by comparing a characteristic friction length with Hubble expansion. We then move to describe the formalism to compute the friction length by evaluating the pressure from particles reflected by the DW surface, and assess the validity regimes of our computation. We then study in detail the pressure induced from ALP--fermion interactions.
We conclude the section with an illustrative case study showing how friction from SM particles can affect the expected GW signal from DWs relevant for Pulsar Timing Array experiments.

\subsection{Domain wall equation of motion and friction length}
\label{sec:friction_eom}

In this subsection we define a condition for friction to be relevant in the evolution of the DW network by comparing with the effect of Hubble expansion. 
To this end, we study the equation of motion of the DW in a FLRW background in the thin
wall approximation (see also \,\cite{Martins:2016ois}), following an analogous derivation to \cite{Vilenkin:1991zk} for the case of cosmic strings.

When the width of the domain wall is much smaller than
the horizon, DWs can effectively be described as
two--dimensional objects whose dynamics are encoded in the following action
\begin{equation}\label{eq:DWaction}
S = -\sigma \int d^3 \zeta \sqrt{\gamma}, 
\end{equation}
where $\gamma$ is the determinant of the induced
metric on the three--dimensional worldvolume,
\begin{equation}
\gamma_{ab} = g_{\mu \nu} \frac{ \partial x^\mu}{\partial \zeta^a}
\frac{ \partial x^\nu}{\partial \zeta^b}.
\end{equation} 
The constant $\sigma$ is the domain wall tension,
which also equals the energy per unit surface of a straight domain wall.
From the action in \eqref{eq:DWaction}, one can derive
the equations of motion for a single domain wall.
The derivation is the easiest in the gauge in which the 
time variable on the worldvolume coincides with the conformal time,
$\zeta_0 = \tau$, with the FLRW metric
\begin{equation}
 \text{d}s^2 = a^2(\tau)(\text{d}\tau^2 -\,\text{d} x^2).
\end{equation}
The space coordinates $\zeta_{1,2}$ can be chosen such that 
$(\dot{x}^\mu, x^\mu_{,1}, x^\mu_{,2})$ is an orthogonal
system, and we have used a shorthand notation for 
$\partial_{\zeta_0} \equiv \partial_\tau \equiv \dot{x}^\mu$,
$\partial_{\zeta^i} x^\mu \equiv x^\mu_{,i}$.
The induced metric $\gamma_{ab}$ is diagonal in this gauge
and given by
\begin{equation}
 \gamma_{ab} = a^2(\tau) \text{diag}( 1- \dot{x}^2, - x^2_{,1},
 -x^2_{,2}),
\end{equation}
where $x$ indicates the three vector corresponding to the spatial
components of $x^\mu$. The action takes the form 
\begin{equation}
 S = -\sigma \int d^3 \zeta a^3(\tau) 
 \sqrt{(1- \dot{x}^2) x^{2}_{,1} x^{2}_{,2}}.
\end{equation}
In the absence of external forces, the equations of motion
following from \eqref{eq:DWaction} are given by
\begin{equation}\label{eq:eoms}
\sigma \frac{1}{\sqrt{\gamma}}
 \frac{\partial}{
 \partial \zeta^a} \left( \sqrt{\gamma} \gamma^{a b} x^\mu_{,b}\right)
 + \sigma \Gamma^\mu_{\nu \sigma} \gamma^{a b} x^\nu_{,a} x^\sigma_{,b} = 0.
\end{equation}
The effect of particle friction can be included in the equations of motion
as a force on the right--hand--side of \eqref{eq:eoms}. 
In the rest frame of the domain wall, this force is proportional
to the velocity of the domain wall itself, as this creates an imbalance
for the scattering particles on the two sides of the wall, as we shall see
below. The correct covariant form of this force turns out to be 
\,\cite{Vilenkin:1991zk}
\begin{equation}\label{eq:force}
 F^\nu = \frac{\sigma}{\ell_\text{f}} (u^\nu - x^\nu_{,a} \gamma^{a b} x^\mu_{,b}
 \, g_{\mu \sigma} u^\sigma),
\end{equation}
where we have parameterized the overall coefficient as $\sigma/\ell_\text{f}$
for future convenience, and $u^\sigma$ is the four velocity
of the thermal bath, which in the case of an expanding Universe and in
the frame of the bath simply reads
\begin{equation}
 u^\sigma = (1/a(\tau),0,0,0).
\end{equation}
Within our gauge for the coordinates, the force reads
\begin{equation}
F^i = -\frac{\sigma}{\ell_\text{f}} \frac{\dot{x}^i}{a(1-\dot{x}^2)}, \quad 
F^0 = - \frac{\sigma}{\ell_\text{f}} \frac{ \dot{x}^2}{a(1-\dot{x}^2)},
\end{equation}
since no contribution comes from the spatial components of $x^{\mu \, \prime}_{1,2}$
when contracted with $u^\sigma$.
Inserting this in the equations of motion, we obtain
\begin{equation}
 \ddot{x} + 
\left( \frac{\dot \epsilon}{\epsilon} + 3 \frac{\dot a}{a} 
+ \frac{1}{\ell_\text{f}} a \right) \dot{x}
= \frac{1}{\epsilon} \frac{\partial}{\partial \zeta^1} (x_{,2}^2 \, x_{,1} / \epsilon)
+  \frac{1}{\epsilon} \frac{\partial}{\partial \zeta^2} ( x_{,1}^2 \, x_{,2} / \epsilon),
\end{equation}
where $\epsilon$ is defined as 
\begin{equation}
 \epsilon = \sqrt{ \frac{x^{2}_{,1} x^{2}_{,2}}{1- \dot{x}^2}}.
\end{equation}
The equation for the time component reads
\begin{equation}\label{eq:eomsi}
\frac{ \partial_\tau ( a \epsilon)}{a^3 \sqrt{(1-\dot{x}^2)x^{\prime 2}_1
x^{\prime 2}_2}} + \frac{ 3 \dot{x}^2-1}{1-\dot{x}^2} \frac{\dot a}{a^3} = -
\frac{1}{\ell_\text{f}} \frac{ \dot{x}^2}{a(1-\dot{x}^2)},
\end{equation}
which simplifies to
\begin{equation}
\label{eq:energy_DW}
\frac{\dot{\epsilon}}{\epsilon} + 
\left( 3 \frac{\dot a}{a} + \frac{1}{\ell_\text{f}} a\right)
\dot{x}^2 = 0.
\end{equation}
The equation above is not an independent one as it follows from the 
spatial equations of motion together with our gauge choice.
It nonetheless helps us bringing \eqref{eq:eomsi}
in the final form 
\begin{equation}
\label{eq:velocity_dw}
\ddot{x} + \left( 3 \frac{\dot a}{a^2} + 
\frac{1}{\ell_\text{f}} \right) (1-\dot{x}^2)
a \dot{x} = \frac{1}{\epsilon} \frac{\partial}{\partial \zeta^1} (x_{,2}^2 \, x_{,1} / \epsilon)
+  \frac{1}{\epsilon} \frac{\partial}{\partial \zeta^2} ( x_{,1}^2 \, x_{,2} / \epsilon).
\end{equation}
We conclude that particle friction enters the equations
of motions only in combination with the Hubble parameter, $H=\dot{a}/a^2$,
through a characteristic damping length scale, $\ell_\text{d}$, given by
\begin{equation}\label{eq:dlength}
 \frac{1}{\ell_\text{d}} = 3 H + \frac{1}{\ell_\text{f}}.
\end{equation}
The condition for friction to be relevant then simply reads 
$3H \lesssim 1/\ell_\text{f}$. 
The friction length can be computed once the physics of the particle
scattering is known. The calculation is best done in the local rest
frame for the domain wall, where the force \eqref{eq:force}
becomes\,\cite{Vilenkin:1991zk}
\begin{equation}\label{eq:Fisimple}
 F^i = -\frac{\sigma}{\ell_\text{f}} \frac{v^i}{\sqrt{1-v^2}},
\end{equation}
and $v$ is the local velocity of the domain wall. The calculation of
$1/\ell_\text{f}$ in this frame can be carried out with standard techniques
involving the scattering probability and the rate of momentum exchange
between the domain wall and the bath particles,
as we shall see in the following section.

\subsection{Pressure from particle reflection}
\label{sec:pfrompf}

We now proceed to compute the friction length induced by interactions of the particles
in the plasma with the domain wall.
In order to evaluate the friction length due to particle
scattering, we need to calculate the net pressure
acting on the domain walls.
We first review the formalism to compute the pressure induced by particles scattering off the wall,
following \cite{Arnold:1993wc}. We then investigate the relevance of friction for the DW evolution 
in a FLRW Universe in some illustrative cases.
We will be mostly concerned with friction from fermions in thermal equilibrium or frozen out. 
For a preliminary investigation of friction, including the one induced by gauge bosons, see \cite{Huang:1985tt}. In the following we shall also neglect thermal corrections to the mass of the scattering particles.

Let us then consider the case of a domain wall
moving with velocity $v$ along the $z$ axis through
a plasma of particles which in the bath rest frame 
follow the standard Fermi--Dirac (FD)
or Bose--Einstein (BE) distribution 
\begin{equation}
    n = \frac{g}{e^{\beta E}\pm1},
\end{equation}
with $\beta=1/T$ and $g$ stands for the total
number of degrees of freedom such as spin, color and flavor. 
Moving from the bath rest frame
to the wall rest frame, the distribution $n$ is modified
as
\begin{equation}
    f(v) = \frac{g}{e^{\gamma(v) \beta(E+ p_z v)}\pm1},
\end{equation}
where $\gamma(v)=1/\sqrt{1-v^2}$, and $z$ is the direction orthogonal to the wall.
Particles interacting with the wall may have a 
momentum--dependent probability $\mathcal R(p)$ of being reflected.
The pressure exerted by particles coming from the right of the wall
is then given by 
\begin{equation}
    P_\text{R} = 2 \int \frac{\text{d}^3 p}{(2\pi)^3} \theta(-p_z)\frac{p_z}{E}
    f(v) \mathcal R(p) p_z
    = 2 \int \frac{\text{d}^2 p}{(2\pi)^3}
    \int_{-\infty}^0 \text{d}p_z 
    \frac{p_z^2}{E}
    f(v) \mathcal R(p) 
\end{equation}
where the factor of 2 takes into account that $\Delta p = 2 p_z$
momentum exchange in case of reflection. 
The pressure from the left side, $P_\text{L}$, is obtained analogously,
leading to the net pressure 
\begin{equation}
    \Delta P = P_\text{R}-P_\text{L}=
   2 \int \frac{\text{d}^2 p}{(2\pi)^3}
    \int_0^\infty \text{d}p_z [f(-v)-f(v)] \frac{p_z^2}{E}\mathcal{R}(p)
\end{equation}
where we have used the fact that $\mathcal R(p) = \mathcal R(-p)$.
This expression may be further simplified by taking into account that
the reflection coefficient only depends on $p_z$. In fact,
the motion parallel to the domain wall cannot affect the dynamics
of the interaction as the domain wall is invariant
for boosts along its tangent space.

The integral over $p_x$ and $p_y$ can be traded
for an integral over the energy
which evaluates to 
\begin{equation}
\label{eq:pressurea}
    \Delta P = \frac{2}{(2\pi)^2}
    \int_0^\infty \text{d}p_z p_z^2 \mathcal{R}(p_z)
    \frac{1}{\beta \gamma a}
    \left[ \,
    2 \beta \gamma p_z v - \text{log}\left(\frac{f(-v)}{f(v)}\right)\right]\bigg|_{E=\sqrt{p_z^2+m^2}}
\end{equation}
with $a=\pm1$. 

The expression in \eqref{eq:pressurea} can be readily used for numerical calculations of the net pressure. However, in order to extract approximate expressions for the
pressure to be compared with the exact results, we shall further simplify 
\eqref{eq:pressurea} by taking the limit of classical statistics, $a \rightarrow 0$, which turns out to be a sensible approximation for the case of Fermi--Dirac distribution:
\begin{equation}\label{eq:DPDM}
     \Delta P \simeq \frac{g}{\pi^2 \beta \gamma}
    \int_0^\infty \text{d}p_z p_z^2 \mathcal{R}(p_z)
    \,\text{exp}\left(-\beta \gamma \sqrt{p_z^2+m^2}\right) \,\text{sinh}\,(\beta \gamma p_z v).
\end{equation}

Finally, as discussed in the previous section,
the average velocity of the domain walls is 
found to be
relativistic but not ultra--relativistic even in the scaling regime,
whereas the velocity is $v\ll1$ when friction dominates. 
It is therefore reasonable to expand
the expression above for small velocities, obtaining
\begin{equation}
\label{eq:DeltaPu}
    \Delta P \simeq v \frac{g}{\pi^2} \int_0^\infty \,\text{d}p_z
    p_z^3 \mathcal R(p_z)\,\text{exp}\left(-\sqrt{m^2+p_z^2}/T\right).
\end{equation}
Notice that this expansion is actually justified only when the main contribution to the integral \eqref{eq:DPDM} comes from momenta such that $p_z \lesssim T/v$.
For thermal distributions this condition is easily ensured by the Boltzmann suppression factor, yielding minor $\mathcal{O}(1)$ corrections as long as $v \lesssim 0.5$\footnote{In temperature regimes such that $T \ll m$ one can expect larger deviations between \eqref{eq:DPDM} and \eqref{eq:DeltaPu}, but this is irrelevant in practice as the overall pressure is utterly small.}.

Once the pressure is evaluated, the friction length 
according to \eqref{eq:Fisimple} is given by
\begin{equation}\label{eq:lffromP}
 \frac{1}{\ell_\text{f}} = \frac{\Delta P}{\sigma \gamma(v) v} 
 \simeq \frac{\Delta P}{\sigma v}
\end{equation}
for non ultra--relativistic walls.

The picture outlined above based on the reflection probability
is strictly speaking only valid when the mean free path of the 
scattering particle is much larger than the width 
of the domain wall. Considering scattering processes with a bath
with density $\sim T^3$
and typical center--of--mass energy $s \sim T^2$ and coupling
constant $\alpha \ll 1$, one obtains that the mean free path $\ell$
is given by $\ell \sim 1/(\alpha^2 T)$. We shall therefore apply
our analysis only at temperatures for which
\begin{equation}
\label{eq: T_mean_free_path}
 T \lesssim \frac{1}{\alpha^2} \delta^{-1},
\end{equation}
where $\delta$ is the domain wall width.

With this formalism, we can now discuss some simple and realistic
interactions that can lead to a sizable pressure.

\subsubsection{Fermions in thermal equilibrium}
\label{sec:fermion_friction}

Let us start by considering the interaction of fermionic degrees of freedom
with the ALP. We shall assume that fermions are in thermal equilibrium.
This can be achieved by assuming some coupling of the fermion with the thermal bath, for instance 
a coupling to photons. In the following we will not specify the nature of this coupling, we simply assume
that it exists and it is of weak-like strength ($\alpha \sim 1/100$).
The case of frozen-out fermion dark matter population will be discussed instead in 
Sec.\,\ref{sec:fermDM}.

Since the details of the domain wall profile play no crucial role
in the fermion reflection (unlike the case of self reflection) we shall use the profile corresponding to the simple cosine potential,
see Sec. \ref{sec:rev},
\begin{equation}\label{eq:cosprofile}
 a(z) = 4 f_a \, \text{arctan}\left(e^{m_a z}\right),
\end{equation}
where $f_a = v_a/N_\text{DW}$.
The interaction Lagrangian contains a pseudo--current
coupled to the ALP profile as 
\begin{equation}\label{eq:apsi}
 \mathcal{L}_{a \psi}= \frac{\kappa}{2 N_\text{DW} f_a}
 \partial_\mu a\,\bar{\psi} \gamma^\mu \gamma_5
 \psi + \bar{\psi}(i \slashed{\partial} - m_f) \psi
\end{equation}
and we shall take $\kappa > 0$ for concreteness. 
For simplicity, we have considered only one fermion coupled
to the ALP, which can be easily generalized to the case of diagonal
couplings if more flavors are present. 

Substituting the profile \eqref{eq:cosprofile} 
in the interaction above, we see that
depending on the spin, particle excitations will see
either a potential well (spin up) or hill (spin down), 
while the opposite is true for anti--particle
excitations. The effective potential seen by the fermion, $V_\text{eff}(z)$,
is then 
\begin{equation}
 V_\text{eff}(z) = m_a \frac{\kappa}{N_\text{DW}} 
 \frac{1}{\text{cosh}(m_a z)}.
\end{equation}
Our strategy in order to evaluate the pressure is to first determine
the reflection coefficient and then perform the momentum integral in
\eqref{eq:DeltaPu}. 
We have evaluated the reflection coefficient fully numerically in Appendix \ref{app:R}.
Here we discuss the main features and provide some analytic approximations 
when possible. 

Notice that in the following we shall evaluate the pressure also at temperatures significantly larger than the ALP mass. This is because the ALP potential remains approximately constant for temperatures below the dark confinement scale $\Lambda$, which is parametrically larger than $m_a$. This temperature range is thus meaningful only for ALPs, as for generic scalars (e.g. $\phi^4$ theory with $\mathbb{Z}_2$ symmetry) the domain wall solution
will no longer exist at temperatures above the relevant mass scale, due to symmetry restoration.

Let us first consider the case $m_f \gg m_a$. 
The degrees of freedom
that see the barrier contribute the most to the pressure as they
give a non--negligible reflection probability at momenta
$p \lesssim \tilde p_z$, where $\tilde p_z$ is the momentum
scale at which the fermion kinetic energy equals the height of the
ALP barrier, $\tilde p_z^2/2 m_f = \kappa/N_\text{DW} m_a$. 
As long as $\kappa/N_\text{DW} \lesssim m_f/m_a$, 
we have that $\tilde p_z \ll m_f$ so
that the particle can be treated as non relativistic
in the kinematic region relevant for the friction.
A good approximation for the reflection coefficient is then
\begin{equation}
    \mathcal{R}(p_z) = \begin{cases}
    1 \quad & p_z < \tilde p_z \\
   0 & p_z > \tilde p_z
\end{cases}    
\end{equation}
for the one spin that sees the potential barrier.
The pressure can be evaluated according
to \eqref{eq:DeltaPu} and for $m_f \gg m_a$ we obtain
\begin{equation}\label{eq:DP1}
    \Delta P(m_f\gg m_a) \simeq 
     g_b \frac{v}{\pi^2} \frac{\kappa^2}{N_\text{DW}^2} 
    m_a^2 m_f^2 e^{-m_f/T},
\end{equation}
where $g_b$ are the degrees of freedom that see the barrier,
$g_b=g/2$ and $g$ the fermionic degrees of freedom.

Let us now discuss the case $m_f \ll m_a$. 
In general, this gives
a smaller pressure with respect to the previous case 
because the fermion mass
is actually related to the ALP--fermion coupling, which vanishes
when $m_f = 0$. The momentum scale at which the fermion kinetic 
energy equals the height of the ALP barrier is now 
$p^\prime_z = \kappa/N_\text{DW} m_a$. 
By assuming that $\kappa/N_\text{DW} \lesssim \mathcal{O}(1)$, 
we have that the fermion wave length at $p^\prime_z$
is much larger than the domain wall width $\delta \sim m_a^{-1}$.
This means that we can map our problem into 
the relativistic scattering
off a potential square\,\cite{Calogeracos:1999yp}, for which we obtain
\begin{equation}\label{eq:R2}
    \mathcal{R}(p_z) = \begin{cases}
    \left(1 + r p_z^2/m_f^2\right)^{-1} \quad & p_z < \nu \\
   0 & p_z > \nu
\end{cases},    
\end{equation}
where
\begin{equation}
r = \frac{1}{\text{sin}^2(\pi \kappa/N_\text{DW})},
\end{equation}
and $\nu \sim m_a/e$ indicates the point where 
the step approximation breaks down. For $p_z > \nu$ 
we have conservatively set $\mathcal{R} = 0$, expecting
an exponential drop with a minor impact for our approximation.
The net pressure is then 
\begin{equation}
    \Delta P = g_b \frac{v}{\pi^2} m_f^4 
    \int_0^{\nu/m_f} \text{d}t \frac{t^3}{1+r t^2}
    e^{-\beta m_f \sqrt{1+t^2}},
\end{equation}
for which we find 
\begin{equation}
\label{eq:DP2}
    \Delta P \simeq g_b \frac{v}{\pi^2} m_f^2 T^2
    \left[ 2 [ h(T/m_f,1)-h(T/m_f,1+1/r) ]
    - \frac{1}{r} e^{-\frac{m_a}{T} r^\prime}(1+ m_a r^\prime/T)
    \right]
\end{equation}
with 
\begin{equation}
    r^\prime =  \sqrt{\frac{\nu^2}{m_a^2}+ \frac{m_f^2}{m_a^2}}
\end{equation}
and
\begin{equation}
h(x,y) = e^{-\sqrt{y}/x}(y + 3 x \sqrt{y} + 3 x^2).
\end{equation}
The pressure approaches a constant value at large temperatures,
\begin{equation}
\label{DP2}
    \Delta P(T \gg m_a) \simeq g_b \,\text{sin}^2\left(\frac{\pi \kappa}{N_\text{DW}}\right) \frac{v}{2 \pi^2} 
    m_f^2 \nu^2,
\end{equation}
whereas for $m_a > T > m_f$ it features a period $\propto T^2$,
\begin{equation}\label{eq:T2}
    \Delta P \approx g_b \,\text{sin}^2\left(\frac{\pi \kappa}{N_\text{DW}}\right) \frac{v}{\pi^2}  m_f^2 T^2,
\end{equation}
and exponential suppression for $T<m_f$.

\begin{figure}
\centering
 \includegraphics[scale=0.4]{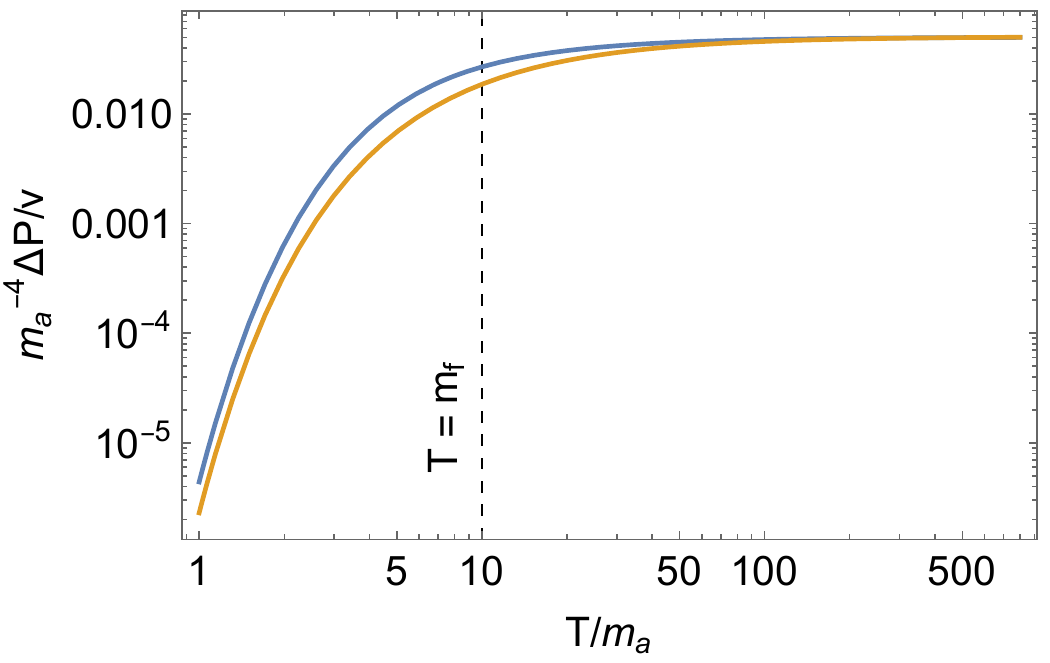}
  \includegraphics[scale=0.4]{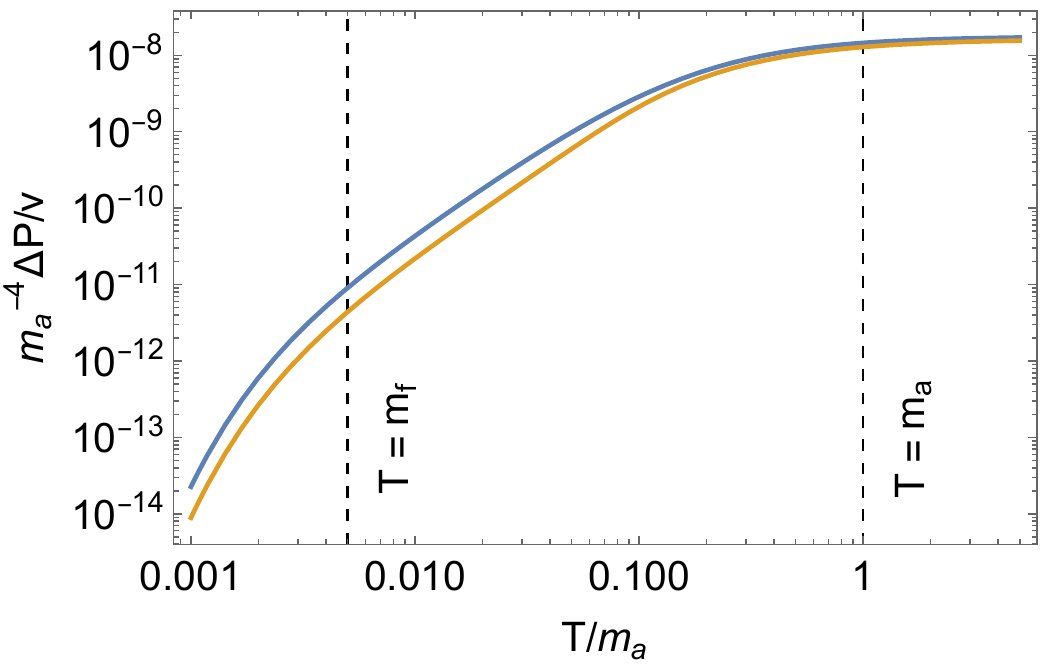}
 \caption{\textbf{Left.} Pressure induced by a fermion coupled via 
 \eqref{eq:apsi} for representative values $m_f = 10 m_a$
 and $\kappa/N_\text{DW} = 0.1$ with the reflection probability evaluated numerically and with the pressure \eqref{eq:pressurea} with $v=0.4$ (blue)
 and according to the approximate formula \eqref{eq:DP1} (orange).
 Dimensionful quantities
 are shown in units of $m_a$.
 \textbf{Right}. Same as the left panel but for
 $m_f = m_a/200$. The blue curve is obtained numerically with $v=0.4$,
 whereas the orange one refers to the approximate
 formula \eqref{eq:DP2} with $\nu/m_a = 1/e$.
  }
 \label{fig:DP}
\end{figure}

The comparison between the net pressure
evaluated numerically and the analytical
approximations made so far is shown in Fig.\,\ref{fig:DP}
fixing $\kappa/N_\text{DW} = 0.1$.
The left panel shows a typical case with $m_f \gg m_a$ where the 
numerical result agrees well with the approximation \eqref{eq:DP1}, up to $\mathcal{O}(1)$ factors:
the pressure reaches a constant value at $T \gg m_f$ and a drop is found at
$T \lesssim m_f$ due to the suppressed abundance in the plasma.
The case $m_f \ll m_a$ is shown in the right panel of Fig.\,\ref{fig:DP}. 
In this case we can identify two drops:
one occurring at $T \lesssim m_a$ (which is independent of how small $\kappa/N_\text{DW}$ is),
and one at $T \lesssim m_f$. Between these two drops, a period of
$\Delta P \propto T^2$ occurs.

\begin{figure}
\centering
 \includegraphics[scale=0.4]{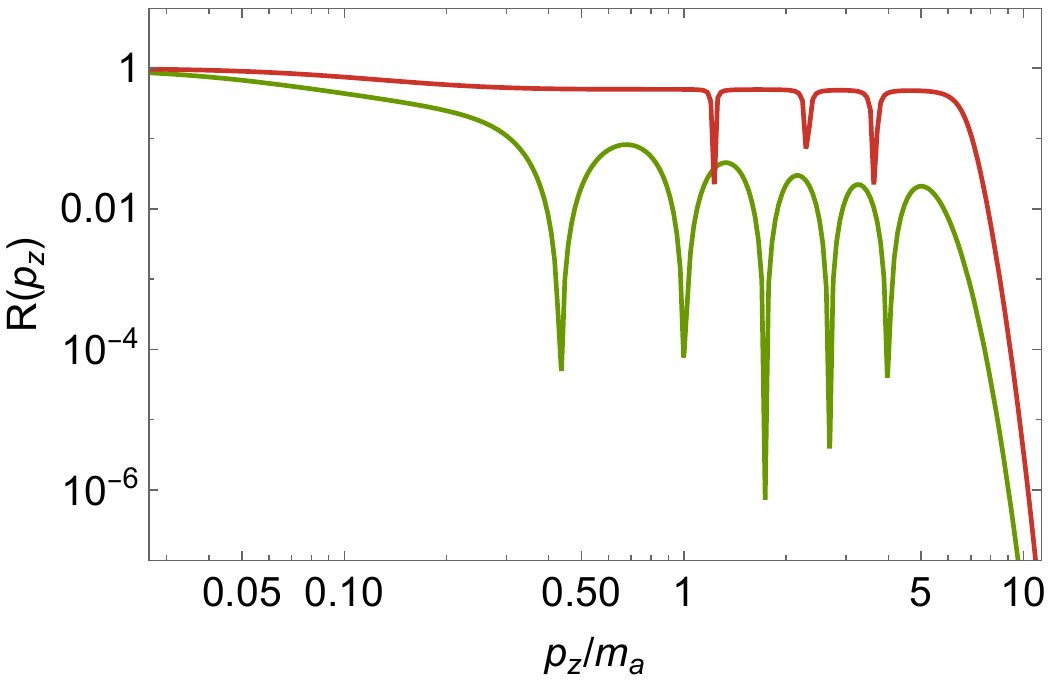}
 \includegraphics[scale=0.4]{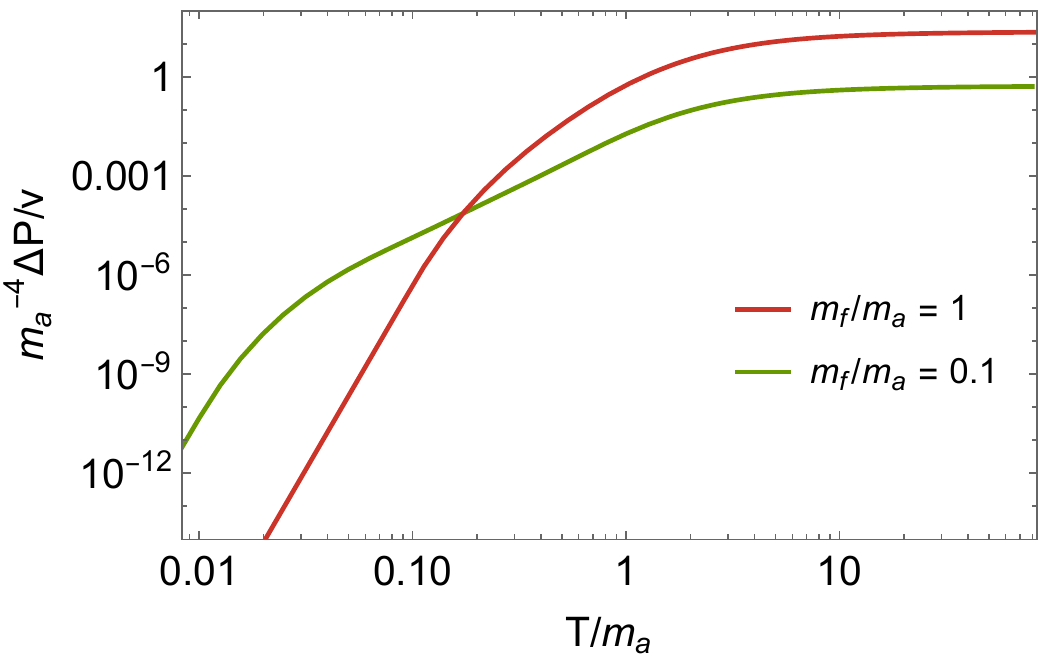}
 \caption{\textbf{Left.} Reflection coefficient, $\mathcal{R}(p_z)$,
 for $m_f/m_a = 1\,(0.1)$ in red (green) and $\kappa/N_\text{DW} = 6$.  The momenta that lead to a sudden drop in $\mathcal{R}$ correspond to 
transmission resonances.
 \textbf{Right}. The resulting pressure as a function of the temperature according to \eqref{eq:pressurea} with $v=0.4$.
}
 \label{fig:PR}
\end{figure}

Finally, let us discuss the case with $\kappa/N_\text{DW} \gtrsim \mathcal{O}(1)$
and $m_f \lesssim m_a$, for which the analytical description becomes challenging. 
This is because the kinematic
region that gives the largest contribution to the pressure coincides 
with the relativistic fermion probing the actual shape of the 
effective potential $V_\text{eff}(z)$. In addition, transmission
resonances characteristic of tunneling in strong fields 
become important\,\cite{Calogeracos:1999yp}. We shall
therefore rely on a numerical calculation
of the reflection coefficient, and the results are shown in Fig.\,\ref{fig:PR}
for $\kappa/N_\text{DW} = 6$ and $m_f/m_a = 1\,(0.1)$ in red (green).
The left panel of Fig.\,\ref{fig:PR} clearly shows transmission resonances
at large momenta $p_z \gtrsim m_a$, where the reflection coefficient is still 
large due to the large coupling $\kappa/N_\text{DW}$. As a consequence,
the behavior of the pressure as a function of the temperature can have
additional features compared to the right panel of Fig.\,\ref{fig:DP},
although two main drops can still be identified: the one 
for $T \lesssim \kappa/N_\text{DW} m_a$ and the one dictated by the thermal abundance
at $T \lesssim m_f$.

In summary, the net pressure from fermions coupled as in \eqref{eq:apsi}
approaches a constant value at high temperatures which scales 
with the fermion and ALP mass as $\sim  m_f^2 m_a^2$. 
The pressure can then feature a period in which $\Delta P 
\propto T^2$ when $m_f \ll m_a$. Notice that this behavior differs from the commonly
assumed $\propto T^4$ due to the specific fermion--ALP interaction. For 
$T<m_f$ the pressure is exponentially suppressed due to the suppressed fermion abundance in the plasma.

We now discuss how the results for the pressure derived in this section can impact
the evolution of the domain walls. In fact,
since a period of friction domination modifies the scaling properties of the network, it can impact 
crucial quantities such as the rate of particle production of ALPs and the expected signal in gravitational
waves. 
According to our condition, friction dominates when the inverse friction length in
\eqref{eq:lffromP} overcomes Hubble, see
the discussion around \eqref{eq:dlength}. We shall now investigate in which part of the 
parameter space this is actually the case and postpone the study of the resulting spectrum
of gravitational waves to Sec.\,\ref{sec: VOS}. 
We shall then define a ``friction region''
as the parameter space for which particle friction overcomes Hubble at any time between
the domain wall formation and the time $t_\text{dom}$ at which the domain walls in scaling regime 
would come to dominate the energy density of the Universe, $t_\text{dom} \sim 1/G\sigma$,
see discussion around \eqref{Tdom}.
There is, however, a further complication due to the fact that above the 
dark confinement scale $\Lambda$ the ALP mass is sensitive to the temperature
in a model dependent way, see Sec.\,\ref{sec:rev}. For this reason, we shall 
check whether the inequality $1/\ell_\text{f} > 3 H$ is ever satisfied only
in the range $t_\Lambda < t < t_\text{dom}$, where $t_\Lambda$ corresponds to the time at which $T = \Lambda$.

For the case $m_f \gg m_a$, we shall set $\kappa/N_\text{DW}=1$ throughout
this analysis. The friction length is then
\begin{equation}
\label{eq:1elef}
 \frac{1}{\ell_\text{f}} =
 \frac{g_b m_a m_f^2}{\gamma \pi^2 f_a^2} e^{-m_f/T},
\end{equation}
where we have used \eqref{eq:DP1}, and $\sigma = \gamma m_a f_a^2$ and $\gamma \simeq 9$. Due to the fast exponential decay for $T<m_f$,
the condition $1/\ell_\text{f} > 3 H$ can be
essentially evaluated at $T \sim m_f$ leading to 
\begin{equation}
 \frac{g_b}{\gamma e \pi^2} \frac{m_a m_f^2}{f_a^2} > 
 3 \pi \sqrt{g_\ast/90} \frac{m_f^2}{M_\text{Pl}}.
\end{equation}
As we can see, the actual value of $m_f$ is irrelevant
as long as $m_f \gg m_a$, and the condition above basically
reads 
\be 
m_a \gtrsim f_a^2/M_\text{Pl},
\ee
which is very far from the standard QCD axion mass scale, but easily realized in ALP models.

\begin{figure}
	\centering
	\includegraphics[width=0.475\textwidth]{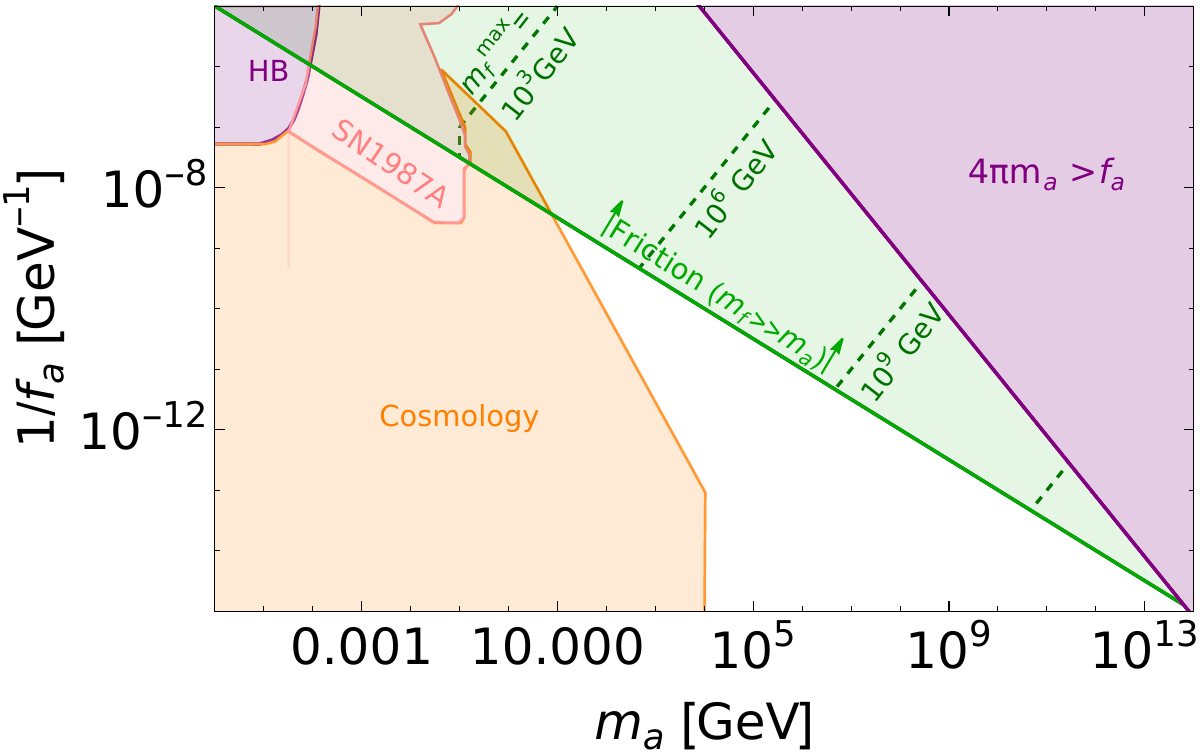}
	\,
	\includegraphics[width=0.475\textwidth]{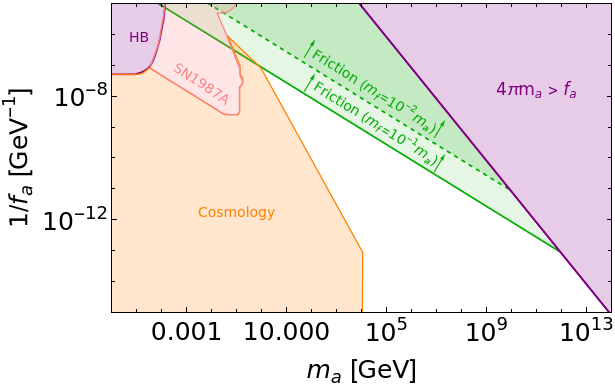}
	\caption{
	The
	$(m_a,1/f_a)$ parameter space displaying the relevance of friction for the case $m_f \gg m_a$ (left) and $m_f < m_a$ (right). 
	We add some ALP constraints coming from cosmology 
	horizontal branch (HB) stars and supernova SN1987A purely for indicative purposes (adapted from \cite{Cadamuro:2011fd,Jaeckel:2015jla,Agrawal_2018,Workman:2022},
	assuming KSVZ-type ALP couplings).
	The purple region is excluded to allow for sufficient scale separation between $m_a$ and $f_a$. 
	These plots were realized with $g_*=106$, $g_b=10$, $\gamma=9$ and
	$\kappa/N_\text{DW} = 1 \,\, (1/2)$ for the case $m_f \gg m_a$ ($m_f < m_a$).
	The green dashed lines indicate contours of $m_f^\text{max}$ relevant only for $m_f \gg m_a$, with $\alpha$ in \eqref{eq:mfmax} taken to be 1/100 as discussed in the text.
	}
	\label{fig:fermionsALP}
\end{figure}

We show the ``friction region'' in Fig.\,\ref{fig:fermionsALP}.
As mentioned above, there is no direct dependence on $m_f$ in this plot.
However, for the friction treatment to be
consistent, we need the fermion to be active below the scale of
dark confinement $\Lambda$, where the ALP parameters are temperature
independent.
In addition, we need to require that the fermion mean free
path is larger than the width of the domain wall,
$m_f < m_a/\alpha^2$, as discussed in the previous section,
with $\alpha \sim 1/100$ for concreteness.
In summary, for given values $(m_a,f_a)$ we shall require
\begin{equation}\label{eq:mfmax}
m_f^\text{max} \equiv \,\text{min}\,(\Lambda,m_a/\alpha^2) > m_f > m_a.
\end{equation}
Our results for $m_f \gg m_a$ are then summarized in the left--hand side of Fig. \ref{fig:fermionsALP}.
\\~

Let us now discuss the case $m_f \ll m_a$.
We can get an estimate for the relevance of friction by
comparing the inverse friction length arising from the 
pressure in \eqref{DP2} with Hubble
evaluated at $T = \nu$.
In fact, note that for $T\lesssim \nu$ both $1/\ell_\text{f}$
and Hubble scale $\propto T^2$,
so that if the inequality is not satisfied at $T=\nu$,
it will not be satisfied at all. Furthermore, we check whether the temperature at which friction dominates, $T=\nu$ , is below the confinement scale, i.e. $\nu<\Lambda$, to ensure the ALP mass is temperature-independent. For this estimate
we shall take $\kappa/N_\text{DW} = 1/2$ such that $r=1$.
We then have
\begin{equation}
 \frac{1}{\ell_\text{f}} = \frac{g_b m_f^2 \nu^2}{2 \gamma \pi^2 m_a f_a^2}
 >  3 \pi \sqrt{g_\ast/90} \frac{\nu^2}{M_\text{Pl}},
\end{equation}
which approximately gives 
\be 
\label{eq:mfineq}
m_a \gg m_f \gtrsim f_a \sqrt{m_a/M_\text{Pl}}.
\ee

We can summarize the results of this section in Fig.\,\ref{fig:fermionsALP},
where the parameter space in which fermion 
friction can be important have been shaded in green.
The left panel shows the case $m_f \gg m_a$,
where the dashed green lines 
indicate contours of $m_f^\text{max}$ in \eqref{eq:mfmax}.
The right panel shows the case $m_f \ll m_a$
for two benchmark values 
$m_f = 0.1 m_a$ and $m_f = 0.01 m_a$.
As we can see, friction can be important for relatively large
(small) values of $m_a$ ($f_a$).
Since the fermion coupling to the ALP is proportional to the mass,
the friction parameter space is reduced for smaller $m_f$.
The implication of this picture for the GW signal will be discussed
in Sec.\,\ref{sec: VOS}.

\subsubsection{Pressure from fermion dark matter after freeze out}
\label{sec:fermDM}
As noticed in \cite{Massarotti:1990qs},
dark matter particles can scatter off the domain wall and impact its evolution.
As we now show, a dark matter fermion
can still give a friction larger than Hubble after freeze--out, which we assume to be instantaneous at $T=T_{\rm fo}$,
if the domain wall tension is small (the same fermion species can nonetheless provide a large friction before freeze out). 

At temperatures $T < T_\text{fo}$ the dark matter velocity
in the bath rest frame becomes increasingly small. Thus, the domain wall
in its rest frame feels a wind of dark matter particles
with momentum $p_f \sim m_f v$. For the reflection
probability to be non--negligible in this regime, we need to require 
that 
\begin{equation}
 \frac{1}{2} m_f v^2 \lesssim m_a
\end{equation}
assuming an $\mathcal{O}(1)$ coupling $\kappa/N_\text{DW}$.
If this condition is satisfied,
all the dark matter particles will be reflected with momentum exchange 
$\Delta p = 2 p_f$.
One can show that domain walls will feel a net pressure given by\,\cite{Massarotti:1990qs}
\begin{equation}
\label{eq:massa}
\Delta P \simeq 2 m_f \mathcal{N} v^2 \simeq 2 \rho_\text{DM} v^2,
\end{equation}
where $\mathcal{N}$ is the fermion number density integrated
over the momentum distribution, and we have introduced the 
dark matter energy density $\rho_\text{DM}$.
We have checked that this result can be consistently obtained with the formalism presented in Sec.\,\ref{sec:pfrompf} when including the appropriate chemical potential $\mu/T$ to describe the fermion population after freeze out. In this case, one can show that the assumption allowing to expand the pressure in the small domain wall velocity -- see discussion below \,\eqref{eq:DeltaPu} -- is never satisfied, and for this reason the resulting pressure in \eqref{eq:massa} is $\propto v^2$ rather than $\propto v$.

\begin{figure}
\centering
 \includegraphics[scale=0.45]{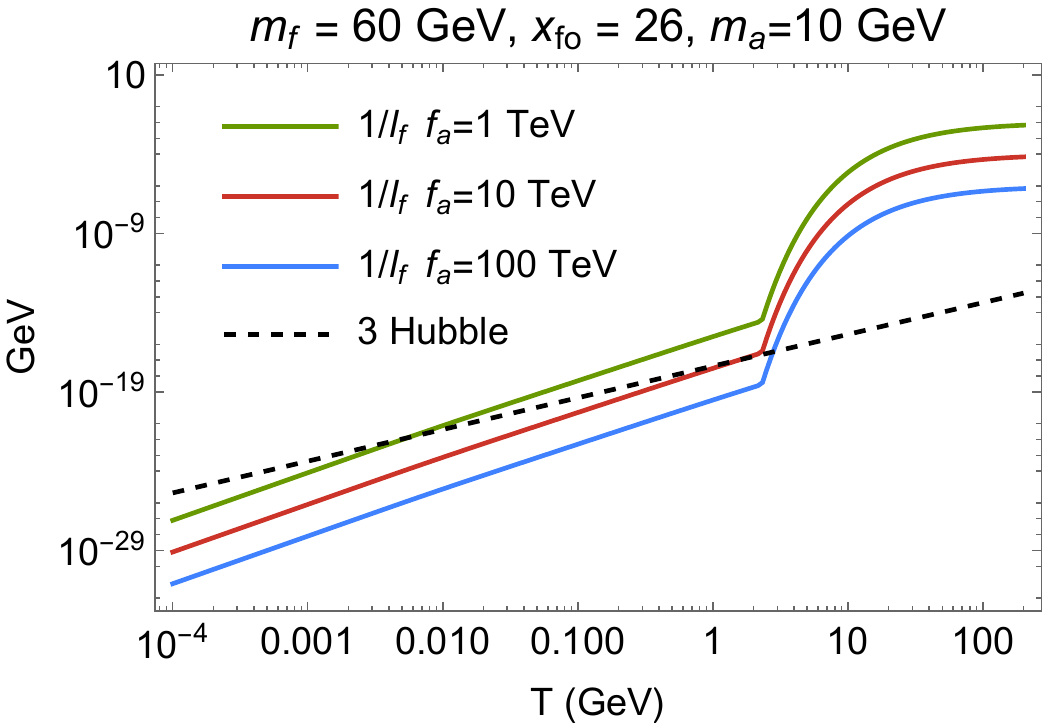}
 \caption{Comparison between $1/\ell_\text{f}$
 and $3H$ for the standard WIMP scenario, $m_f = 60$ GeV,
 $x_\text{fo}= 26$ and $m_a = 10$ GeV, for different values of
 the decay constant. Friction domination requires $1/\ell_\text{f} > 3 H$.
 This figure assumes instantaneous freeze--out at $T_\text{fo} \simeq 2.3$ GeV,
 and we have taken $v=0.5$. For temperatures above freeze-out the friction length is evaluated according to \,\eqref{eq:1elef} for $\kappa/N_\text{DW}=1$.}
 \label{fig:PFO}
\end{figure}

The resulting friction length is 
\begin{equation}\label{eq:DMlf}
 \frac{1}{\ell_\text{f}} \simeq \frac{ 2 \rho_\text{DM} v}{\sigma},
\end{equation}
which is now velocity dependent. Requiring $1/\ell_\text{f} > 3 H$, we find 
that friction domination is possible only for $T > T_\text{min}$,
which we assume in radiation domination,
given by 
\begin{equation}
 T_\text{min} = \frac{\sqrt{\Omega_\text{R}} T_0}{2 M_\text{Pl}^2 \Omega_\text{DM} H_0}
 \frac{1}{v} \sigma \simeq \frac{1}{v} \frac{\sigma}{\text{GeV}^3}\,\text{eV},
\end{equation}
where we have used $\rho_\text{DM} = \rho_c^0 \Omega_\text{DM} (a_0/a)^3$
and $H^2 = \rho_\text{R}/3 M_\text{Pl}^2$, where $\rho_\text{R}$ is the
energy density of radiation.
The consistency of this analysis requires 
to look at dark matter after freeze--out,
namely $T_\text{min} < T_\text{fo} \equiv m_f/x_\text{fo}$. This constraint
can be translated into a constraint on the ALP decay constant $f_a$,
\begin{equation}\label{eq:falim}
 f_a < 30 \,\text{TeV} \sqrt{ \frac{ v m_f}{x_\text{fo} m_a}}
 < 30 \,\text{TeV} \sqrt{\frac{2}{v x_\text{fo}}}
\end{equation}
where we have taken the domain wall tension $\sigma \sim m_a f_a^2$.
The last factor is always less than unity for typical values
$x_\text{fo} \approx 20$ and $v\sim 0.5$, indicating that 
the pressure from a frozen--out fermion can only be relevant 
for rather small decay constants. This is shown in Fig.\,\ref{fig:PFO}
where we plot $1/\ell_\text{f}$ and Hubble for different values 
of the ALP decay constant. At $T_\text{fo} \simeq 2.6$ GeV the exponential
drop stops as number--changing processes freeze out by assumption. 
Friction domination is still possible  
for $T< T_\text{fo}$ as long as $f_a \lesssim 10$ TeV.

\subsubsection{ALPs scattering off ALP domain walls}
\label{selfreflection}

A population of ALPs can scatter
off the domain walls and contribute to the friction as
well. In fact, as opposed to the cosine potential
which leads to identically zero self--reflection,
a realistic ALP potential that takes into account
the possible flavor structure of the confining group
(see e.g.\,\cite{Villadoro}) provides a non--zero
reflection probability, as argued in \cite{Huang:1985tt}.
Identically vanishing self reflection is found also in the case of the simple $\phi^4$ theory.
As DW numerical simulations are usually performed assuming either of these potentials, see e.g.\,\cite{Press:1989yh,Martins:2016ois,Hiramatsu:2012sc,OHare:2021zrq}, 
this effect is yet to be explored.

In the case of a realistic ALP potential, we have checked numerically 
(see Appendix \ref{app:R} for details) that the scattering probability can be
effectively parameterized as 
\begin{equation}
 \mathcal{R}(p_z) = \begin{cases}
                1 \quad & p_z \lesssim m_a \\
                0 \quad & p_z \gtrsim m_a
               \end{cases}.
\end{equation}
In order to estimate the effect of self--reflection,
one should refer to the various axion populations
that may be present after the formation of the DWs. 
These include a population of ALPs produced via the misalignment mechanism,
and ALPs produced from the network of topological defects, namely from cosmic strings and the domain walls themselves, see
e.g.\,\cite{Hiramatsu:2012sc,Gorghetto:2018myk,Gorghetto:2020qws}. 
A detailed study of self friction from these different populations is left for future work.

\subsection{Friction from SM leptons at Pulsar Timing Arrays}
\label{sec:PTA}

Recently, evidence for a possible SGWB signal has been reported by Pulsar Timing Array (PTA) experiments such as NANOGrav \cite{NANOGrav:2020bcs}, European PTA \cite{Chen:2021rqp} and Parkes PTA \cite{Goncharov:2021oub}.
While it is possible that such signal emerges from astrophysical phenomena \cite{Burke-Spolaor:2018bvk}, it is very interesting to explore a possible cosmological origin related to the presence of topological defects, see e.g.\,\cite{Blasi:2020mfx,Ellis:2020ena,Blanco-Pillado:2021ygr,Buchmuller:2020lbh,Bian:2022tju,Chen:2022azo}
for a cosmic string interpretation, and e.g.\,\cite{Bian:2020urb,Wang:2022rjz,Ferreira:2022zzo,Sakharov:2021dim} for DWs.

Here we explore the possibility that friction from SM particles coupled to the ALP may affect the SGWB signal from the DW network in frequencies relevant for the PTA experiments, focussing in particular on NANOGrav, for which the relevant annihilation temperature for the signal peaking around $10^{-9}$ Hz is of $T_\text{NG} \sim 50\,\text{MeV}$--- see \eqref{eq:pheno_freq}. 
We shall assume for concreteness that the ALP couples equally to the SM leptons with an effective Lagrangian as in \eqref{eq:apsi} with coupling $\kappa/N_\text{DW}$,
and that it also couples to photons with coupling strength $E/N_\text{DW}\sim1$ (see Appendix \ref{app: D}, eq.\eqref{eq:axionphoton}).
We instead assume no ALP couplings to gluons and quarks for simplicity,
as the annihilation temperature is close to the QCD crossover making theoretical calculations uncertain in this regime.

\begin{figure}[t]
\includegraphics[scale=.46
]{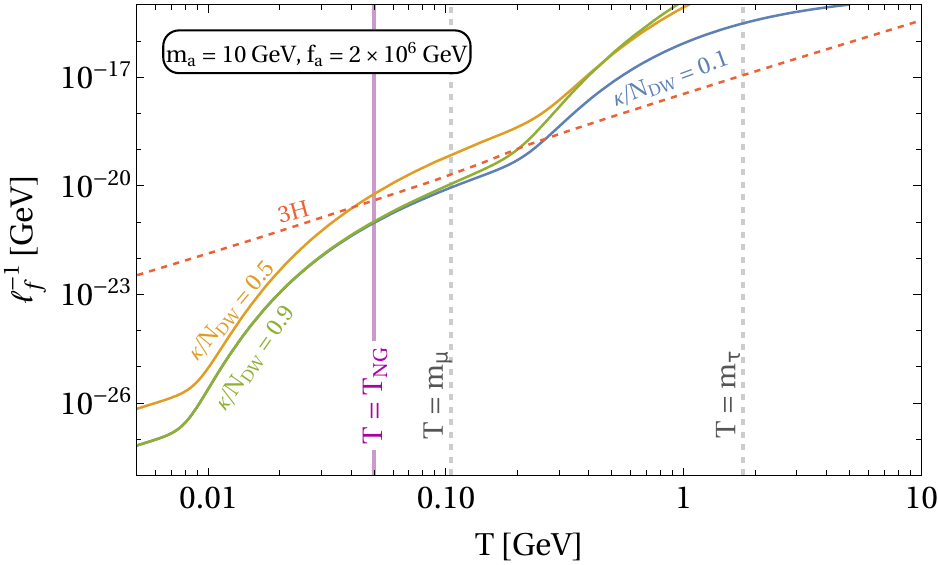}
\includegraphics[scale=.358]{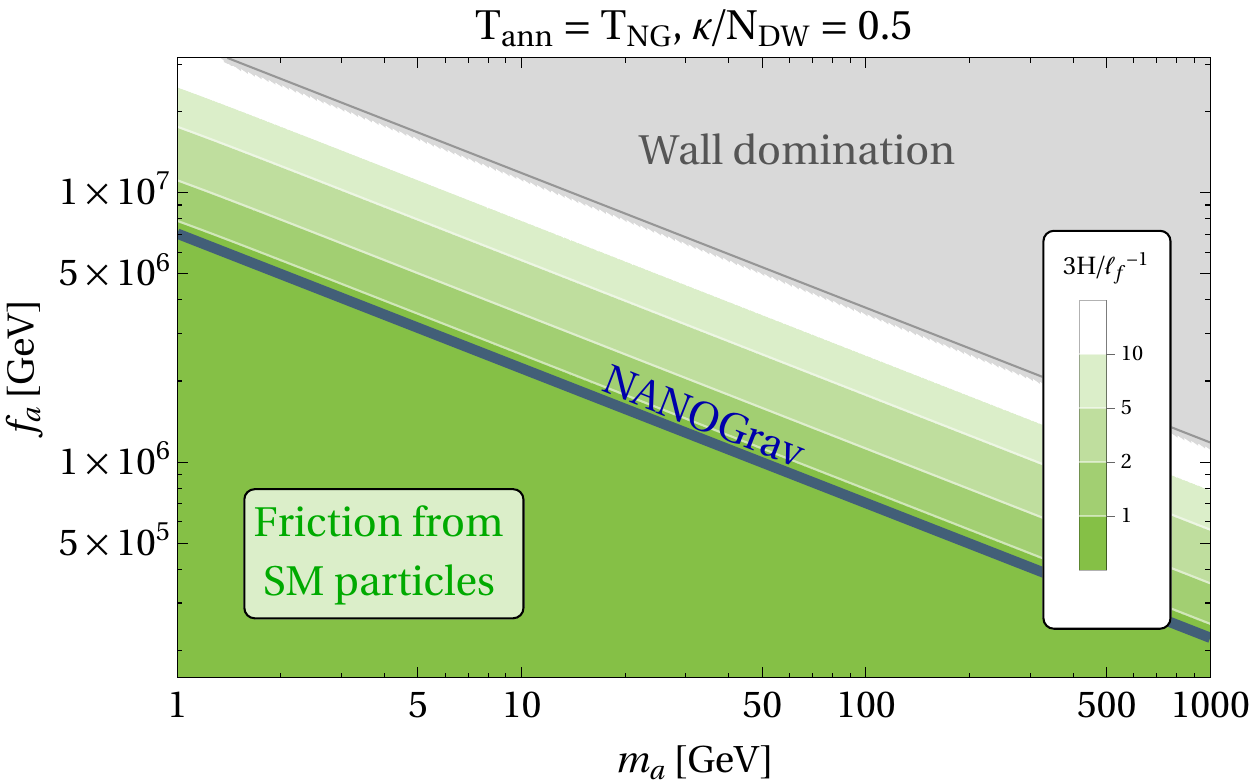}
\caption{\textbf{Left.} Evolution of the Hubble parameter (red, dashed) and the inverse friction length from SM particles (including the electron, muon, tau) for an ALP with mass $m_a = 10$ GeV and decay constant $f_a = 2\times 10^{6}$ GeV as a function of temperature. Different friction lengths are shown for different ALP-fermion couplings $\kappa/N_\text{DW} = 0.1$ (blue), $0.5$ (orange) and $0.9$ (green).
A purple line shows the temperature $T_\text{NG}$ at which NANOGrav is most sensitive ($T_\text{NG}\simeq 50\ \text{MeV}$), corresponding to a peak frequency $f_\text{peak} = 6.10\times 10^{-9}$ Hz \cite{Schmitz:2020syl}. \textbf{Right.} $(m_a,f_a)$ parameter space for fixed annihilation temperature $T_\text{ann} = T_\text{NG}$. The blue line corresponds to a GW signal from ALP DWs as large as the current NANOGrav sensitivity,
whereas the green regions correspond to parameter points for which the ratio $3H/\ell_\text{f}^{-1}$ is 10, 5, 2 and 1 for $\kappa/N_\text{DW} = 0.5$.}
\label{fig:Nanograv}
\end{figure}

In Fig.\,\ref{fig:Nanograv} (left), we show for some representative benchmarks
with $f_a = 2 \cdot 10^{6} \,\text{GeV}$ and $m_a = 10 \,\text{GeV}$ the friction length for leptons 
obtained numerically as a function of the temperature taking equilibrium distributions with negligible chemical potential, 
and compare it to Hubble.
Using the results of \cite{Huang:1985tt}, we have checked that friction from photons is never relevant in the temperature range considered.
We also indicate the annihilation temperature $T_{\text{NG}}$. In the evolution of the lepton friction length from high to low temperatures, we observe that the friction from the $\tau$ decreases because of Boltzmann suppression,
and that the muon provides the main source of friction around $T_{\text{NG}}$;
we can also check that the friction from electrons is always negligible, as in fact the condition \eqref{eq:mfineq} is not satisfied.
As for this benchmark the main contribution
is given by the muon, whose mass is much smaller than the ALP mass, the relevant formula for the pressure is given in \,\eqref{eq:T2}, and its effect is maximized for $\kappa/N_\text{DW} = 1/2$.

From Fig. \ref{fig:Nanograv} (left) we can conclude that
there is a significant range of temperatures, $T \gtrsim T_{\text{NG}}$, where the inverse friction length from leptons is comparable or larger than Hubble, depending on the effective ALP-lepton coupling. In this temperature range we can therefore expect deviations from the scaling regime.
In particular, the standard GW production based on the oscillation of the DWs will be suppressed\,\footnote{New contributions to the GW signal may nonetheless arise from the energy release in the plasma.}. 

In order to estimate the impact of friction from leptons on the ALP parameter space that can be probed at NANOGrav, we perform a further analysis shown in Fig. \ref{fig:Nanograv} (right).
We fix the annihilation temperature at $T_\text{NG}\simeq 50\ \text{MeV}$ and we scan over the ALP parameter space $(m_a,f_a)$. 
With this procedure we identified the band that corresponds to a signal as large as the current NANOGrav sensitivity assuming the standard DW scaling regime,
indicated as a blue solid line in the plot.
As mentioned above, friction can however alter these conclusions if the SM leptons couple strongly enough to the ALP.
In order to quantify this statement we display contours of $3H/\ell_f^{-1}$ evaluated at $T_\text{NG}$, showing that
friction is indeed not negligible for this benchmark scenario.
\medskip

\section{Domain wall evolution with the Velocity One Scale model}
\label{sec: VOS}

In the previous section we have shown how friction from particle scattering can be relevant during the evolution of the ALP domain walls.
In this section we quantify this effect by deriving the
evolution of the energy density, and the resulting SGWB signal.
Previous studies on this subject have assumed simple temperature scalings for the friction (typically $\sim T^4$) \cite{NAKAYAMA2017500}, which is however not realized in the case of fermion scattering off ALP DWs.

\subsection{Friction domination}
In order to describe the dynamics of the DW network we will make use of the velocity-dependent one-scale (VOS) model,
which can be deduced from the equation of motion of one DW in the thin wall approximation \cite{Avelino:2005kn,Martins:2016ois,Avelino:2019wqd,Avelino:2020ubr,Avelino:2022zem}.
Indeed, from the action in \eqref{eq:DWaction} one can derive the equations of motion for a single domain wall
as we reviewed in Sec. \ref{sec:friction_eom}.
These include an equation describing the time evolution of the velocity of the DW \eqref{eq:velocity_dw}, as well
as an equation governing energy conservation \eqref{eq:energy_DW}.
To quantitatively describe the dynamics of a network of domain walls, 
one can resort to the so--called \emph{thermodynamic}
approach which derives relations for averaged quantities,
such as the root--mean--squared velocity of the network,
$v$, and the average energy density $\rho$.
The basic assumption is that there is one single scale $L$ controlling the curvature of the DW 
as well as the average distance between the DWs.
The resulting system of coupled differential equation is the velocity-dependent one-scale (VOS) model
\cite{Martins:2016lzc}
\begin{align}
\label{eq:L}
    & \frac{\text{d}L}{\text{d}t} = H L + v^2 \frac{L}{\ell_\text{d}}
    + c_w v + d[k_0-k(v)]^r,\\
    \label{eq:v}
    & \frac{\text{d}v}{\text{d}t} = (1-v^2)\left(\frac{k(v)}{L}-
    \frac{v}{\ell_\text{d}}\right),
\end{align}
where $t$ is the cosmic time with
$H=\dot{a}(t)/a(t)$,
and $L$ is the characteristic length scale of the network
defined by
\begin{equation}
    L \equiv \frac{\sigma}{\rho_w}.
\end{equation}
The damping length $\ell_\text{d}$ is given in \eqref{eq:dlength}.
The equation governing the velocity is the equivalent of the equation 
we derived in \eqref{eq:velocity_dw} for a single DW, and
indeed, the friction enters in the same combination that we already anticipated in Sec. \ref{sec:friction_eom}.
The equation for the length $L$ is the equivalent of the equation governing the energy density \eqref{eq:energy_DW}, 
with inclusion of curvature terms and a term parameterizing particle production.
The parameters appearing in the VOS equations for the DW network can be calibrated with numerical
simulations, as done in \cite{Martins:2016lzc}, from which we take the numerical values ($c_w = 0$, $d = 0.26$, $r = 1.42$, $k_0 = 1.77$, $q = 3.35$, and $\beta = 1.08$).
Our main conclusions will not be affected by the specific choice of these parameters.

Before delving into the numerical solutions of the VOS equations for some representative benchmark, it is instructive to 
analyze approximated analytical solutions to grasp the main features of the evolution of $L$
and $v$ in the presence of a friction force.
First of all, one can verify that, when Hubble dominates in the damping length \eqref{eq:dlength},
the solution of the VOS equations approaches the scaling regime
\begin{equation}\label{eq:scaleapp}
 L = L_0 t, \quad v = v_0,
\end{equation}
where the constants depend on the exact values of the parameters
in \eqref{eq:L}. For the choices in \cite{Martins:2016lzc}, one obtains
$L_0 \simeq 1.1$ and $v_0 = 0.38$.

On the other hand, we can study the scenario in which friction dominates over Hubble, i.e. if $1/\ell_{\text{d}} \sim 1/\ell_\text{f}$ in \eqref{eq:dlength}.
Assuming a simple power-law time-dependence of the friction length, i.e. $\ell_{\text{f}} \propto t^\lambda$, we
obtain as asymptotic solutions for the characteristic length and for the velocity
\begin{equation}\label{eq:Lapp}
    L(t) \simeq k_0 \sqrt{2 \ell_\text{f} t} 
    \begin{cases}
    \sqrt{1/\lambda} \quad & \lambda \neq 0 \\ 
    \sqrt{\text{log}(t/\ell_\text{f})} \quad & \lambda = 0
    \end{cases}
\end{equation}
and 
\begin{equation}\label{eq:vapp}
    v(t) \simeq (\ell_\text{f}/t)^{1/2}
    \begin{cases}
    \sqrt{2/\lambda}   \quad & \lambda \neq 0 \\ 
    (2 \,\text{log}(t/\ell_\text{f}))^{-1/2} \quad & \lambda = 0
    \end{cases}.
\end{equation}
We note that the energy density of the DW network ($\sim L^{-1}$) decreases faster (slower) than in the scaling regime when
$\lambda > 1$ ($\lambda <1$). Moreover, it decreases equally fast or faster than the radiation energy density as
soon as $\lambda \geq 3$.

\begin{figure}
\centering
 \includegraphics[scale=0.7]{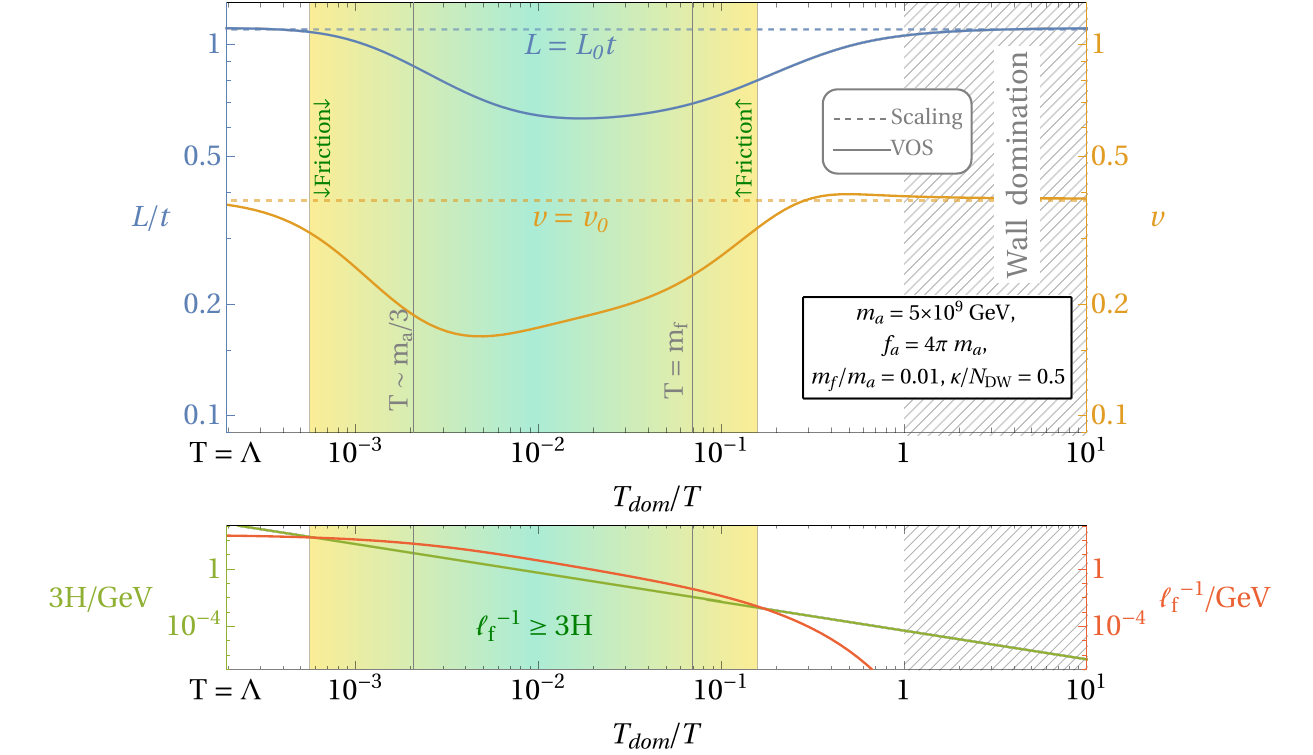}
 \caption{Comparison between the evolution of the scale $L$ (blue) and velocity $v$ (orange) of the DW system in a partially friction dominated regime (green shaded area) and scaling regime for benchmark values $m_a = 5\times 10^9$ GeV, $f_a = 4\pi m_a$, $m_f = 0.01 m_a$, $\kappa/N_\text{DW} = 1/2$ and fermionic degrees of freedom $g =20$, starting at a temperature $T = \Lambda$. The numerically solved VOS equations are indicated by a solid line, whereas the scaling solutions, i.e. $L =L_0 t$ and $v = v_0$, are represented by dashed lines. The green and red lines show the evolution of the Hubble parameter and inverse friction length respectively. The friction dominated regime is determined by the condition $\ell^{-1}_\text{f} \geq 3H$ and gray lines indicate the transitions of the friction length $\ell_\text{f}$ from constant behavior to $\ell_\text{f}^{-1}\sim T^2$ at $T \sim m_a/3$ and exponential decrease at $T = m_f$. Wall domination happens when $T = T_\text{dom}$, as indicated on the right side of the plot.}
 \label{fig:5lines}
\end{figure}

We now proceed to numerically solve the VOS equations in a representative case and 
show the quantitative effect of friction on the evolution of the DW network.
We consider the case of ALP DWs in the presence of a fermion coupled to the ALP as in Sec. \ref{sec:fermion_friction}.
We choose a benchmark with $m_f<m_a$ where the friction can be relevant over a range of temperatures (see Fig. \ref{fig:fermionsALP}) in the interval 
$\sqrt{m_a M_\text{Pl}/f_a^2} \, m_f \gtrsim T \gtrsim  m_f$.
We numerically compute the pressure and the friction, and use it to numerically solve the VOS equations.
The resulting evolution of $L$ and $v$ is displayed in Fig. \ref{fig:5lines}.
We employ as reference temperature $T_{\text{dom}}$, that is the temperature at which the energy density of the DW in scaling would equal the radiation energy density, as given in \eqref{Tdom}.
The DW network should annihilate at a temperature $T_{\text{ann}} > T_{\text{dom}}$, through the introduction of the bias term in order to have a viable cosmology,
as we will discuss later.
For now, we are interested in understanding the evolution of $L$ and $v$,
and we illustrate it as a function of temperature in Fig. \ref{fig:5lines}.

We consider as highest temperature in our analysis the temperature corresponding to the confinement scale $\Lambda$. For higher temperatures, the ALP potential
gets sizable temperature corrections, effectively making $m_a$ and $\sigma$ temperature-dependent, jeopardizing the consistency of our analysis.
If Hubble dominates for even a small temperature window around $T \sim \Lambda$, then the evolution of $L$ and $v$ is attracted to the scaling solution.
In the standard scaling regime, the characteristic scale $L$ grows like $t$ and the velocity is constant.
At $T \sim \sqrt{m_a M_\text{Pl}/f_a^2} \, m_f$, friction becomes dominant with respect to Hubble in the damping length, 
and $1/\ell_{\text{f}}$ is constant. As a consequence, the velocity drops and also $L/t$ drops, approaching the asymptotic behavior of $L/t \sim 1/T$ and $v \sim 1/T$ (up to log terms),
see \eqref{eq:Lapp} and \eqref{eq:vapp}. Then, for $m_a \gtrsim T \gtrsim m_f$, 
the inverse friction length (still dominating over Hubble) features a period of $1/\ell_\text{f} \sim T^2$, and as a consequence $L/t$ and $v$ go towards a different
(and constant in $t$) evolution.
The asymptotic analytic solution 
is not reached since at $T \sim m_f$ the pressure from the fermion drops exponentially, the friction becomes negligible, 
and $L/t$ and $v$ are attracted again to the standard scaling solution which is driven by Hubble damping, i.e. $1/\ell_{\text{d}} \sim 3 H$.

Note that during friction domination, the typical length scale of the DW network grows slower than the Hubble scale. Hence, the energy density
of the DW network grows faster than in scaling. However, the GW emission, which is related to the variation of $L$ during time will usually be suppressed, as we shall see.

\subsection{Gravitational wave emission during friction}\label{sec: GW emission}

Once the evolution of the domain wall network is computed, 
we can derive the emitted gravitational waves.
The estimates we will use to determine the SGWB from DWs are based on several
simplifying assumptions. In particular, we employ the results obtained in numerical simulations that have been
performed for DW network in the scaling regime as reviewed in Section \ref{sec:Gravitational waves from domain wall dynamics}, assuming these are approximately valid also in friction
dominated regimes.
A more precise derivation of the GW signal during friction regimes would require a dedicated numerical simulation, which goes 
beyond the scope of this paper.

In our simplifying assumptions, the GW signal during friction will have the same spectral shape but a different peak amplitude with respect to the scaling regime.
In order to estimate the latter, we  
consider the instantaneous emission power in gravitational waves at a certain $t$, which is assumed to be the one from the quadrupole formula,
\begin{equation}\label{eq:P}
 P_\text{gw} = k G \left( \frac{ d^3 I}{dt^3} \right)^2,
\end{equation}
with the quadrupole of the domain wall network estimated as (we denoted explicitly the time dependence of $L(t)$)
\begin{equation}\label{eq:quad}
 I(t) = \sigma L(t)^4
\end{equation}
where here $L(t)$ is determined by solving the VOS equations.
The factor $k$ will be fixed by comparing with the numerical
simulations in the scaling regime.
The energy density in gravitational waves is
estimated as
\begin{equation}\label{eq:rhoGW}
 \rho_\text{gw}(t) = \frac{ P_\text{gw} H^{-1}}{H^{-3}} \frac{H^{-3}}{L^3}
 = \frac{ P_\text{gw} H^{-1}}{L^3}.
\end{equation}
The $H^{-1}$ factor in the numerator is the duration of the emission,
corresponding to one Hubble time.
The $H^{-3}/L^3$ factor corresponds to the number of regions
of size $L$ that are emitting in one Hubble volume, so that the final
$1/L^3$ can be thought of as the number density of the emitters.

The calibration factor $k$ in \eqref{eq:P} will be fixed by requiring
that the GW energy density \eqref{eq:rhoGW}
is in agreement with  numerical simulations 
where friction is neglected and the standard scaling
regime is found, as in \eqref{eq:Omegadfpeak}. In the scaling regime, we have $L = L_0 t$ and substituting this in \eqref{eq:rhoGW} using $H = 1/(2t)$ in a radiation dominated Universe leads to $\rho_\text{gw} = 1152 kG L_0^5\sigma^2$. Furthermore, requiring that the integrated GW energy density per logarithmic frequency (see \eqref{eq: GW spectrum}) is equal to $\rho_\text{gw}$ in \eqref{eq:rhoGW},
\begin{equation}
\int d\ln f \frac{d\rho_\text{gw}(t)}{d\ln f} = \rho_\text{gw}(t)\,,
\end{equation}
gives
\begin{equation}\label{eq:peak_and_full}
\left(\frac{d\rho_\text{gw}(t)}{d\ln f}\right)_\text{peak}= \frac{3}{4}\rho_\text{gw}(t)
\end{equation}
for the assumed spectral shape in \eqref{eq:Omegadf}. The calibration factor $k$ can then be readily determined, resulting in
\begin{equation}
\label{eq:calibration}
k = \frac{\Tilde{\epsilon}_\text{gw}\mathcal{A}^2}{864\ L_0^5}\,.
\end{equation}
Once the calibration factor is determined, we can use the previous equations to compute the GW peak amplitude
as a function of $L(t)$, which is obtained by solving the VOS equations.

\begin{figure}
\centering
 \includegraphics[scale=0.7]{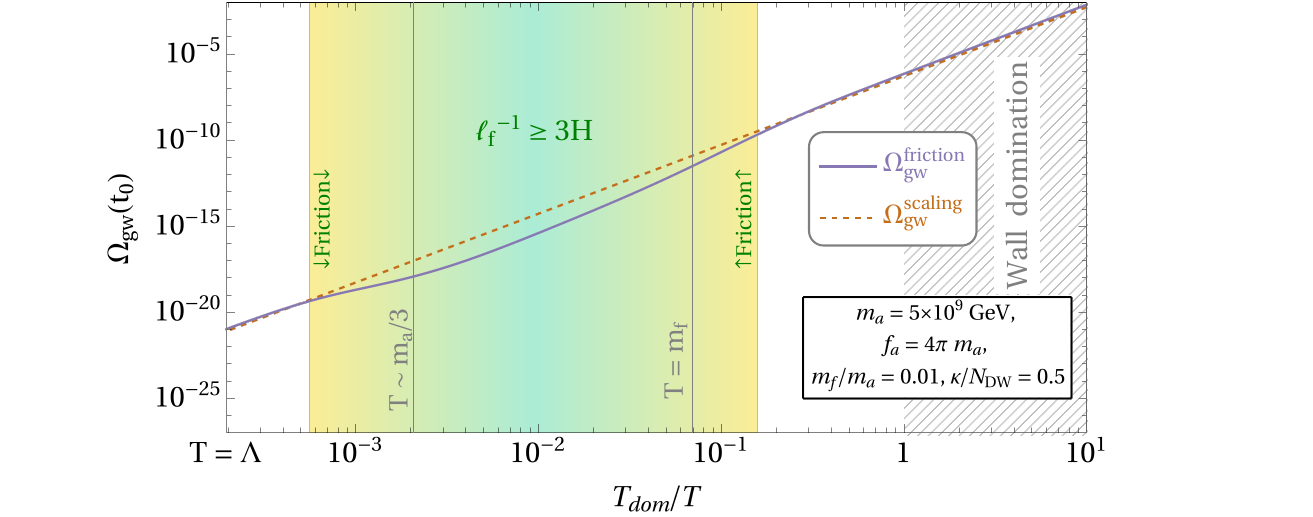}
 \caption{Comparison between the  evolution of the redshifted GW peak amplitude $\Omega_\text{gw}$ at present time $t_0$ when a friction dominated regime (green shaded area) and scaling regime have occurred for similar benchmark values as Fig. \ref{fig:5lines}, starting at a temperature $T = \Lambda$. The evolution under a friction dominated regime is indicated by a solid purple line, whereas the scaling evolution is shown by a dashed brown line. Similar lines and regions as in Fig. \ref{fig:5lines} are shown.}
 \label{fig:5thline}
\end{figure}

With this procedure in Fig. \ref{fig:5thline} we show the (redshifted) instantaneous normalized energy density of gravitational waves \eqref{eq:OmegaGW} (evaluated at the peak frequency\footnote{Note that the peak frequency moves from $H(t)$ in a scaling regime to $1/L(t)$ in a friction dominated regime, as it is related to the scale of the domain wall.})
for the same benchmark as considered in Fig. \ref{fig:5lines}.
We note that
during the scaling regime with $L\propto t$, the later the emission, the
stronger the signal in gravitational waves. 
This implies that in the scaling regime, the 
spectrum today is essentially dominated by
the last emission before the DW network annihilates. 
This results in a single peak with the amplitude \eqref{eq:OmegaGW}
evaluated at the annihilation time $t = t_\text{ann}$.
In general, this is also the case if there is a regime of friction domination.
\footnote{
In Appendix \ref{App:arche} we explore the possibility that, in the presence of a significant friction,
the dominant and visible GW signal is actually generated at a time prior to $t_\text{ann}$, when the DW network was still in scaling regime,
and we show that this typically requires some tuning in the choice of the parameters and of the annihilation time.
}

The importance of taking into account friction when studying GW signatures from DW is reflected in the general observation that the amplitude of the SGWB signal 
will be weaker than the corresponding signal assuming scaling (for the same given $T_\text{ann}$). Indeed, using the asymptotics given in eqs.\eqref{eq:Lapp} and \eqref{eq:vapp} one finds that the mean square velocity scales like $v \sim L/t$. Plugging this back in eq.\eqref{eq:rhoGW} and ignoring higher order derivatives, one finds $\rho_\text{gw, friction} \sim G \sigma^2 v^5$.
This should be compared to the standard result in the scaling regime, $\rho_\mathrm{gw} \sim G \sigma^2$ which shows that the GW signal will be damped because of the velocity suppression. Note however that the GW signal returns quite fast to scaling at the end of the friction dominated regime, as it is visible in Figure \ref{fig:5thline}.

In general, one should also keep in mind that if friction dominates, there would be a significant release of energy of the DW network in the plasma. This could be 
a further source of GW, as it occurs in first order phase transitions with sound waves and turbulence contributions, which might further modify the expected GW spectrum.
In addition, in scenarios in which friction provides a very strong suppression of the GW emission as estimated from the quadrupole formula \,\eqref{eq:P}, the contribution from the dynamics intrinsically associated to the annihilation of the DW network may become the dominant one. We leave these investigations for future work.

\section{Exploring the ALP parameter space with gravitational waves}
\label{sec:ALPpheno}

In this section, we study the GW signal in the ALP parameter space and the possible impact of friction in some concrete benchmark.
We consider ALP effective theories with $N_{\text{DW}}>1$, keeping the axion mass $m_a$, the axion decay constant $f_a$,
and the bias as fundamental parameters. 
This class of models can also include the QCD axion, or generalizations thereof.
For generic ALPs, possibly originating in hidden sectors, the couplings to the SM are model dependent and can be considered as free parameters.

Hence, we first investigate the GW signal from ALP DWs without imposing other phenomenological requirements.
We explore the ALP parameter space and we show that a certain quality is required in order for the ALP DW signal to be detectable by current and future GW experiments.
Then, we select one type of higher-dimensional Planck suppressed operators as bias, 
and show the detectability of GW signatures in the broad ALP parameter space.

\subsection{Axion quality and gravitational waves}

\begin{figure}
\centering
\includegraphics[scale=0.6]{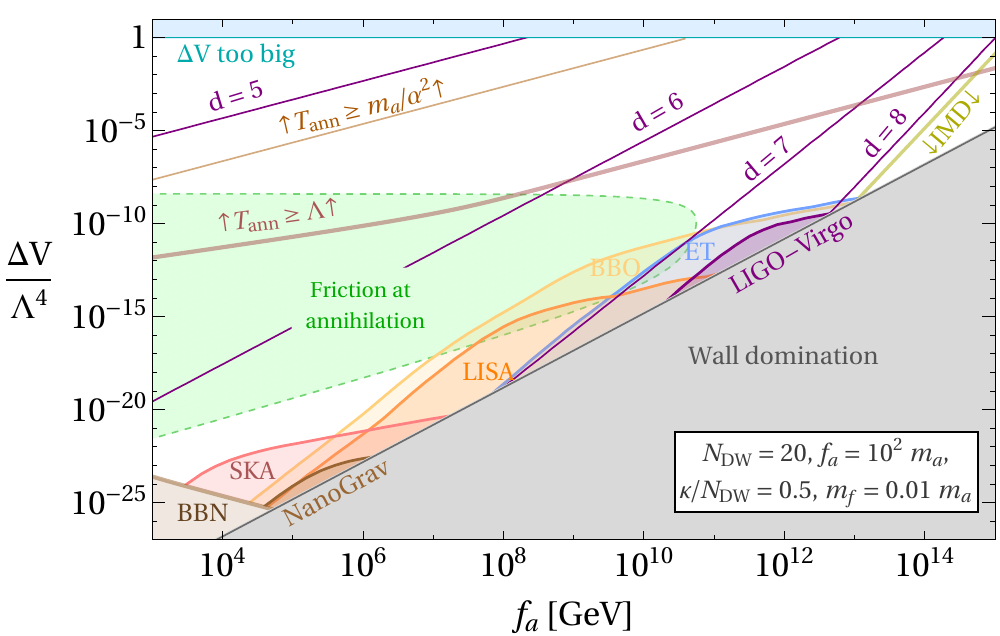}
  \caption{$(f_a,\Delta V/\Lambda^4)$ parameter space with $\Lambda = \sqrt{m_a f_a}$ for fixed ratios $f_a/m_a = 100$, $\kappa/N_\text{DW} = 0.5$, $g = 20$ and $m_f = 10^{-2}m_a$. Solid contours bound the sensitivity regions of ET (blue), BBO (yellow), LIGO-Virgo (purple), LISA (orange), SKA (pink) and NanoGrav (brown), for which the sensitivity curves were taken from \cite{Schmitz:2020syl}, except for LIGO-Virgo A+ taken from \cite{KAGRA:2021kbb}. The green dashed contour encompasses the friction region defined as satisfying the condition $3H_\text{ann}\leq \ell_\text{f}^{-1}$. The cyan and grey regions are forbidden due to the bias being too large or the walls dominating the Universe, respectively. The parameter values for which $T_\text{ann}>\Lambda$ are bounded from below by the dark pink line. Fermions with a mean free path smaller than the DW width are in the top left corner, above the brown line at which condition \eqref{eq: T_mean_free_path} is satisfied. The unviable brown region ``BBN'' indicates where the BBN constraints would be violated. The transition to the IMD era is marked by the yellow line. Purple contours indicating the dimension $d$ of higher dimensional operators given in \eqref{eq: quality_HDO} ranging from $d = 5$ to $8$ and $N_\text{DW} = 20$ are shown as well.}
\label{fig: biasplot1}
\end{figure}

In order to visualize the reach of gravitational wave experiments in a two dimensional parameter space, 
we fix $m_a/f_a$ to a representative value and
we display our results in the $f_a$ vs $\Delta V$ plane in Fig. \ref{fig: biasplot1}. We employ as variable the dimensionless ratio $(\Delta V/\Lambda^4)$, which
gives an estimate of the \emph{quality} of the axion potential.
For every point we solve the VOS equations and we determine the evolution of $L(t)$, which can deviate from scaling when friction is relevant,
and we obtain the corresponding SGWB signal.
The different regions and lines in the plots are derived as follows:
\begin{itemize}
\item In the upper part of Fig. \ref{fig: biasplot1}, a forbidden region called ``$\Delta V$ too big'' is shown where the bias becomes too large, i.e. $\Delta V > \Lambda^4$, effectively spoiling the (approximate) discrete symmetry such that DWs do not form.

\item To investigate the impact of friction, we consider an axion coupled to a fermion in thermal equilibrium with the Standard Model bath, as 
studied in Sec. \ref{sec:fermion_friction}.
For concreteness, 
we set the fermion mass w.r.t. the axion mass to $m_f/m_a = 10^{-2}$, set $\kappa/N_\text{DW} = 1/2$ and took $g = 20$.
The friction region is defined as the region in which the Hubble parameter at annihilation is smaller than the inverse friction length, i.e. $3H_\text{ann} \leq \ell_\text{f}^{-1}$.

\item
The DWs annihilate at a time $t_\text{ann}$ when the pressure from the bias $p_V \sim \Delta V$
becomes comparable to the DW tension force, $p_T \sim \sigma/L$, resulting in
\be
\frac{\sigma}{L(t_\text{ann})} = \Delta V\,.
\label{eq: tann_eq}
\ee
Note that this is the condition setting the annihilation time also in the case of friction dominated regime. Using the analytic scaling in \eqref{eq:Lapp} and \eqref{eq:vapp}, one can indeed 
show that during friction this condition corresponds to balancing the bias with the friction pressure, i.e. $\Delta P \sim \Delta V$.
In order to be consistent with BBN constraints, the DWs need to annihilate above $T_\text{ann} \sim 1$ MeV.
This determines the brown unviable region in Fig. \ref{fig: biasplot1}.
We do not consider constraints deriving from particles produced at DW annihilation, whose properties and lifetime can be model dependent.

\item The DW energy density $\rho_{w} = \sigma/L$ will dominate the Universe, i.e. be larger than the critical energy density, at a time $t_\text{dom}$
\begin{equation}
\label{eq: tdom_eq}
\frac{\sigma}{L(t_\text{dom})} = 3H^2(t_\text{dom})M_\text{Pl}^2\,.
\end{equation}  
This equation together with \eqref{eq: tann_eq} 
defines a viable region of parameter space where the DWs annihilate before dominating the Universe, i.e. $t_\text{ann} \lesssim t_\text{dom}$. 
 In the right bottom corner, the forbidden region ``Wall domination'' is found where the DWs do not annihilate before dominating the energy density of the Universe, i.e. $t_\text{ann}\geq t_\text{dom}$.

\item 
We also assume that the ALP couples to photons  as can arise e.g. if the fermion is charged under the SM $U(1)_Y$
\footnote{For concreteness we set $E/N_\text{DW} = 1$, with $E$ the electromagnetic anomaly coefficient, see Appendix \ref{app: D}.}. 
The latter is a minimal possibility to keep the fermion in thermal equilibrium with the bath.
On the other hand, we assume there is no coupling between the axion and other SM particles.
Under these assumptions, we find that the ALPs decay prominently into fermions (see Appendix \ref{app: D}).
For every parameter value, we also check whether an intermediate matter dominated (IMD) epoch is established. Indeed, the ALP population produced by the DWs might dominate the Universe if they do not decay fast. 
This intermediate period alters the peak amplitude and frequency as discussed in \cite{ZambujalFerreira:2021cte} and reviewed in Appendix \ref{app: D}. 

\item In order for our computation to be consistent, we have to take into account two constraints on $T_{\text{ann}}$,
which are also shown in the plot.
First we must have that $T_{\text{ann}} < \Lambda$ in order for the tension to be temperature independent.
The change of slope of the line $T_\text{ann} = \Lambda$ in the friction region is due to the fact that during friction the annihilation temperature (time) decreases (increases) for fixed parameter values. 
Second, the condition on the mean free path implies that $T \lesssim m_a/\alpha^2$ as discussed around \eqref{eq: T_mean_free_path}.
Above these two lines our assumptions are not valid and our computation should be improved.
However, we note that the interesting region for the SGWB signal lies below these lines, within the allowed region.

\item The region of parameter space that can be probed by the GW experiments correspond to parameter values for which the associated GW spectra exceed the so--called power-law integrated sensitivity curves of various GW experiments. The experiments we consider are the future Einstein Telescope (ET) \cite{Punturo:2010zz,Hild:2010id, Sathyaprakash:2012jk,Maggiore:2019uih}, LISA \cite{LISA:2017pwj,Baker:2019nia}, BBO \cite{Corbin:2005ny, Harry:2006fi}, SKA \cite{Janssen:2014dka}, NANOGrav \cite{NANOGrav:2020bcs} and the LIGO-Virgo observatories \cite{Harry:2010zz,LIGOScientific:2014pky,VIRGO:2014yos} (more specifically with the upgrade of the advanced LIGO facilities known as A+ \cite{LIGO-Virgo}). The form of the GW spectrum during the scaling regime is found from numerical simulations \cite{Hiramatsu:2010yz,Hiramatsu:2013qaa,Saikawa:2017hiv}, for which its redshifted shape is given by \eqref{eq:Omegadf} and \eqref{eq:OmegaGW}. We assume the spectral shape is unaltered during friction domination. As the spectrum today is dominated by the last emission at annihilation, we evaluate both the peak amplitude and peak frequency at $t_\text{ann}$.\footnote{
In principle,
due to friction effects, it is possible that amplitudes at earlier times are greater than at $t_\text{ann}$ as explained in Sec. \ref{sec: GW emission}. 
However, as previously mentioned, this will not happen for generic choices of the model parameters. In this analysis, the GW amplitude 
at $t = t_\text{ann}$ is found to dominate.}
\end{itemize}

By combining all previous constraints,
Fig. \ref{fig: biasplot1} shows the region of the parameter space of ALP models with $N_{\text{DW}}>1$ that can be probed by current and future GW experiments.
Note that in Fig. \ref{fig: biasplot1}  we do not display ALP constraints arising from ALP couplings with photons or other SM particles, which are generically model dependent, while we will reintroduce them in the next subsection.

As expected, friction can affect part of the parameter space within the experimental reach. For this illustrative benchmark the signal generated in a friction dominated regime can be probed at ET, LISA and BBO (notice that friction is here not relevant at the annihilation time for PTAs as opposed to Fig.\,\ref{fig:Nanograv} because of the concrete choice of $f_a$ and $m_a$).

Fig. \ref{fig: biasplot1} was obtained by keeping the bias term as a free input variable. However, we can identify lines corresponding to specific choices of the bias term in this plot.
In particular, for the case
of higher-dimensional operators of the form \eqref{bias_MP} that explicitly break the discrete symmetry of the axion potential creating a bias, the quality can be estimated as
\begin{equation}
\label{eq: quality_HDO}
\frac{\Delta V}{\Lambda^4} \simeq \frac{f_a^{d-2}N_\text{DW}^d}{m_a^2 M_\text{Pl}^{d-4}}\,,
\end{equation}
where $d$ is the dimension of the operator. In Fig. \ref{fig: biasplot1}, we show purple lines that correspond to a higher dimensional operator with dimension $d$ for $N_\text{DW} = 20$ \footnote{The value $d = 6$ for $N_\text{DW} = 20$ is purely indicative, since for those values the degenerate minima are actually not lifted since $d$ and $N_\text{DW}$ are not co-prime, as discussed in the next section.}. 

In general, from Fig. \ref{fig: biasplot1} we see that in order to have a sizable SGWB, a certain \emph{quality} is needed in the ALP potential. A large bias, such as a dimension-five Planck suppressed operator, 
would make the DW network decay fast after its creation, i.e. at temperatures significantly larger than $T_{\text{dom}}$, implying a small GW signal (see \eqref{eq:pheno_omega}).
This conclusion does not change by reducing the ratio $m_a/f_a$ with respect to the value used in Fig. \ref{fig: biasplot1}, which makes the needed quality for the GW signal to be observable even higher (see also \cite{Gelmini:2022nim}).

\subsection{Bias from Planck suppressed operators}

\begin{figure}[t]
\centering
\includegraphics[scale=.8]{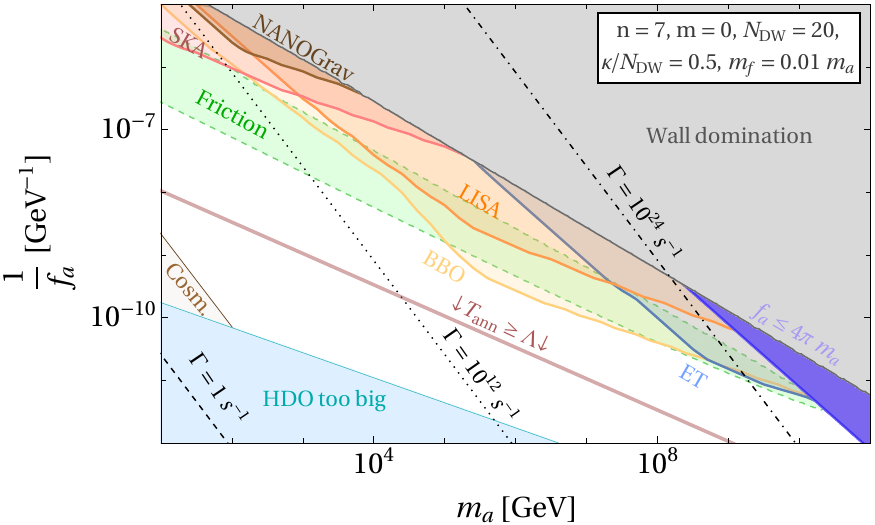}
\caption{Regions of ALP parameter space which are detectable by GW experiments for a 7 dimensional HDO with $N_\text{DW} = 20$, $\kappa/N_\text{DW} = 0.5$, $g = 20$ and $m_f = 0.01 m_a$. Similar contours as in Fig. \ref{fig: biasplot1} are displayed. In the cyan region, the HDO becomes larger than the axion potential \eqref{eq:pote}, whereas the dark blue region is excluded to allow for sufficient scale separation between $m_a$ and $f_a$. Contours of constant decay width are shown for  $1\ \text{s}^{-1}$ (dashed), $10^{12}\ \text{s}^{-1}$ (dotted) and $10^{24}\ \text{s}^{-1}$ (dot-dashed) (see \eqref{eq: decay_widths}). The constraints on the ALP parameter space from the photon coupling  (Cosm.) are shown for completeness as discussed in the text.
}
\label{fig: constraintplotsmallmf}
\end{figure}

We then consider a representative case where the bias is induced by higher-dimensional operators (HDOs) suppressed by the Planck scale (omitting $O(1)$ numbers), as in \eqref{bias_MP}, namely
\begin{align}\label{eq: HDO}
V_{M_\text{Pl}} &= -e^{-i\delta}\frac{(\Phi^\dagger\Phi)^m \Phi^n}{M_\text{Pl}^{2m + n-4}} + h.c. = -\frac{(f_a N_\text{DW})^4}{2}\left(\frac{f_a N_\text{DW}}{\sqrt{2}M_\text{Pl}}\right)^{d-4}\cos\left(\frac{n}{N_\text{DW}}\frac{a}{f_a} - \delta\right)\,,
\end{align}
with $d = 2m + n$ and $\delta$ a CP  violating phase. For temperatures below $\Lambda$, the total ALP potential is given by the sum of \eqref{eq:pote} and \eqref{eq: HDO}. An energy bias between the consecutive minima is then created if the degeneracy is lifted due to the extra symmetry breaking terms, which happens when $n = 1$ or is co-prime with $N_\text{DW}$. We take the bias as the energy difference between two consecutive minima, which is of the order

\begin{equation}
\Delta V = \frac{(f_a N_\text{DW})^4}{2}\left(\frac{f_a N_\text{DW}}{\sqrt{2}M_\text{Pl}}\right)^{d-4}\left[1 - \cos\left(\frac{2n\pi}{N_\text{DW}}\right)\right]\,.
\end{equation}

Given the previous discussion about the quality of the $U(1)$ symmetry, one expects that a detectable GW signal can only be obtained for operators with a large dimensionality. 
Indeed, in Fig. \ref{fig: biasplot1} 
it can be seen that dimension $5$
operators generically lead to small GW detectability regions in the ALP parameter space.

In Fig. \ref{fig: constraintplotsmallmf}, as an illustrative case, we consider the case where the first non-trivial Planck suppressed operator is of dimension $7$. 
This can be easily enforced in appropriate UV completions of the $U(1)$ breaking dynamics,
along the lines of model building strategies which have ben proposed to solve the quality problem of the QCD axion. 

In Fig. \ref{fig: constraintplotsmallmf}, we show the result of our numerical analysis on the standard ALP parameter space $1/f_a$ vs $m_a$ plane. 
The meaning of the various regions and lines is analogous to Fig. \ref{fig: biasplot1}.
As before, we assume a fermion mass $m_f = 10^{-2}m_a$. 
It can be seen that the GW detectors can probe significant parts of the ALP parameter space which is not accessible by other experiments.
The region where annihilation occurs during a friction regime is denoted by the green shaded region, in which the effect of friction on the GW sensitivity is clearly visible (most noticeable for BBO).

In the bottom left corner of Fig. \ref{fig: constraintplotsmallmf}, the magnitude of the HDO becomes of the same or higher order as the magnitude of the axion potential, i.e. $O(V_{M_\text{Pl}})\geq O(V(a))$, which coincides with the forbidden cyan region in Fig. \ref{fig: biasplot1}. A darker blue region excludes parameter values for which the scale separation between $f_a$ and $m_a$ does not satisfy $f_a \geq 4\pi m_a$. The LIGO-Virgo sensitivity region resides within this excluded area.

In the parameter space displayed, the ALP decay into fermions, which dominates the one into photons, is always sufficiently fast to avoid an intermediate matter dominated phase.
The assumed coupling to photons implies cosmological constraints, that are visible in Fig. \ref{fig: constraintplotsmallmf} in a small corner (taken from \cite{Workman:2022}, where we assumed that $E/N_\text{DW} \sim 1$ for concreteness).\footnote{Some of these bounds could be relaxed due to the fast decay of the ALP into the dark fermions.}

Our results show that interesting (and otherwise untestable) regions of parameter space of ALP models can be probed by the SGWB generated by the DW network when the bias is induced by higher dimensional Planck suppressed operators,
and that the signal could span over a variety of GW experiments.

\section{Conclusions and outlook}

In this paper we have investigated the effect of particle friction on the dynamics of DWs arising in axion-like particle models.
In particular, in Sec.\,\ref{sec:3} we have performed a quantitative computation of a realistic fermion--induced pressure on the wall within the ALP effective theory, and identified
the regions of parameter space where friction dominates over the Hubble expansion in determining the evolution of the DW network.
Our findings show that friction can be important in the early stages of the DW evolution (i.e. after formation), but also down to low temperatures where the GW emission from the DW becomes sizable.
Indeed, we emphasized that friction from SM (or BSM) fermions can affect the dynamics of the DW and the resulting SGWB signal 
in regions which are accessible at current and future GW experiments.
As an illustrative case study, in Sec.\,\ref{sec:PTA} we have considered scenarios where the SGWB signal from ALP DWs would peak at frequencies relevant for PTA experiments. We showed that if the ALP couples to the SM leptons, the 
induced friction (particularly from the muon) does affect the DW evolution at the annihilation temperature and above, potentially modifying the resulting SGWB signature.

While determining the relevance of friction relies only on the particle physics of the scattering in comparison with the Hubble expansion, understanding the implications for the GW signal requires to solve for the consequent evolution of the network.
In order to provide an estimate of the SGWB signal, in Sec.\,\ref{sec: VOS} and
\,\ref{sec:ALPpheno} we have then employed the VOS approximation to describe the DW network evolution. We showed that friction will generically reduce the average
velocity of the DW network, and thus the amplitude of the SGWB in agreement with previous work\,\cite{NAKAYAMA2017500}. According to the VOS result, the GW signal during a period of friction domination
is suppressed with respect to the one in the scaling regime only if there is a large hierarchy between Hubble and the inverse friction length, $3H/\ell^{-1}_\text{f} \ll1$.

By considering as a case study the friction induced by a fermion in the thermal plasma coupled to the ALP, we explored how friction can impact the experimental reach of several GW experiments.
During this investigation we also noticed how, independently of friction, a sufficiently high quality of the underlying ALP $U(1)$ symmetry is needed for the GW signature to be sizable.\\

There are several interesting directions that we have identified along our analysis that can be studied to further clarify the role of friction.
First, in order to estimate the effect of friction 
we have employed a simple VOS approximation.
A study through dedicated
numerical simulations would be needed to corroborate our conclusions.
In addition, when friction is sizable,  
it is possible that new mechanisms that generate GWs from the DW interaction with the plasma (such as sound waves or turbulence) become important.

Secondly, there is another case where friction can be naturally relevant for the DW evolution, namely scenarios in which the bias in the potential is generated by a strong dynamics. This is for instance the case of heavy ALPs coupling with gluons, where the QCD induced potential acts as a bias and implies annihilation at temperatures relevant for PTA experiments. The friction on the DW is here induced by strongly coupled QCD, and it would be interesting to estimate it thoroughly.

\section*{Acknowledgments}
SB is thankful to Marco Gorghetto and Fabrizio Rompineve for useful discussions.
AM is thankful to Diego Redigolo and Paolo Panci for useful discussions.
We thank Mairi Sakellariadou and Miguel Vanvlasselaer for useful comments.

All authors are supported in part by the Strategic Research
Program High-Energy Physics of the Research Council
of the Vrije Universiteit Brussel and by the iBOF ``Unlocking the Dark Universe with Gravitational Wave Observations: from Quantum Optics to Quantum Gravity'' of the Vlaamse Interuniversitaire Raad. SB and AM are supported in part by the ``Excellence
of Science - EOS'' - be.h project n.30820817. SB, KT and AR are supported by FWO-Vlaanderen through grant numbers 12B2323N, 1179522N and 1152923N respectively.
AS is supported in part by the FWO-Vlaanderen through the project G006119N.


\begin{appendices}

\section{Dark QCD ALP domain walls}
\label{darkqcd}
The ALP potential considered previously in \eqref{eq:pote} is a simplification with respect to what can happen in realistic scenarios.
For instance, in QCD, the axion domain wall is more complicated because of the mixing between the axion and the pions \cite{Huang:1985tt}.
This is reflected in the effective potential for the axion which differs
from \eqref{eq:pote} \cite{Villadoro}.

For concreteness, in order to discuss interesting dynamics that can emerge when the scalar potential does not take the simple form in \eqref{eq:pote},
we will
consider an ALP model where the effective potential is in form equivalent to the one of QCD, 
but for a dark QCD-like theory. 
Hence the ALP potential reads 
\be
\label{villadoro}
V(a) = -\tilde \Lambda^4 \sqrt{1-\frac{4 \tilde m_u \tilde m_d}{(\tilde m_u +\tilde m_d)^2} \sin^2 \left(\frac{a}{2 f_a} \right)}
\ee
where $\tilde \Lambda$ is the strong coupling scale of the dark QCD-like gauge group, and $ q_{}\equiv \frac{4 \tilde m_u \tilde m_d}{(\tilde m_u +\tilde m_d)^2} $ is an order $1$ number depending on the dark up and down quark masses, that we assume as the one
of QCD, that is $q = 0.9$.
The axion mass induced by this potential is 
\be
\label{eq:mafa}
m_a^2 = \frac{q}{4} \frac{\tilde \Lambda^4}{f_a^2}\,.
\ee

\begin{figure}[h]
\centering
 \includegraphics[scale=0.6]{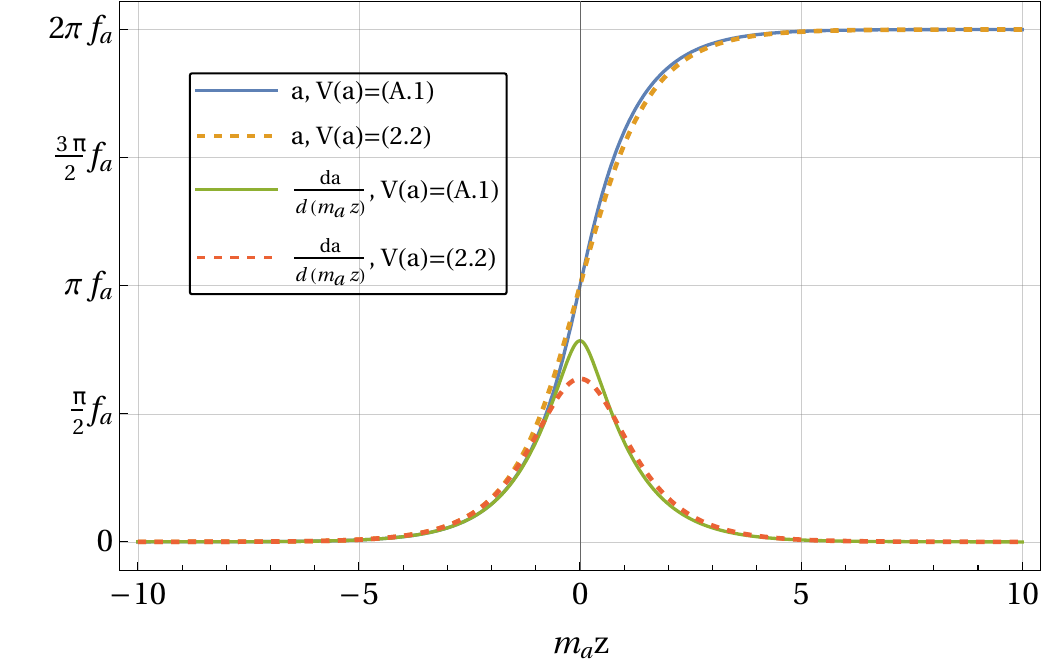}
 \caption{DW profile for the dark QCD ALP model \eqref{villadoro} (blue, solid). In addition, the first derivative w.r.t. $m_a z$ of the profile is shown as well (green, solid). For comparison, the DW profile and its first derivative for the cosine potential \eqref{eq:pote} are given by the orange and red dashed lines respectively.}
 \label{fig:DWprofile}
\end{figure}

The domain wall solution cannot be found analytically as in the case of the potential \eqref{eq:pote}, but can be easily determined numerically starting from the EOM,
\be
\label{eq:ALP EOM}
\frac{d^2a}{dz^2} = \frac{dV(a)}{da}\,,\quad \text{with}\quad \frac{dV(a)}{da} = 2m_a^2 f_a \frac{\sin\left(\frac{a}{2f_a}\right)\cos\left(\frac{a}{2f_a}\right)}{\sqrt{1-q\sin^2\left(\frac{a}{2f_a}\right)}}\,,
\ee
as it is shown in Fig. \ref{fig:DWprofile}.
However, the domain wall tension can be found analytically and it is given by \cite{Villadoro}
\be
\sigma = \int_0^{2\pi f_a} da \ \sqrt{2\left[V(a) - V(0)\right]}\,,
\ee
where $V(a)$ is the dark QCD ALP potential \eqref{villadoro} and $V(0)$ has been added in order to have a positive tension. Using the substitution $x = \sin^2(a/2f_a)$ and that the integrand is even, the tension reduces to
\begin{align}
\sigma &= 2\sqrt{2}\ \tilde \Lambda^2 f_a \int_0^1 \frac{\left[1-(1-qx)^{1/2}\right]^{1/2}}{\sqrt{x(1-x)}}dx \nonumber \\
&= 2\sqrt{2}\ \tilde \Lambda^2 f_a \int_0^1 \frac{\left[\left(\frac{1+(1-qx)^{1/2}}{1+(1-qx)^{1/2}}\right)\left[1-(1-qx)^{1/2}\right]\right]^{1/2}}{\sqrt{x(1-x)}}dx \nonumber \\
&= 2\sqrt{2q}\ \tilde \Lambda^2 f_a \int_0^1 \frac{1}{\sqrt{(1-x)\left[1+(1-qx)^{1/2}\right]}}dx\,.
\end{align}
Using the identity \eqref{eq:mafa} yields
\be
\sigma = 8 m_a f_a^2 \mathcal{E}[q]
\ee
where 
\be
\mathcal{E}[q] \equiv \int_{0}^1 \frac{dx}{\sqrt{2(1-x)[1+(1-qx)^{\frac{1}{2}}]}}
\ee
is an $O(1)$ number depending on the dark up and down quark mass ratio. For a case as QCD, it is $\mathcal{E}[0.9]\sim 1.12$
and hence $\sigma \simeq 9 m_a f_a^2$. The width of the domain wall, as it is visible in Fig. \ref{fig:DWprofile}, is of order $\delta \sim m_a^{-1}$
as in the previous case.

One interesting aspect of the ALP potential in \eqref{villadoro} is that it implies a non-vanishing self-reflection coefficient
(contrarily to the case in \eqref{eq:pote})
and hence friction on the domain wall induced by the ALP population itself as suggested in \cite{Huang:1985tt}. 
We discuss the implications of this in Appendix \ref{app:ALP reflection}.

\section{Reflection coefficient}
\label{app:R}

The reflection coefficient represents the probability of a particle being reflected once it strikes upon a potential barrier. This scattering problem is described by a one-dimensional time-independent Schr\"odinger-like equation
\be
\frac{d^2\psi}{dx^2} + (k^2-V(x))\psi = 0\,,
\ee
where $k$ is the wave number. Three regions are considered: the left and right sides of the barrier where the potential is zero and the region inside the barrier. The wave function will have the form
\be
\label{eq:3regions}
\psi(x) = \left\{
    \begin{array}{ll}
       e^{ik(x-a)} + Re^{-ik(x-a)}, & x < a \\
        c_1\psi_1(x) + c_2\psi_2(x), & x \in [a,b]\\
        Te^{ik(x-b)}, & x > b
    \end{array}
\right.,    
\quad \text{for} \quad
V(x) = \left\{
    \begin{array}{lll}
        0, & x < a \\
        V_0(x), & x \in [a,b]\\
        0, & x > b
    \end{array}
\right.,
\ee 
where $a$ and $b$ are the boundaries of the region where the potential barrier $V_0(x)$ is defined and $\psi_1(x)$ and $\psi_2(x)$ are the linearly independent solutions on $[a,b]$ with $c_1$ and $c_2$ constants. The incident wave amplitude is taken to be unity, whereas $R$ and $T$ are the reflection and transmission amplitudes for which the modulus gives the reflection and transmission probability respectively. Following \citep{cma/1418919772}, we assume that the independent solutions $\psi_1$ and $\psi_2$ satisfy the conditions of the Cauchy problem, i.e. $\psi_1(a) = \psi_2'(a) = 1$ and $\psi_2(a) = \psi_1'(a) = 0$, where a prime denotes a derivative w.r.t. $x$. Applying these conditions to \eqref{eq:3regions} gives the reflection amplitude,
\be
R = \frac{\psi_1'(b) + k^2\psi_2(b) + ik\left[\psi_2'(b) - \psi_1(b)\right]}{k^2\psi_2(b) - \psi_1'(b) + ik\left[\psi_1(b) + \psi_2'(b)\right]}\,.
\ee

\subsection{ALP self--reflection}
\label{app:ALP reflection}
The self-reflection of the ALP particles against the ALP DWs is found to be zero for the cosine potential \eqref{eq:pote} \citep{Vilenkin:2000jqa}. In contrast, the potential \eqref{villadoro} in the dark QCD ALP model allows friction to occur as argued by \citep{Huang:1985tt}.

\begin{figure}[h]
\centering
 \includegraphics[scale=0.5]{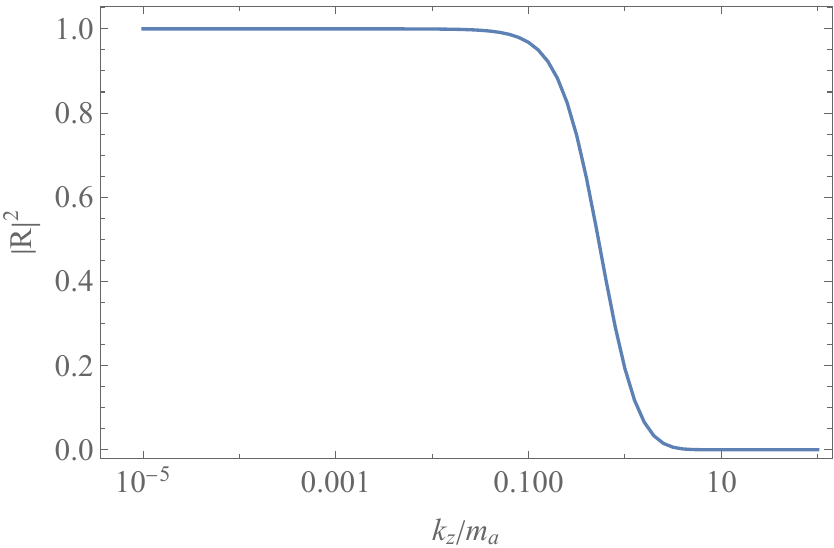}
 \caption{Reflection coefficient $|R|^2$ of an ALP against the ALP DW as a function of the momentum $k_z$ in units of the axion mass $m_a$.}
 \label{fig:RALP}
\end{figure}

One performs a linear perturbation about the DW solution $a_\text{DW}$ that satisfies the EOM \eqref{eq:ALP EOM}, with a plane wave ansatz for the perturbation, $a = a_\text{DW} + \delta a(z)e^{-iEt + ik_xx + ik_yy}$.
A Taylor expansion up to first order results in
\begin{equation}
\frac{d^2 \delta a}{dz^2} + k_z^2\delta a - \left(\frac{d^2V(a_\text{DW})}{da^2} - m_a^2\right)\delta a = 0\,,
\end{equation}
where
\begin{equation}
\frac{d^2V(a_\text{DW})}{da^2}= - m_a^2\frac{(1-q)\sin^4\left(\frac{a_\text{DW}}{2f_a}\right) - \cos^4\left(\frac{a_\text{DW}}{2f_a}\right)}{\left[\cos^2\left(\frac{a_\text{DW}}{2f_a}\right) + (1-q)\sin^2\left(\frac{a_\text{DW}}{2f_a}\right)\right]^{3/2}}\,,
\end{equation}
and $k_z^2 = E^2 - k_x^2-k_y^2 - m_a^2$. Following the procedure outlined above to calculate the reflection coefficient of the ALP against the DW, we obtain a non-zero reflection as shown in Fig.\ref{fig:RALP}.

\subsection{Fermion reflection}

Results for fermion scattering with $\mathbb{Z}_2$ DWs were presented in \cite{Campanelli:2004si} and in the case of ALP DWs, qualitative arguments have been put forward in \citep{Huang:1985tt}. In the following, we present the procedure used to determine the reflection coefficient for fermions scattering off the ALP DW from the simple cosine potential.

The reflection coefficient for a fermion is found from its current, $J^\mu = \bar\psi\gamma^\mu\psi$. More specifically, the ratio of the reflected and incoming currents provide the reflection probability. As the current gives an expression in terms of the components of the fermion field, we can determine the reflection of the components themselves following the procedure in Appendix \ref{app:R}. To this end, we consider a fermion moving perpendicular to the wall along the $z$ direction parameterized as $\psi(z) = (\alpha, \beta, \gamma, \delta)^Te^{-iEt}$, where the components depend on $z$, the energy $E$ and momentum $k_z$ of the fermion. We will work in the Weyl basis for which the current along the $z$ direction reads
\be
J^z = -|\alpha|^2 + |\beta|^2 + |\gamma|^2 - |\delta|^2\,.
\ee
From the interaction Lagrangian \eqref{eq:apsi}, the EOM of the fermion is given by
\be
\label{eq:fermionEOM}
 \left(E\gamma^0 + i\gamma^3\partial_z - m_f + \frac{\kappa}{2N_\text{DW}f_a}\frac{da}{dz}\gamma^3\gamma_5\right) \psi = 0\,,
\ee
which for the fermion components reads
\begin{align}
\label{eq:1ODE}
i\alpha' = E\alpha - m_f\gamma + \frac{\kappa}{2v_a}\frac{da}{dz}\alpha\,, \quad
        i\beta' = -E\beta + m_f\delta + \frac{\kappa}{2v_a}\frac{da}{dz}\beta\,, \nonumber\\[6pt]
        i\gamma' = -E\gamma + m_f\alpha - \frac{\kappa}{2v_a}\frac{da}{dz}\gamma\,, \quad
        i\delta' = E\delta - m_f\beta - \frac{\kappa}{2v_a}\frac{da}{dz}\delta\,.
\end{align}
Here a prime denotes a derivative w.r.t. $z$ and we used the identity $v_a = f_a N_\text{DW}$. Taking again the derivative yields
\begin{align}
\label{eq:2ODE}
\alpha'' = \left(m_f^2 - \left[E + \frac{\kappa}{2v_a}\frac{da}{dz}\right]^2 - i\frac{\kappa}{2v_a}\frac{d^2a}{dz^2}\right)\alpha\,, \quad
        \beta'' = \left(m_f^2 - \left[E - \frac{\kappa}{2v_a}\frac{da}{dz}\right]^2 - i\frac{\kappa}{2v_a}\frac{d^2a}{dz^2}\right)\beta\,, \nonumber\\[6pt]
       \gamma'' = \left(m_f^2 - \left[E + \frac{\kappa}{2v_a}\frac{da}{dz}\right]^2 + i\frac{\kappa}{2v_a}\frac{d^2a}{dz^2}\right)\gamma\,, \quad
        \delta'' = \left(m_f^2 - \left[E - \frac{\kappa}{2v_a}\frac{da}{dz}\right]^2 + i\frac{\kappa}{2v_a}\frac{d^2a}{dz^2}\right)\delta\,.
\end{align}
These are four Schr\"odinger-like equations, thus allowing us to calculate the reflection probability of each component separately. However, one should take the relationships between the components as shown in \eqref{eq:1ODE} into account. Therefore, we first calculate the reflection amplitude $R_\alpha$ of $\alpha$ and then use \eqref{eq:1ODE} outside the wall where $da/dz \rightarrow 0$ to obtain the correct reflection amplitude of $\gamma$. A similar approach is applied for $\beta$ and $\delta$. In order to see how the reflection amplitudes relate to one another, we first consider the free wave solutions outside the wall.

\paragraph{Free fermion solutions outside the wall}
The free fermion solutions are obtained from the free Dirac equation, i.e. \eqref{eq:fermionEOM} without the ALP part. The solutions in the Weyl representation for a fermion moving along the $z$ direction are found to be
\begin{align}\label{eq: Weyl spinors along z}
    u_1 = N\begin{pmatrix} E+m_f-k_z\\[.2cm]0\\[.2cm]E+m_f+k_z\\[.2cm]0\end{pmatrix}, &\quad 
    u_2 = N\begin{pmatrix} 0\\[.2cm]E+m_f+k_z\\[.2cm]0\\[.2cm]E+m_f-k_z\end{pmatrix}, \nonumber \\[6pt]
    u_3 = N\begin{pmatrix} E-m_f-k_z\\[.2cm]0\\[.2cm]-E+m_f-k_z\\[.2cm]0\end{pmatrix}, &\quad 
    u_4 = N\begin{pmatrix} 0\\[.2cm]E-m_f+k_z\\[.2cm]0\\[.2cm]-E+m_f+k_z\end{pmatrix},
\end{align}
where $N = \frac{1}{\sqrt{2(|E|+m_f)}}$ is a normalization constant\footnote{The free spinors are normalized to $2|E|$ particles per unit volume.} and we used the fact that the fermion moves along the $z$ direction ($k_x = k_y = 0$). $u_1$ and $u_2$ are the spin-up and spin-down\footnote{With spin-up and spin-down, we mean spin along the positive and negative $z$ direction respectively.} positive energy solutions, while $u_3$ and $u_4$ are the spin-up and spin-down negative energy solutions respectively. It is observed that the spin-up spinors correspond to $\beta = \delta = 0$ and the spin-down spinors with $\alpha = \gamma = 0$.

The fermion coupling to the ALP has the form $\partial_\mu a\,\bar{\psi} \gamma^\mu \gamma_5\psi$ and since $a = a(z)$, one finds that the fermion part is non-vanishing for expressions of the form $\Bar{u}_i\gamma^3\gamma_5 u_i$ with $i = \{1, 2, 3, 4\}$. Hence, an incoming free spinor remains the same once interacting with the wall. Consequently, if the incoming fermion is a $u_1$ spinor with non-zero components $\alpha$ and $\gamma$,  then the reflected spinor consists again of the reflected components $\alpha$ and $\gamma$. Both the incoming and reflected currents for a $u_1$ spinor thus simplify to $J_{u_1}^z = -|\alpha|^2 + |\gamma|^2$. A similar approach is applied for the other spinors.

\paragraph{Reflection of the entire fermion}
Inspired by \eqref{eq:3regions}, we write the components as incoming and reflected waves on the left side of the wall and transmitted waves on the right side. For instance, the first component reads
\be
\label{eq:alpha free}
 \alpha = \left\{
    \begin{array}{ll}
       e^{ik_z z} + R_\alpha e^{-ik_z z}, & \quad\text{left side of wall} \\
        T_\alpha e^{ik_z z}, & \quad\text{right side of wall}
    \end{array}
\right..
\ee
$R_\alpha$ is then calculated as outlined above. Similarly for the third component $\gamma$, we have an incoming, reflected and transmitted wave, but with different amplitudes,
\begin{equation}
\label{eq:gamma free}
   \gamma = \left\{
    \begin{array}{ll}
       Ae^{ik_z z} + B e^{-ik_z z}, & \quad\text{left side of wall} \\
        C  e^{ik_z z}, & \quad\text{right side of wall}
    \end{array}
\right..
\end{equation}
The amplitudes of $\gamma$ are such that they satisfy the relations in \eqref{eq:1ODE} outside the wall,
\begin{equation}
\label{eq:A, B, C}
A = \frac{E+k_z}{m_f}, \quad B = R_\alpha\left(\frac{E-k_z}{m_f}\right), \quad C = T_\alpha\left(\frac{E+k_z}{m_f}\right).
\end{equation}
The same ratios are obtained from the ratio of the first and third components of the $u_1$ and $u_3$ free spinors. In addition, by writing $\delta$ similarly as in \eqref{eq:alpha free} and $\beta$ as in \eqref{eq:gamma free}, the ratios in \eqref{eq:A, B, C} are found again with $R_\alpha \rightarrow R_\delta$, showing how the non-zero components of the $u_2$ and $u_4$ spinors reflect.

\begin{figure}[h!]
    \centering
    \includegraphics[scale = 0.5]{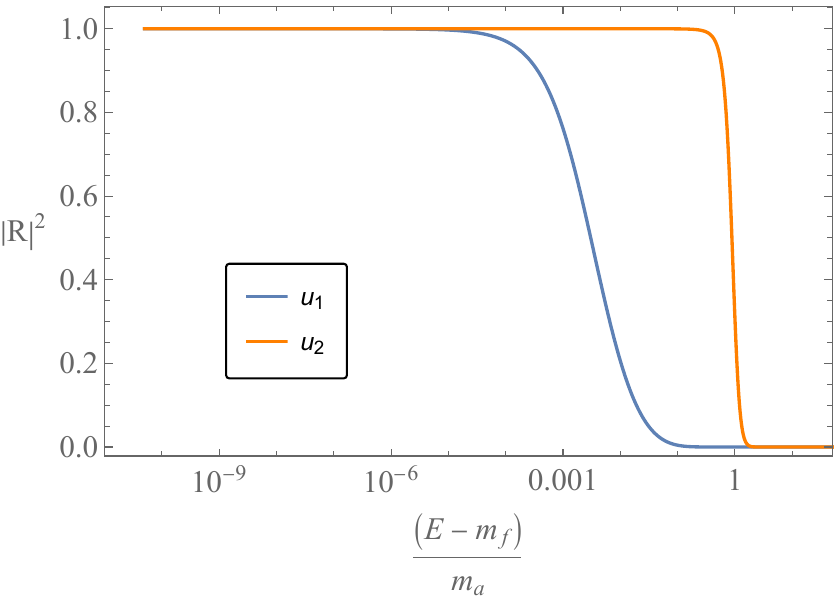}
    \caption{Reflection probability of the $u_1$ (blue) and $u_2$ (orange) spinors as a function of the kinetic energy of the fermion in units of the axion mass $m_a$. The mass of the fermion is $m_f = m_a$ and $\kappa/N_\text{DW} = 1$.}
    \label{fig:Rfermion}
\end{figure}

The incoming and reflected currents of a $u_1$ spinor (given by the incoming and reflected parts of $J^z_{u_1} = -|\alpha|^2 + |\gamma|^2$ as discussed above) are then found to be
\begin{equation}
\left.J^z_{u_1}\right|_{inc.} = \left(\frac{E+k_z}{m_f}\right)^2 - 1\,, \quad \left.J^z_{u_1}\right|_{refl.} = |R_\alpha|^2\left[\left(\frac{E-k_z}{m_f}\right)^2-1\right]\,,
\end{equation}
from which we obtain the reflection coefficient of the entire $u_1$ spinor,
\begin{equation}
|R_{u_1}|^2 = -\frac{\left.J^z_{u_1}\right|_{refl.}}{\left.J^z_{u_1}\right|_{inc.}} = |R_\alpha|^2\left(\frac{E-k_z}{E+k_z}\right).
\end{equation}
A similar expression is found for $u_2$ with $R_{\alpha} \rightarrow R_\delta$. The reflection coefficients for both a $u_1$ and $u_2$ spinor are shown in Fig. \ref{fig:Rfermion}. It is observed that there is a difference in the reflection behavior of $u_1$ and $u_2$. This can be explained by the fact that depending on the spin, the fermion sees a potential hill or well, as can be seen from \eqref{eq:fermionEOM}. Indeed, for a positive energy fermion, the spin-up spinor $u_1$ reflects on a potential well, whereas the spin-down spinor $u_2$ reflects on a potential hill. For transmission to happen, $u_2$ must therefore have a larger kinetic energy than $u_1$. 

A similar behavior is found for the negative energy states. In this case, one considers the spinors $u_3$ and $u_4$ for which $E = -\sqrt{k_z^2 + m_f^2}$. The main difference is that the spin-up negative energy state $u_3$ reflects on a potential hill and the spin-down negative energy state $u_4$ on a potential well. Whenever a positive energy fermion sees a hill, the negative version sees a well and vice versa. The overall global picture remains the same and we will therefore only consider the positive energy states. The final total reflection probability we use in our work for fermions is then taken to be
\begin{equation}
\mathcal{R} = \frac{|R_{u_1}|^2 + |R_{u_2}|^2}{2}\,.
\end{equation}

\section{Strong friction regime}
\label{App:arche}

In the main text we observed that friction generically reduces the amplitude of the GW signal emitted by DWs.
One can then wonder if, in the case of a DW network experiencing a very long period of friction, 
the dominant GW signal could have actually been generated at earlier times, when the network was still in the scaling regime.

We can easily estimate when this can happen by using the simple power-law assumption for friction that we 
derived previously in \eqref{eq:Lapp}.
In a friction dominated regime, if $\ell_{\text{f}} \sim t^{\lambda}$ and assuming $\lambda >0$, one gets that $L(t) \sim t^{\frac{1}{2} (1+\lambda)}$.
Inserting this into the formula for the gravitational wave emission, we deduce that $\Omega_\text{gw} \sim t^{-\frac{1}{2} (1-5 \lambda )}$.
We then conclude that the normalized gravitational wave energy is a growing function with time as soon as $\lambda > 1/5$. 

The standard scaling regime can be formally recovered by taking $\lambda=1$
(in this case $L \propto t$), confirming that 
the emission is stronger at later times.
The intermediate period of friction $\propto T^2$ that can arise
in the case of a light fermion, see e.g.\,\eqref{eq:T2},
gives $\lambda = 1$.
A possible period with
$1/\ell_\text{f}\propto T^4 \propto t^{-2}$ would give $\lambda = 2$ so that the early emission would be again overruled by the late emission. The case of a cold component 
scattering off the DW as discussed in  
Sec.\,\ref{sec:fermDM} leads to the same conclusion.

\begin{figure}
\centering
 \includegraphics[scale=0.7]{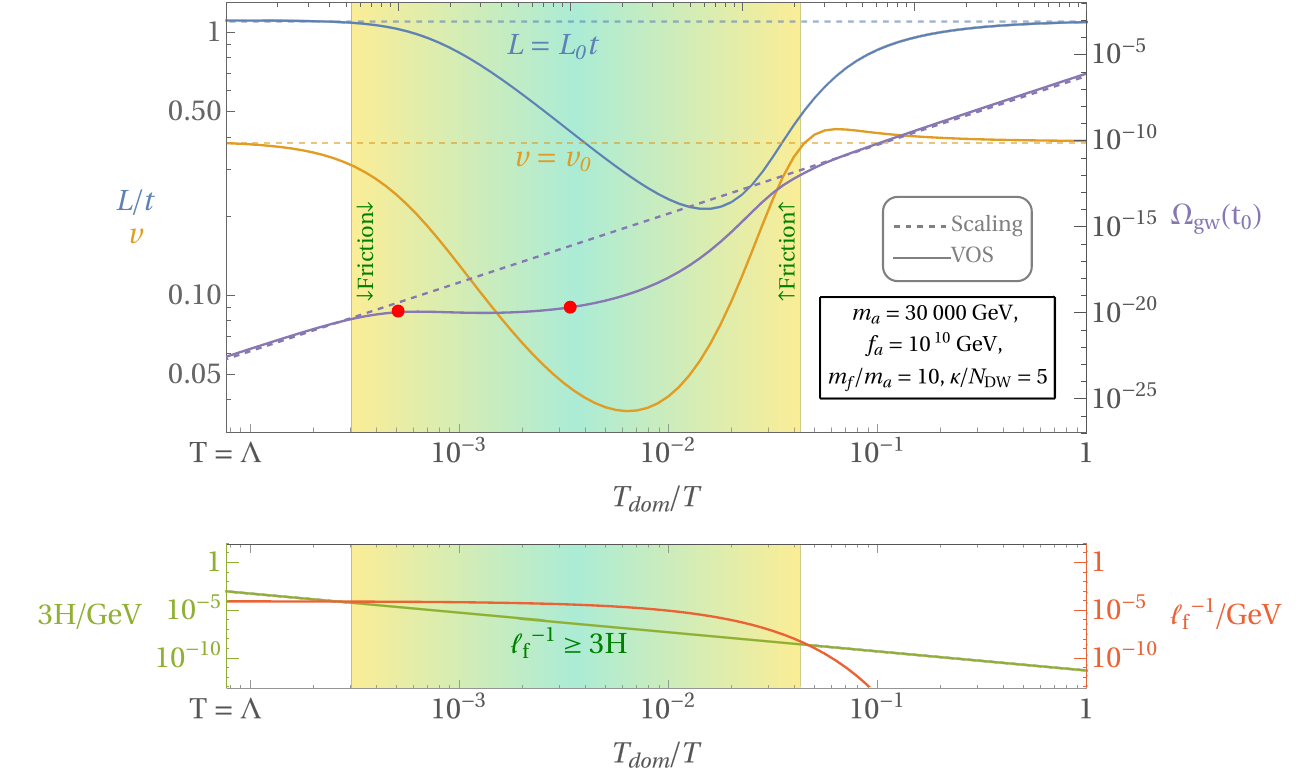}
 \caption{Combination of Fig. \ref{fig:5lines} and \ref{fig:5thline} for benchmark values $m_a = $ 30 TeV, $f_a = 10^{10}$ GeV, $m_f = 10 m_a$ and $\kappa/N_\text{DW} = 5$. The red dots indicate a possible scenario in which early and late emission would equally contribute to the final GW signal.}
 \label{fig:5lines_bis}
\end{figure}

A possibly more interesting case concerns the period in which
friction can dominate while being temperature (and time) 
independent, $\lambda=0$. This corresponds e.g. to the scattering of fermions in thermal equilibrium for temperatures $T\gg m_a$:
the longer the period in which friction dominates with constant $1/\ell_\text{f}$, the more likely it is for the early emission to dominate over the late emission. 
In Fig. \ref{fig:5lines_bis} we show an illustrative case for this interesting possibility.
If annihilation of the DW network occurs at
$T/T_\text{dom} \lesssim 2 \times 10^{-3}$, the GW amplitude is dominated by the signal emitted at the end of the scaling regime at earlier times,
at around $T/T_\text{dom} \sim 5 \times 10^{-4} $
(the absolute size of this signal is still very small, $\Omega_\text{gw} \sim 10^{-18}$, at the border of the BBO sensitivity).
However, as soon as annihilation occurs at lower temperatures,
the GW signal quickly becomes dominated by the late emission (while still being in a friction dominated regime).
This also shows that having the strongest GW signal from emission prior to the network annihilation would require some tuning in the choice of the bias.

\section{Matter domination}\label{app: D}

The energy of the DW network is transferred to both mildly relativistic axions and GWs \cite{Hiramatsu:2012sc}. An intermediate matter dominated era (IMD) is therefore plausible if the axions do not decay before their energy density $\rho_a$ exceeds the critical density of the Universe $\rho_c$. The altered evolution of the Universe should then be taken into account in the resulting GW spectrum, as elaborated in \cite{ZambujalFerreira:2021cte}.

Part of the energy of the walls goes into  axion production such that $\rho_a = \xi \rho_{w}$, where $\xi$ is the fraction of the DW energy transferred to the axions, for which we assume $\xi = 1$ in our work. After annihilation of the DW network, the axion population scales as $\rho_a \sim a^{-3}$, and an IMD therefore happens when $\rho_a(t) = \rho_c(t)$ or
\begin{equation}
\xi \rho_{w}(t_\text{ann})\left(\frac{a(t_\text{ann})}{a(t)}\right)^3 = 3H^2(t)M_\text{Pl}^2\,.
\end{equation}
As the annihilation of the DW occurs when $\rho_{w} = \Delta V$, the time at which the IMD phase begins is given by
\begin{equation}\label{eq: tm}
t_m = \frac{9}{16}\frac{M_\text{Pl}^4}{\xi^2}t_\text{ann}^{-3}\Delta V^{-2}
\end{equation} 
in a radiation dominated Universe.

However, if the axions were to decay before $t_m$, an IMD will not happen. Therefore, we compare $t_m$ w.r.t. the time at which the decay rate of axions becomes efficient, i.e. when $\Gamma \sim H$, with $\Gamma$ the axion decay rate. Depending on whether the axion decays into photons or fermions, the decay rates are given by \cite{Bauer:2017ris}
\begin{equation}
\label{eq: decay_widths}
\Gamma_{a\rightarrow \gamma\gamma} = \frac{\alpha^2 m_a^3}{64\pi^3 f_a^2}c_\gamma^2\,, \quad
\Gamma_{a\rightarrow ff} = \frac{m_a m_f^2}{8\pi f_a^2}c_{f}^2\sqrt{1-\frac{4m_f^2}{m_a^2}} \,,
\end{equation}
where $\alpha = 1/137$ is the fine-structure constant and we assumed the Lagrangian density describing the interaction between photons and the ALP is given by
\begin{equation}\label{eq:axionphoton}
\mathcal{L}_{a\gamma} = \frac{\alpha}{4\pi}\frac{E}{N_\text{DW}}\frac{a}{f_a}F\tilde{F}\,,
\end{equation}
where $E$ is the anomaly with respect to the gauge current. $c_\gamma$ and $c_f$ can be traced back to the couplings of the ALP, for which we have
\begin{equation}
c_\gamma = E/N_\text{DW}\,, \quad c_f = \kappa/N_\text{DW}\,.
\end{equation}

In a radiation dominated Universe, the decay rates become efficient at times
\begin{equation}\label{eq: tdecay}
t_{a\rightarrow\gamma\gamma} = \frac{32\pi^3}{c_\gamma^2 \alpha^2}\frac{f_a^2}{m_a^3}\,,\quad t_{a\rightarrow f f} = \frac{4 \pi}{c_f^2}\frac{f_a^2}{m_a m_f^2}\left(1-\frac{4 m_f^2}{m_a^2}\right)^{-1/2}\,,
\end{equation}
and the axion therefore starts to decay at a time
\begin{equation}\label{eq: tdec}
t_\text{dec} = \text{min}\left(t_{a\rightarrow\gamma\gamma}, t_{a\rightarrow f f}\right)\,.
\end{equation}

If  $t_m \lesssim t_\text{dec}$, an IMD epoch would alter the GW spectrum due to a different behavior of the scale factor between $t_m$ and $t_\text{dec}$. Both the peak frequency and amplitude depend on the ratio of the scale factor today and at the time of annihilation\footnote{We approximate the total signal by considering the spectrum at the time of annihilation, as this corresponds to the largest contribution to the signal (see the discussion at the end of Section \ref{sec:Gravitational waves from domain wall dynamics}).}, $a(t_\text{ann})/a(t_0)$, which we can rewrite as
\begin{equation}
\frac{a(t_\text{ann})}{a(t_0)}= \left(\frac{a(t_\text{ann})}{a(t_m)}\right)\left(\frac{H(t_\text{dec})}{H(t_{m})}\right)^{2/3}\left(\frac{a(t_\text{dec})}{a(t_0)}\right)\,.
\end{equation}
In the second factor, we used the fact that for a matter dominated era one has $a \sim H^{-2/3}$. The peak frequency and amplitude are then given by substituting the new ratio of the scaling factor in \eqref{eq:OmegaGW} and \eqref{eq:pheno_freq},
\begin{align}
\left.f_{\text{peak},m}(t_\text{ann})\right|_{t_0}&= \left.f_\text{peak}(t_\text{ann})\right|_{t_0}\left(\frac{t_m}{t_\text{dec}}\right)^{1/6}\,,\\
\left.\Omega_{\text{gw},m}(t_\text{ann}, f)\right|_{t_0} &= \left.\Omega_\text{gw}(t_\text{ann},f)\right|_{t_0}\left(\frac{t_m}{t_\text{dec}}\right)^{2/3}\,,
\end{align}
where we used $H \sim t^{-1}$ and $a\sim t^{1/2}$ in a matter and radiation dominated Universe, respectively.

\end{appendices}


\bibliographystyle{JHEP}
\bibliography{bibfr}

\providecommand{\href}[2]{#2}\begingroup\raggedright\begin{thebibliography}{100}

\bibitem{LIGOScientific:2016aoc}
{\bf LIGO Scientific, Virgo}, B.~P. Abbott et~al., {\it {Observation of
  Gravitational Waves from a Binary Black Hole Merger}},  {\em Phys. Rev.
  Lett.} {\bf 116} (2016), no.~6 061102,
  [\href{http://arxiv.org/abs/1602.03837}{{\tt arXiv:1602.03837}}].

\bibitem{Christensen:2018iqi}
N.~Christensen, {\it {Stochastic Gravitational Wave Backgrounds}},  {\em Rept.
  Prog. Phys.} {\bf 82} (2019), no.~1 016903,
  [\href{http://arxiv.org/abs/1811.08797}{{\tt arXiv:1811.08797}}].

\bibitem{Kibble:1976sj}
T.~W.~B. Kibble, {\it {Topology of Cosmic Domains and Strings}},  {\em J. Phys.
  A} {\bf 9} (1976) 1387--1398.

\bibitem{Vilenkin:2000jqa}
A.~Vilenkin and E.~P.~S. Shellard, {\em {Cosmic Strings and Other Topological
  Defects}}.
\newblock Cambridge University Press, 7, 2000.

\bibitem{Zeldovich:1974uw}
Y.~B. Zeldovich, I.~Y. Kobzarev, and L.~B. Okun, {\it {Cosmological
  Consequences of the Spontaneous Breakdown of Discrete Symmetry}},  {\em Zh.
  Eksp. Teor. Fiz.} {\bf 67} (1974) 3--11.

\bibitem{Hiramatsu:2012sc}
T.~Hiramatsu, M.~Kawasaki, K.~Saikawa, and T.~Sekiguchi, {\it {Axion cosmology
  with long-lived domain walls}},  {\em JCAP} {\bf 01} (2013) 001,
  [\href{http://arxiv.org/abs/1207.3166}{{\tt arXiv:1207.3166}}].

\bibitem{Hiramatsu:2013qaa}
T.~Hiramatsu, M.~Kawasaki, and K.~Saikawa, {\it {On the estimation of
  gravitational wave spectrum from cosmic domain walls}},  {\em JCAP} {\bf 02}
  (2014) 031, [\href{http://arxiv.org/abs/1309.5001}{{\tt arXiv:1309.5001}}].

\bibitem{Saikawa:2017hiv}
K.~Saikawa, {\it {A review of gravitational waves from cosmic domain walls}},
  {\em Universe} {\bf 3} (2017), no.~2 40,
  [\href{http://arxiv.org/abs/1703.02576}{{\tt arXiv:1703.02576}}].

\bibitem{Gelmini:2021yzu}
G.~B. Gelmini, A.~Simpson, and E.~Vitagliano, {\it {Gravitational waves from
  axionlike particle cosmic string-wall networks}},  {\em Phys. Rev. D} {\bf
  104} (2021), no.~6 061301, [\href{http://arxiv.org/abs/2103.07625}{{\tt
  arXiv:2103.07625}}].

\bibitem{Gelmini:2020bqg}
G.~B. Gelmini, S.~Pascoli, E.~Vitagliano, and Y.-L. Zhou, {\it {Gravitational
  wave signatures from discrete flavor symmetries}},  {\em JCAP} {\bf 02}
  (2021) 032, [\href{http://arxiv.org/abs/2009.01903}{{\tt arXiv:2009.01903}}].

\bibitem{Gelmini:2022nim}
G.~B. Gelmini, A.~Simpson, and E.~Vitagliano, {\it {Catastrogenesis: DM, GWs,
  and PBHs from ALP string-wall networks}},
  \href{http://arxiv.org/abs/2207.07126}{{\tt arXiv:2207.07126}}.

\bibitem{Craig:2020bnv}
N.~Craig, I.~Garcia~Garcia, G.~Koszegi, and A.~McCune, {\it {P not PQ}},
  \href{http://arxiv.org/abs/2012.13416}{{\tt arXiv:2012.13416}}.

\bibitem{ZambujalFerreira:2021cte}
R.~Zambujal~Ferreira, A.~Notari, O.~Pujol\`as, and F.~Rompineve, {\it {High
  Quality QCD Axion at Gravitational Wave Observatories}},
  \href{http://arxiv.org/abs/2107.07542}{{\tt arXiv:2107.07542}}.

\bibitem{Babichev:2021uvl}
E.~Babichev, D.~Gorbunov, S.~Ramazanov, and A.~Vikman, {\it {Gravitational
  shine of dark domain walls}},  {\em JCAP} {\bf 04} (2022), no.~04 028,
  [\href{http://arxiv.org/abs/2112.12608}{{\tt arXiv:2112.12608}}].

\bibitem{Barman:2022yos}
B.~Barman, D.~Borah, A.~Dasgupta, and A.~Ghoshal, {\it {Probing High Scale
  Dirac Leptogenesis via Gravitational Waves from Domain Walls}},
  \href{http://arxiv.org/abs/2205.03422}{{\tt arXiv:2205.03422}}.

\bibitem{Wu:2022tpe}
Y.~Wu, K.-P. Xie, and Y.-L. Zhou, {\it {Classification of Abelian domain
  walls}},  \href{http://arxiv.org/abs/2205.11529}{{\tt arXiv:2205.11529}}.

\bibitem{Borah:2022wdy}
D.~Borah and A.~Dasgupta, {\it {Probing Left-Right Symmetry via Gravitational
  Waves from Domain Walls}},  \href{http://arxiv.org/abs/2205.12220}{{\tt
  arXiv:2205.12220}}.

\bibitem{Wu:2022stu}
Y.~Wu, K.-P. Xie, and Y.-L. Zhou, {\it {Collapsing domain walls beyond $Z_2$}},
   \href{http://arxiv.org/abs/2204.04374}{{\tt arXiv:2204.04374}}.

\bibitem{Ferreira:2022zzo}
R.~Z. Ferreira, A.~Notari, O.~Pujolas, and F.~Rompineve, {\it {Gravitational
  Waves from Domain Walls in Pulsar Timing Array Datasets}},
  \href{http://arxiv.org/abs/2204.04228}{{\tt arXiv:2204.04228}}.

\bibitem{Fornal:2022qim}
B.~Fornal and E.~Pierre, {\it {Asymmetric Dark Matter from Gravitational
  Waves}},  \href{http://arxiv.org/abs/2209.04788}{{\tt arXiv:2209.04788}}.

\bibitem{Peccei:1977hh}
R.~D. Peccei and H.~R. Quinn, {\it {CP Conservation in the Presence of
  Instantons}},  {\em Phys. Rev. Lett.} {\bf 38} (1977) 1440--1443.

\bibitem{Peccei:1977ur}
R.~D. Peccei and H.~R. Quinn, {\it {Constraints Imposed by CP Conservation in
  the Presence of Instantons}},  {\em Phys. Rev. D} {\bf 16} (1977) 1791--1797.

\bibitem{Weinberg:1977ma}
S.~Weinberg, {\it {A New Light Boson?}},  {\em Phys. Rev. Lett.} {\bf 40}
  (1978) 223--226.

\bibitem{Wilczek:1977pj}
F.~Wilczek, {\it {Problem of Strong $P$ and $T$ Invariance in the Presence of
  Instantons}},  {\em Phys. Rev. Lett.} {\bf 40} (1978) 279--282.

\bibitem{Svrcek:2006yi}
P.~Svrcek and E.~Witten, {\it {Axions In String Theory}},  {\em JHEP} {\bf 06}
  (2006) 051, [\href{http://arxiv.org/abs/hep-th/0605206}{{\tt
  hep-th/0605206}}].

\bibitem{Arvanitaki:2009fg}
A.~Arvanitaki, S.~Dimopoulos, S.~Dubovsky, N.~Kaloper, and J.~March-Russell,
  {\it {String Axiverse}},  {\em Phys. Rev. D} {\bf 81} (2010) 123530,
  [\href{http://arxiv.org/abs/0905.4720}{{\tt arXiv:0905.4720}}].

\bibitem{Holdom:1982ex}
B.~Holdom and M.~E. Peskin, {\it {Raising the Axion Mass}},  {\em Nucl. Phys.
  B} {\bf 208} (1982) 397--412.

\bibitem{Holdom:1985vx}
B.~Holdom, {\it {Strong QCD at High-energies and a Heavy Axion}},  {\em Phys.
  Lett. B} {\bf 154} (1985) 316. [Erratum: Phys.Lett.B 156, 452 (1985)].

\bibitem{Flynn:1987rs}
J.~M. Flynn and L.~Randall, {\it {A Computation of the Small Instanton
  Contribution to the Axion Potential}},  {\em Nucl. Phys. B} {\bf 293} (1987)
  731--739.

\bibitem{Rubakov:1997vp}
V.~A. Rubakov, {\it {Grand unification and heavy axion}},  {\em JETP Lett.}
  {\bf 65} (1997) 621--624, [\href{http://arxiv.org/abs/hep-ph/9703409}{{\tt
  hep-ph/9703409}}].

\bibitem{Choi:1998ep}
K.~Choi and H.~D. Kim, {\it {Small instanton contribution to the axion
  potential in supersymmetric models}},  {\em Phys. Rev. D} {\bf 59} (1999)
  072001, [\href{http://arxiv.org/abs/hep-ph/9809286}{{\tt hep-ph/9809286}}].

\bibitem{Berezhiani:2000gh}
Z.~Berezhiani, L.~Gianfagna, and M.~Giannotti, {\it {Strong CP problem and
  mirror world: The Weinberg-Wilczek axion revisited}},  {\em Phys. Lett. B}
  {\bf 500} (2001) 286--296, [\href{http://arxiv.org/abs/hep-ph/0009290}{{\tt
  hep-ph/0009290}}].

\bibitem{Hook:2014cda}
A.~Hook, {\it {Anomalous solutions to the strong CP problem}},  {\em Phys. Rev.
  Lett.} {\bf 114} (2015), no.~14 141801,
  [\href{http://arxiv.org/abs/1411.3325}{{\tt arXiv:1411.3325}}].

\bibitem{Fukuda:2015ana}
H.~Fukuda, K.~Harigaya, M.~Ibe, and T.~T. Yanagida, {\it {Model of visible QCD
  axion}},  {\em Phys. Rev. D} {\bf 92} (2015), no.~1 015021,
  [\href{http://arxiv.org/abs/1504.06084}{{\tt arXiv:1504.06084}}].

\bibitem{Dimopoulos:2016lvn}
S.~Dimopoulos, A.~Hook, J.~Huang, and G.~Marques-Tavares, {\it {A collider
  observable QCD axion}},  {\em JHEP} {\bf 11} (2016) 052,
  [\href{http://arxiv.org/abs/1606.03097}{{\tt arXiv:1606.03097}}].

\bibitem{Agrawal:2017ksf}
P.~Agrawal and K.~Howe, {\it {Factoring the Strong CP Problem}},  {\em JHEP}
  {\bf 12} (2018) 029, [\href{http://arxiv.org/abs/1710.04213}{{\tt
  arXiv:1710.04213}}].

\bibitem{Gaillard:2018xgk}
M.~K. Gaillard, M.~B. Gavela, R.~Houtz, P.~Quilez, and R.~Del~Rey, {\it {Color
  unified dynamical axion}},  {\em Eur. Phys. J. C} {\bf 78} (2018), no.~11
  972, [\href{http://arxiv.org/abs/1805.06465}{{\tt arXiv:1805.06465}}].

\bibitem{Preskill:1982cy}
J.~Preskill, M.~B. Wise, and F.~Wilczek, {\it {Cosmology of the Invisible
  Axion}},  {\em Phys. Lett. B} {\bf 120} (1983) 127--132.

\bibitem{Abbott:1982af}
L.~F. Abbott and P.~Sikivie, {\it {A Cosmological Bound on the Invisible
  Axion}},  {\em Phys. Lett. B} {\bf 120} (1983) 133--136.

\bibitem{Dine:1981rt}
M.~Dine, W.~Fischler, and M.~Srednicki, {\it {A Simple Solution to the Strong
  CP Problem with a Harmless Axion}},  {\em Phys. Lett. B} {\bf 104} (1981)
  199--202.

\bibitem{Dine:1982ah}
M.~Dine and W.~Fischler, {\it {The Not So Harmless Axion}},  {\em Phys. Lett.
  B} {\bf 120} (1983) 137--141.

\bibitem{Arias:2012az}
P.~Arias, D.~Cadamuro, M.~Goodsell, J.~Jaeckel, J.~Redondo, and A.~Ringwald,
  {\it {WISPy Cold Dark Matter}},  {\em JCAP} {\bf 06} (2012) 013,
  [\href{http://arxiv.org/abs/1201.5902}{{\tt arXiv:1201.5902}}].

\bibitem{Ringwald:2012hr}
A.~Ringwald, {\it {Exploring the Role of Axions and Other WISPs in the Dark
  Universe}},  {\em Phys. Dark Univ.} {\bf 1} (2012) 116--135,
  [\href{http://arxiv.org/abs/1210.5081}{{\tt arXiv:1210.5081}}].

\bibitem{Marsh:2015xka}
D.~J.~E. Marsh, {\it {Axion Cosmology}},  {\em Phys. Rept.} {\bf 643} (2016)
  1--79, [\href{http://arxiv.org/abs/1510.07633}{{\tt arXiv:1510.07633}}].

\bibitem{Bauer:2017ris}
M.~Bauer, M.~Neubert, and A.~Thamm, {\it {Collider Probes of Axion-Like
  Particles}},  {\em JHEP} {\bf 12} (2017) 044,
  [\href{http://arxiv.org/abs/1708.00443}{{\tt arXiv:1708.00443}}].

\bibitem{Brivio:2017ije}
I.~Brivio, M.~B. Gavela, L.~Merlo, K.~Mimasu, J.~M. No, R.~del Rey, and
  V.~Sanz, {\it {ALPs Effective Field Theory and Collider Signatures}},  {\em
  Eur. Phys. J. C} {\bf 77} (2017), no.~8 572,
  [\href{http://arxiv.org/abs/1701.05379}{{\tt arXiv:1701.05379}}].

\bibitem{CidVidal:2018blh}
X.~Cid~Vidal, A.~Mariotti, D.~Redigolo, F.~Sala, and K.~Tobioka, {\it {New
  Axion Searches at Flavor Factories}},  {\em JHEP} {\bf 01} (2019) 113,
  [\href{http://arxiv.org/abs/1810.09452}{{\tt arXiv:1810.09452}}]. [Erratum:
  JHEP 06, 141 (2020)].

\bibitem{Bauer:2020jbp}
M.~Bauer, M.~Neubert, S.~Renner, M.~Schnubel, and A.~Thamm, {\it {The
  Low-Energy Effective Theory of Axions and ALPs}},
  \href{http://arxiv.org/abs/2012.12272}{{\tt arXiv:2012.12272}}.

\bibitem{Goh:2008xz}
H.-S. Goh and M.~Ibe, {\it {R-axion detection at LHC}},  {\em JHEP} {\bf 03}
  (2009) 049, [\href{http://arxiv.org/abs/0810.5773}{{\tt arXiv:0810.5773}}].

\bibitem{Bellazzini:2017neg}
B.~Bellazzini, A.~Mariotti, D.~Redigolo, F.~Sala, and J.~Serra, {\it {$R$-axion
  at colliders}},  {\em Phys. Rev. Lett.} {\bf 119} (2017), no.~14 141804,
  [\href{http://arxiv.org/abs/1702.02152}{{\tt arXiv:1702.02152}}].

\bibitem{Belyaev:2015hgo}
A.~Belyaev, G.~Cacciapaglia, H.~Cai, T.~Flacke, A.~Parolini, and H.~Ser\^odio,
  {\it {Singlets in composite Higgs models in light of the LHC 750 GeV diphoton
  excess}},  {\em Phys. Rev. D} {\bf 94} (2016), no.~1 015004,
  [\href{http://arxiv.org/abs/1512.07242}{{\tt arXiv:1512.07242}}].

\bibitem{DiLuzio:2020wdo}
L.~Di~Luzio, M.~Giannotti, E.~Nardi, and L.~Visinelli, {\it {The landscape of
  QCD axion models}},  {\em Phys. Rept.} {\bf 870} (2020) 1--117,
  [\href{http://arxiv.org/abs/2003.01100}{{\tt arXiv:2003.01100}}].

\bibitem{Choi:2020rgn}
K.~Choi, S.~H. Im, and C.~Sub~Shin, {\it {Recent Progress in the Physics of
  Axions and Axion-Like Particles}},  {\em Ann. Rev. Nucl. Part. Sci.} {\bf 71}
  (2021) 225--252, [\href{http://arxiv.org/abs/2012.05029}{{\tt
  arXiv:2012.05029}}].

\bibitem{Sikivie:1982qv}
P.~Sikivie, {\it {Of Axions, Domain Walls and the Early Universe}},  {\em Phys.
  Rev. Lett.} {\bf 48} (1982) 1156--1159.

\bibitem{Barr:1992qq}
S.~M. Barr and D.~Seckel, {\it {Planck scale corrections to axion models}},
  {\em Phys. Rev. D} {\bf 46} (1992) 539--549.

\bibitem{Kamionkowski:1992ax}
M.~Kamionkowski and J.~March-Russell, {\it {Are textures natural?}},  {\em
  Phys. Rev. Lett.} {\bf 69} (1992) 1485--1488,
  [\href{http://arxiv.org/abs/hep-th/9201063}{{\tt hep-th/9201063}}].

\bibitem{Kamionkowski:1992mf}
M.~Kamionkowski and J.~March-Russell, {\it {Planck scale physics and the
  Peccei-Quinn mechanism}},  {\em Phys. Lett. B} {\bf 282} (1992) 137--141,
  [\href{http://arxiv.org/abs/hep-th/9202003}{{\tt hep-th/9202003}}].

\bibitem{Holman:1992us}
R.~Holman, S.~D.~H. Hsu, T.~W. Kephart, E.~W. Kolb, R.~Watkins, and L.~M.
  Widrow, {\it {Solutions to the strong CP problem in a world with gravity}},
  {\em Phys. Lett. B} {\bf 282} (1992) 132--136,
  [\href{http://arxiv.org/abs/hep-ph/9203206}{{\tt hep-ph/9203206}}].

\bibitem{Berezhiani:1992pq}
Z.~G. Berezhiani, R.~N. Mohapatra, and G.~Senjanovic, {\it {Planck scale
  physics and solutions to the strong CP problem without axion}},  {\em Phys.
  Rev. D} {\bf 47} (1993) 5565--5570,
  [\href{http://arxiv.org/abs/hep-ph/9212318}{{\tt hep-ph/9212318}}].

\bibitem{Ghigna:1992iv}
S.~Ghigna, M.~Lusignoli, and M.~Roncadelli, {\it {Instability of the invisible
  axion}},  {\em Phys. Lett. B} {\bf 283} (1992) 278--281.

\bibitem{Senjanovic:1993uz}
G.~Senjanovic, {\it {Discrete symmetries, strong CP problem and gravity}},  in
  {\em {4th Hellenic School on Elementary Particle Physics}}, 5, 1993.
\newblock \href{http://arxiv.org/abs/hep-ph/9311371}{{\tt hep-ph/9311371}}.

\bibitem{Dobrescu:1996jp}
B.~A. Dobrescu, {\it {The Strong CP problem versus Planck scale physics}},
  {\em Phys. Rev. D} {\bf 55} (1997) 5826--5833,
  [\href{http://arxiv.org/abs/hep-ph/9609221}{{\tt hep-ph/9609221}}].

\bibitem{Banks:2010zn}
T.~Banks and N.~Seiberg, {\it {Symmetries and Strings in Field Theory and
  Gravity}},  {\em Phys. Rev. D} {\bf 83} (2011) 084019,
  [\href{http://arxiv.org/abs/1011.5120}{{\tt arXiv:1011.5120}}].

\bibitem{NAKAYAMA2017500}
K.~Nakayama, F.~Takahashi, and N.~Yokozaki, {\it Gravitational waves from
  domain walls and their implications},  {\em Physics Letters B} {\bf 770}
  (2017) 500--506.

\bibitem{Arnold:1993wc}
P.~B. Arnold, {\it {One loop fluctuation - dissipation formula for bubble wall
  velocity}},  {\em Phys. Rev. D} {\bf 48} (1993) 1539--1545,
  [\href{http://arxiv.org/abs/hep-ph/9302258}{{\tt hep-ph/9302258}}].

\bibitem{Abel:1995wk}
S.~A. Abel, S.~Sarkar, and P.~L. White, {\it {On the cosmological domain wall
  problem for the minimally extended supersymmetric standard model}},  {\em
  Nucl. Phys. B} {\bf 454} (1995) 663--684,
  [\href{http://arxiv.org/abs/hep-ph/9506359}{{\tt hep-ph/9506359}}].

\bibitem{Huang:1985tt}
M.~C. Huang and P.~Sikivie, {\it {The Structure of Axionic Domain Walls}},
  {\em Phys. Rev. D} {\bf 32} (1985) 1560.

\bibitem{NANOGrav:2020bcs}
{\bf NANOGrav}, Z.~Arzoumanian et~al., {\it {The NANOGrav 12.5 yr Data Set:
  Search for an Isotropic Stochastic Gravitational-wave Background}},  {\em
  Astrophys. J. Lett.} {\bf 905} (2020), no.~2 L34,
  [\href{http://arxiv.org/abs/2009.04496}{{\tt arXiv:2009.04496}}].

\bibitem{Chen:2021rqp}
S.~Chen et~al., {\it {Common-red-signal analysis with 24-yr high-precision
  timing of the European Pulsar Timing Array: inferences in the stochastic
  gravitational-wave background search}},  {\em Mon. Not. Roy. Astron. Soc.}
  {\bf 508} (2021), no.~4 4970--4993,
  [\href{http://arxiv.org/abs/2110.13184}{{\tt arXiv:2110.13184}}].

\bibitem{Goncharov:2021oub}
B.~Goncharov et~al., {\it {On the Evidence for a Common-spectrum Process in the
  Search for the Nanohertz Gravitational-wave Background with the Parkes Pulsar
  Timing Array}},  {\em Astrophys. J. Lett.} {\bf 917} (2021), no.~2 L19,
  [\href{http://arxiv.org/abs/2107.12112}{{\tt arXiv:2107.12112}}].

\bibitem{Vilenkin:1984ib}
A.~Vilenkin, {\it {Cosmic Strings and Domain Walls}},  {\em Phys. Rept.} {\bf
  121} (1985) 263--315.

\bibitem{Vilenkin:1982ks}
A.~Vilenkin and A.~E. Everett, {\it {Cosmic Strings and Domain Walls in Models
  with Goldstone and PseudoGoldstone Bosons}},  {\em Phys. Rev. Lett.} {\bf 48}
  (1982) 1867--1870.

\bibitem{Villadoro}
G.~Grilli~di Cortona, E.~Hardy, J.~Pardo~Vega, and G.~Villadoro, {\it {The QCD
  axion, precisely}},  {\em JHEP} {\bf 01} (2016) 034,
  [\href{http://arxiv.org/abs/1511.02867}{{\tt arXiv:1511.02867}}].

\bibitem{PhysRevLett.60.257}
D.~P. Bennett and F.~m. c.~R. Bouchet, {\it Evidence for a scaling solution in
  cosmic-string evolution},  {\em Phys. Rev. Lett.} {\bf 60} (Jan, 1988)
  257--260.

\bibitem{PhysRevD.40.973}
A.~Albrecht and N.~Turok, {\it Evolution of cosmic string networks},  {\em
  Phys. Rev. D} {\bf 40} (Aug, 1989) 973--1001.

\bibitem{PhysRevLett.64.119}
B.~Allen and E.~P.~S. Shellard, {\it Cosmic-string evolution: A numerical
  simulation},  {\em Phys. Rev. Lett.} {\bf 64} (Jan, 1990) 119--122.

\bibitem{Martins:2018dqg}
C.~J. A.~P. Martins, {\it {Scaling properties of cosmological axion strings}},
  {\em Phys. Lett. B} {\bf 788} (2019) 147--151,
  [\href{http://arxiv.org/abs/1811.12678}{{\tt arXiv:1811.12678}}].

\bibitem{Gorghetto:2018myk}
M.~Gorghetto, E.~Hardy, and G.~Villadoro, {\it {Axions from Strings: the
  Attractive Solution}},  {\em JHEP} {\bf 07} (2018) 151,
  [\href{http://arxiv.org/abs/1806.04677}{{\tt arXiv:1806.04677}}].

\bibitem{Hindmarsh:2019csc}
M.~Hindmarsh, J.~Lizarraga, A.~Lopez-Eiguren, and J.~Urrestilla, {\it {Scaling
  Density of Axion Strings}},  {\em Phys. Rev. Lett.} {\bf 124} (2020), no.~2
  021301, [\href{http://arxiv.org/abs/1908.03522}{{\tt arXiv:1908.03522}}].

\bibitem{PhysRevD.35.1138}
A.~Vilenkin and T.~Vachaspati, {\it Radiation of goldstone bosons from cosmic
  strings},  {\em Phys. Rev. D} {\bf 35} (Feb, 1987) 1138--1140.

\bibitem{Davis:1989nj}
R.~L. Davis and E.~P.~S. Shellard, {\it {DO AXIONS NEED INFLATION?}},  {\em
  Nucl. Phys. B} {\bf 324} (1989) 167--186.

\bibitem{Yamaguchi:1998gx}
M.~Yamaguchi, M.~Kawasaki, and J.~Yokoyama, {\it {Evolution of axionic strings
  and spectrum of axions radiated from them}},  {\em Phys. Rev. Lett.} {\bf 82}
  (1999) 4578--4581, [\href{http://arxiv.org/abs/hep-ph/9811311}{{\tt
  hep-ph/9811311}}].

\bibitem{Hagmann:2000ja}
C.~Hagmann, S.~Chang, and P.~Sikivie, {\it {Axion radiation from strings}},
  {\em Phys. Rev. D} {\bf 63} (2001) 125018,
  [\href{http://arxiv.org/abs/hep-ph/0012361}{{\tt hep-ph/0012361}}].

\bibitem{Gorghetto:2020qws}
M.~Gorghetto, E.~Hardy, and G.~Villadoro, {\it {More axions from strings}},
  {\em SciPost Phys.} {\bf 10} (2021), no.~2 050,
  [\href{http://arxiv.org/abs/2007.04990}{{\tt arXiv:2007.04990}}].

\bibitem{OHare:2021zrq}
C.~A.~J. O'Hare, G.~Pierobon, J.~Redondo, and Y.~Y.~Y. Wong, {\it {Simulations
  of axionlike particles in the postinflationary scenario}},  {\em Phys. Rev.
  D} {\bf 105} (2022), no.~5 055025,
  [\href{http://arxiv.org/abs/2112.05117}{{\tt arXiv:2112.05117}}].

\bibitem{Vachaspati:1984gt}
T.~Vachaspati and A.~Vilenkin, {\it {Gravitational Radiation from Cosmic
  Strings}},  {\em Phys. Rev. D} {\bf 31} (1985) 3052.

\bibitem{Blanco-Pillado:2011egf}
J.~J. Blanco-Pillado, K.~D. Olum, and B.~Shlaer, {\it {Large parallel cosmic
  string simulations: New results on loop production}},  {\em Phys. Rev. D}
  {\bf 83} (2011) 083514, [\href{http://arxiv.org/abs/1101.5173}{{\tt
  arXiv:1101.5173}}].

\bibitem{Blanco-Pillado:2013qja}
J.~J. Blanco-Pillado, K.~D. Olum, and B.~Shlaer, {\it {The number of cosmic
  string loops}},  {\em Phys. Rev. D} {\bf 89} (2014), no.~2 023512,
  [\href{http://arxiv.org/abs/1309.6637}{{\tt arXiv:1309.6637}}].

\bibitem{Ringeval:2005kr}
C.~Ringeval, M.~Sakellariadou, and F.~Bouchet, {\it {Cosmological evolution of
  cosmic string loops}},  {\em JCAP} {\bf 02} (2007) 023,
  [\href{http://arxiv.org/abs/astro-ph/0511646}{{\tt astro-ph/0511646}}].

\bibitem{Lorenz:2010sm}
L.~Lorenz, C.~Ringeval, and M.~Sakellariadou, {\it {Cosmic string loop
  distribution on all length scales and at any redshift}},  {\em JCAP} {\bf 10}
  (2010) 003, [\href{http://arxiv.org/abs/1006.0931}{{\tt arXiv:1006.0931}}].

\bibitem{Gorghetto:2021fsn}
M.~Gorghetto, E.~Hardy, and H.~Nicolaescu, {\it {Observing invisible axions
  with gravitational waves}},  {\em JCAP} {\bf 06} (2021) 034,
  [\href{http://arxiv.org/abs/2101.11007}{{\tt arXiv:2101.11007}}].

\bibitem{Daverio:2015nva}
D.~Daverio, M.~Hindmarsh, M.~Kunz, J.~Lizarraga, and J.~Urrestilla, {\it
  {Energy-momentum correlations for Abelian Higgs cosmic strings}},  {\em Phys.
  Rev. D} {\bf 93} (2016), no.~8 085014,
  [\href{http://arxiv.org/abs/1510.05006}{{\tt arXiv:1510.05006}}]. [Erratum:
  Phys.Rev.D 95, 049903 (2017)].

\bibitem{Hindmarsh:2017qff}
M.~Hindmarsh, J.~Lizarraga, J.~Urrestilla, D.~Daverio, and M.~Kunz, {\it
  {Scaling from gauge and scalar radiation in Abelian Higgs string networks}},
  {\em Phys. Rev. D} {\bf 96} (2017), no.~2 023525,
  [\href{http://arxiv.org/abs/1703.06696}{{\tt arXiv:1703.06696}}].

\bibitem{Barr:1986hs}
S.~M. Barr, K.~Choi, and J.~E. Kim, {\it {Axion Cosmology in Superstring
  Models}},  {\em Nucl. Phys. B} {\bf 283} (1987) 591--604.

\bibitem{Shellard:1986in}
E.~P.~S. Shellard, {\it {AXIONIC DOMAIN WALLS AND COSMOLOGY}},  in {\em {27th
  Liege International Astrophysical Colloquium on Origin and Ea Early History
  of the Universe}}, 1986.

\bibitem{BARR1987591}
S.~Barr, K.~Choi, and J.~E. Kim, {\it Some aspects of axion cosmology in
  unified and superstring models},  {\em Nuclear Physics B} {\bf 283} (1987)
  591--604.

\bibitem{Chang:1998tb}
S.~Chang, C.~Hagmann, and P.~Sikivie, {\it {Studies of the motion and decay of
  axion walls bounded by strings}},  {\em Phys. Rev. D} {\bf 59} (1999) 023505,
  [\href{http://arxiv.org/abs/hep-ph/9807374}{{\tt hep-ph/9807374}}].

\bibitem{Ryden:1989vj}
B.~S. Ryden, W.~H. Press, and D.~N. Spergel, {\it {THE EVOLUTION OF NETWORKS OF
  DOMAIN WALLS AND COSMIC STRINGS}}, .

\bibitem{Hindmarsh:1996xv}
M.~Hindmarsh, {\it {Analytic scaling solutions for cosmic domain walls}},  {\em
  Phys. Rev. Lett.} {\bf 77} (1996) 4495--4498,
  [\href{http://arxiv.org/abs/hep-ph/9605332}{{\tt hep-ph/9605332}}].

\bibitem{Garagounis:2002kt}
T.~Garagounis and M.~Hindmarsh, {\it {Scaling in numerical simulations of
  domain walls}},  {\em Phys. Rev. D} {\bf 68} (2003) 103506,
  [\href{http://arxiv.org/abs/hep-ph/0212359}{{\tt hep-ph/0212359}}].

\bibitem{Oliveira:2004he}
J.~C. R.~E. Oliveira, C.~J. A.~P. Martins, and P.~P. Avelino, {\it {The
  Cosmological evolution of domain wall networks}},  {\em Phys. Rev. D} {\bf
  71} (2005) 083509, [\href{http://arxiv.org/abs/hep-ph/0410356}{{\tt
  hep-ph/0410356}}].

\bibitem{Avelino:2005pe}
P.~P. Avelino, J.~C. R.~E. Oliveira, and C.~J. A.~P. Martins, {\it
  {Understanding domain wall network evolution}},  {\em Phys. Lett. B} {\bf
  610} (2005) 1--8, [\href{http://arxiv.org/abs/hep-th/0503226}{{\tt
  hep-th/0503226}}].

\bibitem{Leite:2011sc}
A.~M.~M. Leite and C.~J. A.~P. Martins, {\it {Scaling Properties of Domain Wall
  Networks}},  {\em Phys. Rev. D} {\bf 84} (2011) 103523,
  [\href{http://arxiv.org/abs/1110.3486}{{\tt arXiv:1110.3486}}].

\bibitem{Vilenkin:1981zs}
A.~Vilenkin, {\it {Gravitational Field of Vacuum Domain Walls and Strings}},
  {\em Phys. Rev. D} {\bf 23} (1981) 852--857.

\bibitem{Coulson:1995nv}
D.~Coulson, Z.~Lalak, and B.~A. Ovrut, {\it {Biased domain walls}},  {\em Phys.
  Rev. D} {\bf 53} (1996) 4237--4246.

\bibitem{Larsson:1996sp}
S.~E. Larsson, S.~Sarkar, and P.~L. White, {\it {Evading the cosmological
  domain wall problem}},  {\em Phys. Rev. D} {\bf 55} (1997) 5129--5135,
  [\href{http://arxiv.org/abs/hep-ph/9608319}{{\tt hep-ph/9608319}}].

\bibitem{Avelino:2008qy}
P.~P. Avelino, C.~J. A.~P. Martins, and L.~Sousa, {\it {Dynamics of Biased
  Domain Walls and the Devaluation Mechanism}},  {\em Phys. Rev. D} {\bf 78}
  (2008) 043521, [\href{http://arxiv.org/abs/0805.4013}{{\tt
  arXiv:0805.4013}}].

\bibitem{Correia:2014kqa}
J.~R. C. C.~C. Correia, I.~S. C.~R. Leite, and C.~J. A.~P. Martins, {\it
  {Effects of Biases in Domain Wall Network Evolution}},  {\em Phys. Rev. D}
  {\bf 90} (2014), no.~2 023521, [\href{http://arxiv.org/abs/1407.3905}{{\tt
  arXiv:1407.3905}}].

\bibitem{Correia:2018tty}
J.~R. C. C.~C. Correia, I.~S. C.~R. Leite, and C.~J. A.~P. Martins, {\it
  {Effects of biases in domain wall network evolution. II. Quantitative
  analysis}},  {\em Phys. Rev. D} {\bf 97} (2018), no.~8 083521,
  [\href{http://arxiv.org/abs/1804.10761}{{\tt arXiv:1804.10761}}].

\bibitem{STAUFFER19791}
D.~Stauffer, {\it Scaling theory of percolation clusters},  {\em Physics
  Reports} {\bf 54} (1979), no.~1 1--74.

\bibitem{Gelmini:1988sf}
G.~B. Gelmini, M.~Gleiser, and E.~W. Kolb, {\it {Cosmology of Biased Discrete
  Symmetry Breaking}},  {\em Phys. Rev. D} {\bf 39} (1989) 1558.

\bibitem{Lalak:1992px}
Z.~Lalak and B.~A. Ovrut, {\it {Domain walls, percolation theory and Abell
  clusters}},  {\em Phys. Rev. Lett.} {\bf 71} (1993) 951--954.

\bibitem{Kawasaki:2014sqa}
M.~Kawasaki, K.~Saikawa, and T.~Sekiguchi, {\it {Axion dark matter from
  topological defects}},  {\em Phys. Rev. D} {\bf 91} (2015), no.~6 065014,
  [\href{http://arxiv.org/abs/1412.0789}{{\tt arXiv:1412.0789}}].

\bibitem{Martins:2016ois}
C.~J. A.~P. Martins, I.~Y. Rybak, A.~Avgoustidis, and E.~P.~S. Shellard, {\it
  {Extending the velocity-dependent one-scale model for domain walls}},  {\em
  Phys. Rev. D} {\bf 93} (2016), no.~4 043534,
  [\href{http://arxiv.org/abs/1602.01322}{{\tt arXiv:1602.01322}}].

\bibitem{Vilenkin:1991zk}
A.~Vilenkin, {\it {Cosmic string dynamics with friction}},  {\em Phys. Rev. D}
  {\bf 43} (1991) 1060--1062.

\bibitem{Calogeracos:1999yp}
A.~Calogeracos and N.~Dombey, {\it {History and physics of the Klein paradox}},
   {\em Contemp. Phys.} {\bf 40} (1999) 313--321,
  [\href{http://arxiv.org/abs/quant-ph/9905076}{{\tt quant-ph/9905076}}].

\bibitem{Cadamuro:2011fd}
D.~Cadamuro and J.~Redondo, {\it {Cosmological bounds on pseudo Nambu-Goldstone
  bosons}},  {\em JCAP} {\bf 02} (2012) 032,
  [\href{http://arxiv.org/abs/1110.2895}{{\tt arXiv:1110.2895}}].

\bibitem{Jaeckel:2015jla}
J.~Jaeckel and M.~Spannowsky, {\it {Probing MeV to 90 GeV axion-like particles
  with LEP and LHC}},  {\em Phys. Lett. B} {\bf 753} (2016) 482--487,
  [\href{http://arxiv.org/abs/1509.00476}{{\tt arXiv:1509.00476}}].

\bibitem{Agrawal_2018}
P.~Agrawal and K.~Howe, {\it Factoring the strong {CP} problem},  {\em Journal
  of High Energy Physics} {\bf 2018} (dec, 2018).

\bibitem{Workman:2022}
{\bf Particle Data Group}, R.~Workman et~al., {\it {Review of Particle
  Physics}}, . to be published (2022).

\bibitem{Massarotti:1990qs}
A.~Massarotti, {\it {Evolution of light domain walls interacting with dark
  matter. 1.}},  {\em Phys. Rev. D} {\bf 43} (1991) 346--352.

\bibitem{Press:1989yh}
W.~H. Press, B.~S. Ryden, and D.~N. Spergel, {\it {Dynamical Evolution of
  Domain Walls in an Expanding Universe}},  {\em Astrophys. J.} {\bf 347}
  (1989) 590--604.

\bibitem{Burke-Spolaor:2018bvk}
S.~Burke-Spolaor et~al., {\it {The Astrophysics of Nanohertz Gravitational
  Waves}},  {\em Astron. Astrophys. Rev.} {\bf 27} (2019), no.~1 5,
  [\href{http://arxiv.org/abs/1811.08826}{{\tt arXiv:1811.08826}}].

\bibitem{Blasi:2020mfx}
S.~Blasi, V.~Brdar, and K.~Schmitz, {\it {Has NANOGrav found first evidence for
  cosmic strings?}},  {\em Phys. Rev. Lett.} {\bf 126} (2021), no.~4 041305,
  [\href{http://arxiv.org/abs/2009.06607}{{\tt arXiv:2009.06607}}].

\bibitem{Ellis:2020ena}
J.~Ellis and M.~Lewicki, {\it {Cosmic String Interpretation of NANOGrav Pulsar
  Timing Data}},  {\em Phys. Rev. Lett.} {\bf 126} (2021), no.~4 041304,
  [\href{http://arxiv.org/abs/2009.06555}{{\tt arXiv:2009.06555}}].

\bibitem{Blanco-Pillado:2021ygr}
J.~J. Blanco-Pillado, K.~D. Olum, and J.~M. Wachter, {\it {Comparison of cosmic
  string and superstring models to NANOGrav 12.5-year results}},  {\em Phys.
  Rev. D} {\bf 103} (2021), no.~10 103512,
  [\href{http://arxiv.org/abs/2102.08194}{{\tt arXiv:2102.08194}}].

\bibitem{Buchmuller:2020lbh}
W.~Buchmuller, V.~Domcke, and K.~Schmitz, {\it {From NANOGrav to LIGO with
  metastable cosmic strings}},  {\em Phys. Lett. B} {\bf 811} (2020) 135914,
  [\href{http://arxiv.org/abs/2009.10649}{{\tt arXiv:2009.10649}}].

\bibitem{Bian:2022tju}
L.~Bian, J.~Shu, B.~Wang, Q.~Yuan, and J.~Zong, {\it {Searching for cosmic
  string induced stochastic gravitational wave background with the Parkes
  Pulsar Timing Array}},  \href{http://arxiv.org/abs/2205.07293}{{\tt
  arXiv:2205.07293}}.

\bibitem{Chen:2022azo}
Z.-C. Chen, Y.-M. Wu, and Q.-G. Huang, {\it {Search for the Gravitational-wave
  Background from Cosmic Strings with the Parkes Pulsar Timing Array Second
  Data Release}},  {\em Astrophys. J.} {\bf 936} (2022), no.~1 20,
  [\href{http://arxiv.org/abs/2205.07194}{{\tt arXiv:2205.07194}}].

\bibitem{Bian:2020urb}
L.~Bian, R.-G. Cai, J.~Liu, X.-Y. Yang, and R.~Zhou, {\it {Evidence for
  different gravitational-wave sources in the NANOGrav dataset}},  {\em Phys.
  Rev. D} {\bf 103} (2021), no.~8 L081301,
  [\href{http://arxiv.org/abs/2009.13893}{{\tt arXiv:2009.13893}}].

\bibitem{Wang:2022rjz}
D.~Wang, {\it {Novel Physics with International Pulsar Timing Array: Axionlike
  Particles, Domain Walls and Cosmic Strings}},
  \href{http://arxiv.org/abs/2203.10959}{{\tt arXiv:2203.10959}}.

\bibitem{Sakharov:2021dim}
A.~S. Sakharov, Y.~N. Eroshenko, and S.~G. Rubin, {\it {Looking at the NANOGrav
  signal through the anthropic window of axionlike particles}},  {\em Phys.
  Rev. D} {\bf 104} (2021), no.~4 043005,
  [\href{http://arxiv.org/abs/2104.08750}{{\tt arXiv:2104.08750}}].

\bibitem{Schmitz:2020syl}
K.~Schmitz, {\it {New Sensitivity Curves for Gravitational-Wave Signals from
  Cosmological Phase Transitions}},  {\em JHEP} {\bf 01} (2021) 097,
  [\href{http://arxiv.org/abs/2002.04615}{{\tt arXiv:2002.04615}}].

\bibitem{Avelino:2005kn}
P.~P. Avelino, C.~J. A.~P. Martins, and J.~C. R.~E. Oliveira, {\it {One-scale
  model for domain wall network evolution}},  {\em Phys. Rev. D} {\bf 72}
  (2005) 083506, [\href{http://arxiv.org/abs/hep-ph/0507272}{{\tt
  hep-ph/0507272}}].

\bibitem{Avelino:2019wqd}
P.~P. Avelino, {\it {Parameter-free velocity-dependent one-scale model for
  domain walls}},  {\em Phys. Rev. D} {\bf 101} (2020), no.~2 023514,
  [\href{http://arxiv.org/abs/1910.07011}{{\tt arXiv:1910.07011}}].

\bibitem{Avelino:2020ubr}
P.~P. Avelino, {\it {Comparing parametric and non-parametric velocity-dependent
  one-scale models for domain wall evolution}},  {\em JCAP} {\bf 04} (2020)
  012, [\href{http://arxiv.org/abs/2001.06318}{{\tt arXiv:2001.06318}}].

\bibitem{Avelino:2022zem}
P.~P. Avelino, D.~Gr\"uber, and L.~Sousa, {\it {Analytical scaling solutions
  for the evolution of cosmic domain walls in a parameter-free
  velocity-dependent one-scale model}},
  \href{http://arxiv.org/abs/2203.16173}{{\tt arXiv:2203.16173}}.

\bibitem{Martins:2016lzc}
C.~J. A.~P. Martins, I.~Y. Rybak, A.~Avgoustidis, and E.~P.~S. Shellard, {\it
  {Stretching and Kibble scaling regimes for Hubble-damped defect networks}},
  {\em Phys. Rev. D} {\bf 94} (2016), no.~11 116017,
  [\href{http://arxiv.org/abs/1612.08863}{{\tt arXiv:1612.08863}}]. [Erratum:
  Phys.Rev.D 95, 039902 (2017)].

\bibitem{Punturo:2010zz}
M.~Punturo et~al., {\it {The Einstein Telescope: A third-generation
  gravitational wave observatory}},  {\em Class. Quant. Grav.} {\bf 27} (2010)
  194002.

\bibitem{Hild:2010id}
S.~Hild et~al., {\it {Sensitivity Studies for Third-Generation Gravitational
  Wave Observatories}},  {\em Class. Quant. Grav.} {\bf 28} (2011) 094013,
  [\href{http://arxiv.org/abs/1012.0908}{{\tt arXiv:1012.0908}}].

\bibitem{Sathyaprakash:2012jk}
B.~Sathyaprakash et~al., {\it {Scientific Objectives of Einstein Telescope}},
  {\em Class. Quant. Grav.} {\bf 29} (2012) 124013,
  [\href{http://arxiv.org/abs/1206.0331}{{\tt arXiv:1206.0331}}]. [Erratum:
  Class.Quant.Grav. 30, 079501 (2013)].

\bibitem{Maggiore:2019uih}
M.~Maggiore et~al., {\it {Science Case for the Einstein Telescope}},  {\em
  JCAP} {\bf 03} (2020) 050, [\href{http://arxiv.org/abs/1912.02622}{{\tt
  arXiv:1912.02622}}].

\bibitem{LISA:2017pwj}
{\bf LISA}, P.~Amaro-Seoane et~al., {\it {Laser Interferometer Space Antenna}},
   \href{http://arxiv.org/abs/1702.00786}{{\tt arXiv:1702.00786}}.

\bibitem{Baker:2019nia}
J.~Baker et~al., {\it {The Laser Interferometer Space Antenna: Unveiling the
  Millihertz Gravitational Wave Sky}},
  \href{http://arxiv.org/abs/1907.06482}{{\tt arXiv:1907.06482}}.

\bibitem{Corbin:2005ny}
V.~Corbin and N.~J. Cornish, {\it {Detecting the cosmic gravitational wave
  background with the big bang observer}},  {\em Class. Quant. Grav.} {\bf 23}
  (2006) 2435--2446, [\href{http://arxiv.org/abs/gr-qc/0512039}{{\tt
  gr-qc/0512039}}].

\bibitem{Harry:2006fi}
G.~M. Harry, P.~Fritschel, D.~A. Shaddock, W.~Folkner, and E.~S. Phinney, {\it
  {Laser interferometry for the big bang observer}},  {\em Class. Quant. Grav.}
  {\bf 23} (2006) 4887--4894. [Erratum: Class.Quant.Grav. 23, 7361 (2006)].

\bibitem{Janssen:2014dka}
G.~Janssen et~al., {\it {Gravitational wave astronomy with the SKA}},  {\em
  PoS} {\bf AASKA14} (2015) 037, [\href{http://arxiv.org/abs/1501.00127}{{\tt
  arXiv:1501.00127}}].

\bibitem{Harry:2010zz}
{\bf LIGO Scientific}, G.~M. Harry, {\it {Advanced LIGO: The next generation of
  gravitational wave detectors}},  {\em Class. Quant. Grav.} {\bf 27} (2010)
  084006.

\bibitem{LIGOScientific:2014pky}
{\bf LIGO Scientific}, J.~Aasi et~al., {\it {Advanced LIGO}},  {\em Class.
  Quant. Grav.} {\bf 32} (2015) 074001,
  [\href{http://arxiv.org/abs/1411.4547}{{\tt arXiv:1411.4547}}].

\bibitem{VIRGO:2014yos}
{\bf VIRGO}, F.~Acernese et~al., {\it {Advanced Virgo: a second-generation
  interferometric gravitational wave detector}},  {\em Class. Quant. Grav.}
  {\bf 32} (2015), no.~2 024001, [\href{http://arxiv.org/abs/1408.3978}{{\tt
  arXiv:1408.3978}}].

\bibitem{LIGO-Virgo}
L.~Barsotti, L.~McCuller, M.~Evans, and P.~Fritschel.
  \url{https://dcc.ligo.org/LIGO-T1800042/public}.

\bibitem{Hiramatsu:2010yz}
T.~Hiramatsu, M.~Kawasaki, and K.~Saikawa, {\it {Gravitational Waves from
  Collapsing Domain Walls}},  {\em JCAP} {\bf 05} (2010) 032,
  [\href{http://arxiv.org/abs/1002.1555}{{\tt arXiv:1002.1555}}].

\bibitem{KAGRA:2021kbb}
{\bf KAGRA, Virgo, LIGO Scientific}, R.~Abbott et~al., {\it {Upper limits on
  the isotropic gravitational-wave background from Advanced LIGO and Advanced
  Virgo\textquoteright{}s third observing run}},  {\em Phys. Rev. D} {\bf 104}
  (2021), no.~2 022004, [\href{http://arxiv.org/abs/2101.12130}{{\tt
  arXiv:2101.12130}}].

\bibitem{cma/1418919772}
V.~Rabinovich and F.~U. Altamirano, {\it {Application of the SPPS Method to the
  One-dimensional Quantum Scattering}},  {\em Communications in Mathematical
  Analysis} {\bf 17} (2014), no.~2 295 -- 310.

\bibitem{Campanelli:2004si}
L.~Campanelli, {\it {Scattering of Dirac and Majorana fermions off domain
  walls}},  {\em Phys. Rev. D} {\bf 70} (2004) 116008,
  [\href{http://arxiv.org/abs/hep-ph/0408078}{{\tt hep-ph/0408078}}].

\end{thebibliography}\endgroup


\end{document}